\documentclass[12pt,preprint]{aastex}
\newcommand{\bdv}[1]{\mbox{\boldmath$#1$}}

\def\au{{\rm au}}

\def\masyr{{\rm mas}\,{\rm yr}^{-1}}
\def\kpc{{\rm kpc}}
\def\mas{{\rm mas}}

\def\muas{\mu{\rm as}}

\def\max{{\rm max}}

\def\rel{{\rm rel}}

\def\eff{{\rm eff}}

\def\hel{{\rm hel}}

\def\e{{\rm E}}
\def\bpi{{\bdv\pi}}
\def\bmu{{\bdv\mu}}

\def\bgamma{{\bdv\gamma}}

\begin{document}
\title{Systematic KMTNet Planetary Anomaly Search. VIII.  Complete Sample of 2019 Subprime Field Planets}

\author{
Youn Kil Jung$^{1,2}$, 
Weicheng Zang$^{3,4}$,
Hanyue Wang$^{4}$,
Cheongho Han$^{5}$, 
Andrew Gould$^{6,7}$, 
Andrzej Udalski$^{8}$,
(Lead Authors)\\
Michael D. Albrow$^{9}$, 
Sun-Ju Chung$^{1,4}$, 
Kyu-Ha Hwang$^{1}$, 
Yoon-Hyun Ryu$^{1}$, 
In-Gu Shin$^{4}$, 
Yossi Shvartzvald$^{10}$, 
Hongjing Yang$^{3}$, 
Jennifer C. Yee$^{4}$, 
Sang-Mok Cha$^{1,11}$, 
Dong-Jin Kim$^{1}$,
Seung-Lee Kim$^{1}$, 
Chung-Uk Lee$^{1}$,
Dong-Joo Lee$^{1}$,
Yongseok Lee$^{1,11}$, 
Byeong-Gon Park$^{1,2}$, 
Richard W. Pogge$^{7}$ \\
(The KMTNet Collaboration)\\
Przemek Mr{\'o}z$^{8}$,
Micha{\l} K. Szyma{\'n}ski$^{8}$,
Jan Skowron$^{8}$,
Radek Poleski$^{8}$,
Igor Soszy{\'n}ski$^{8}$,  
Pawe{\l} Pietrukowicz$^{8}$,
Szymon Koz{\l}owski$^{8}$,
Krzysztof Ulaczyk$^{12}$,
Krzysztof A. Rybicki$^{8,10}$,
Patryk Iwanek$^{8}$,
Marcin Wrona$^{8}$\\
(The OGLE Collaboration)\\
Grant Christie$^{13}$,
Jonathan Green$^{14}$,
Steve Hennerley$^{14}$,
Andrew Marmont$^{14}$,
Shude Mao$^{3}$,
Dan Maoz$^{15}$,
Jennie McCormick$^{16}$,
Tim Natusch$^{13,17}$,
Matthew T.~Penny$^{18}$,
Ian Porritt$^{19}$,
Wei Zhu$^{3}$\\
(The Tsinghua and $\mu$FUN Follow-Up Teams)\\
}

%----------------------------------------------------------------
\affil{$^{1}$Korea Astronomy and Space Science Institute, Daejon
34055, Republic of Korea}

\affil{$^{2}$Korea University of Science and Technology, Korea, 
(UST), 217 Gajeong-ro, Yuseong-gu, Daejeon, 34113, Republic of Korea}

\affil{$^{3}$ Department of Astronomy,
Tsinghua University, Beijing 100084, China}

\affil{$^{4}$ Center for Astrophysics $|$ Harvard \& Smithsonian, 60 Garden
St., Cambridge, MA 02138, USA}

\affil{$^{5}$Department of Physics, Chungbuk National University,
Cheongju 28644, Republic of Korea}

\affil{$^{6}$Max-Planck-Institute for Astronomy, K\"{o}nigstuhl 17,
69117 Heidelberg, Germany}

\affil{$^{7}$Department of Astronomy, Ohio State University, 140 W.
18th Ave., Columbus, OH 43210, USA}

\affil{$^{8}$Astronomical Observatory, University of Warsaw, 
Al.~Ujazdowskie~4, 00-478~Warszawa, Poland}

\affil{$^{9}$University of Canterbury, Department of Physics and
Astronomy, Private Bag 4800, Christchurch 8020, New Zealand}

\affil{$^{10}$Department of Particle Physics and Astrophysics, 
Weizmann Institute of Science, Rehovot 76100, Israel}

\affil{$^{11}$School of Space Research, Kyung Hee University,
Yongin, Kyeonggi 17104, Republic of Korea}

\affil{$^{12}$Department of Physics, University of Warwick, 
Gibbet Hill Road, Coventry, CV4~7AL,~UK}

\affil{$^{13}$Auckland Observatory, Auckland, New Zealand}

\affil{$^{14}$Kumeu Observatory, Kumeu, New Zealand}

\affil{$^{15}$School of Physics and Astronomy, Tel-Aviv University,
Tel-Aviv 6997801, Israel}

\affil{$^{16}$Farm Cove Observatory, Centre for Backyard Astrophysics,
Pakuranga, Auckland, New Zealand}

\affil{$^{17}$Institute for Radio Astronomy and Space Research (IRASR),
AUT University, Auckland, New Zealand}

\affil{$^{18}$Department of Physics and Astronomy, Louisiana State University,
Baton Rouge, LA 70803, USA}

\affil{$^{19}$Turitea Observatory, Palmerston, New Zealand}

\begin{abstract}

We complete the publication of all microlensing planets (and
``possible planets'') identified by the uniform approach of the
KMT AnomalyFinder system in the
21 KMT subprime fields during the 2019 observing season, namely
KMT-2019-BLG-0298,
KMT-2019-BLG-1216,
KMT-2019-BLG-2783,
OGLE-2019-BLG-0249, and
OGLE-2019-BLG-0679 (planets),
as well as
OGLE-2019-BLG-0344, and
KMT-2019-BLG-0304 (possible planets).  The five planets have 
mean log mass-ratio measurements of $(-2.6,-3.6,-2.5,-2.2,-2.3)$, 
median mass estimates of $(1.81,0.094,1.16,7.12,3.34)\, M_{\rm Jup}$, 
and median distance estimates of $(6.7,2.7,5.9,6.4,5.6)\,\kpc$, respectively. 
The main scientific interest of these planets is that they complete the AnomalyFinder 
sample for 2019, which has a total of 25 planets that are likely to enter
the statistical sample.  We find statistical consistency with the previously
published 33 planets from the 2018 AnomalyFinder analysis according to
an ensemble of five tests.  Of the 58 planets from 2018-2019, 23 were newly
discovered by AnomalyFinder.  Within statistical precision, half
of all the planets have caustic crossings while half do not
(as predicted by \citealt{zhu14}), an equal number of detected planets
result from major-image and minor-image light-curve perturbations, and
an equal number come from KMT prime fields versus subprime fields.

\end{abstract}


\section{{Introduction}
\label{sec:intro}}

We present the analysis of all planetary events that were
identified by the KMTNet AnomalyFinder algorithm \citep{ob191053,af2}
and occurred during the 2019 season within the 21 subprime KMTNet fields,
covering $\sim 84\,{\rm deg}^2$ that lie in the periphery of the richest
microlensing region of the Galactic bulge, and which are observed with
cadences $\Gamma=0.2$--$1\,{\rm hr}^{-1}$.
This work follows the publications of
complete samples of the 2018 prime \citep{ob180383,kb190253,2018prime},
and subprime \citep{2018subprime} AnomalyFinder events, 
the 2019 prime \citep{ob191053,kb190253,af2} events,
as well as a complete sample of all events from 2016-2019 with
planet-host mass ratios $q<10^{-4}$ \citep{logqlt-4}.
%There are also 6 prime fields that contiguously cover $\sim 13\,{\rm deg}^2$ of the Galactic bulge with cadences $\Gamma=2$--$4\,{\rm hr}^{-1}$.
The above references are
(ignoring duplicates) Papers I, IV, II, III, V, VI, and VII, in the
AnomalyFinder series.  The locations
and cadences of the KMTNet fields are shown in Figure~12 of \citet{eventfinder}.
Our immediate goal, which we expect to achieve within a year, is to publish
all AnomalyFinder planets from 2016-2019.  Over the longer term, we plan
to apply AnomalyFinder to all subsequent KMT seasons, beginning 2021.

%2019 subprime has 1895 events
For the 2019 subprime fields, the AnomalyFinder %\citep{af2}
identified a total of 182 anomalous events (from an underlying sample of 1895
events), which were classified as
``planet'' (9), ``planet/binary'' (10), ``binary/planet'' (18),
``binary'' (136), and ``finite source'' (9).
Among the 136 in the ``binary'' classification, 56 were judged by
eye to be unambiguously non-planetary in nature.  Among the 9 in the
``planet'' classification, 4 were previously published (including 2
AnomalyFinder
discoveries), while one had been recognized but remained unpublished.
Among the 10 in the ``planet/binary'' classification,
3 were either published planets (1) or had been recognized by eye (2),
and among the 18 in the ``binary/planet'' classification, 
one was a previously published planet.  Among the 136
classified as ``binary'', 1 (a two-planet system) was published.
%Among the 9 classified as ``finite source'', 1 (KMT-2019-BLG-0414) was previously published as a ``possible planet'' because there is an alternate, orbiting binary-source (xallarap) solution that is disfavored by only $\Delta\chi^2=4$ \citep{kb190414}.
Thus, in total, the AnomalyFinder recovered 9 planets that had been previously
found by eye, including 6 that were published
and 3 others that had not been published.  The latter are
KMT-2019-BLG-1216, OGLE-2019-BLG-0249, and OGLE-2019-BLG-0679.
% recovered, published here
%1216, 0679/2688, 0249/0109 = planet, planet/binary, planet/binary

Our overall goal is to present full analyses of all events with mass ratios
$q<0.03$.  To this end, we carry out systematic investigations of all
of the AnomalyFinder candidates (other than the 56 classified by eye
as unambiguously non-planetary) using end-of-season pipeline data.
Any (unpublished) event that is found to have a viable solution
with $q<0.06$ is then reanalyzed based on tender loving care (TLC)
rereductions.  If there are viable planetary solutions ($q<0.03$),
then we report a detailed analysis regardless of whether the planetary
interpretation is decisively favored.  If the TLC analysis leads to
viable solutions with $0.03<q<0.05$ then we report briefly on the analysis
but do not present all details.  In the 2019 subprime sample, there
was one such event, KMT-2019-BLG-0967, with $q=0.040\pm 0.004$.  This
event also has a competing solution in which the anomaly is generated
by a binary source rather than a low-mass companion, so it could not be
included in the final sample even if the sample boundary were moved upward.
Finally, we note that one event, OGLE-2019-BLG-1352, that was selected 
as ``finite source'' and hence required detailed investigation, may
well have strong evidence of a planet based on extensive follow-up data.
However, we find that, even if so, the signal from survey-only data
is not strong enough to claim a planet.  Hence, we do not include it
in the present paper.

\section{{Observations}
\label{sec:obs}}

The description of the observations is nearly identical to that in 
\citet{2018prime} and \citet{2018subprime}.
The KMTNet data are taken from three identical 1.6m telescopes,
each equipped with cameras of 4 deg$^2$ \citep{kmtnet} 
and located in Australia (KMTA),
Chile (KMTC), and South Africa (KMTS).  When available, our general policy is
to include Optical Gravitational Lensing Experiment (OGLE) and 
Microlensing Observations in Astrophysics (MOA) data in the analysis.
However, none of the 7 events analyzed here were alerted by MOA.
OGLE data were taken using their 1.3m telescope
with $1.4\,{\rm deg}^2$ field of view at Las Campanas Observatory in Chile
\citep{ogle-iv}.
For the light-curve analysis, we use only the $I$-band data.

As in those papers, Table~\ref{tab:names}
gives basic
observational information about each event.  Column~1 gives the
event names in the order of discovery (if discovered by multiple teams),
which enables cross identification.  The nominal cadences are given in column 2,
and column 3 shows the first discovery date.  The remaining four columns
show the event coordinates in the equatorial and galactic systems.
Events with OGLE names were originally discovered by the OGLE Early
Warning System \citep{ews1,ews2}.  Events with KMT names and discovery dates
were first found by the KMT AlertFinder system \citep{alertfinder},
while those listed as ``post-season'' were found by the KMT
EventFinder system \citep{eventfinder}.

Two events, OGLE-2019-BLG-0249 and OGLE-2019-BLG-0679, were observed by
{\it Spitzer} as part of a large-scale microlensing program \citep{yee15},
but these data will be analyzed elsewhere.  As is generally the case,
these {\it Spitzer} observations were supported by ground-based observations.
In the case of OGLE-2019-BLG-0679, these observations consisted of 21 epochs,
spread over 72 days,
of $I/H$ observations on the ANDICAM camera at the SMARTS 1.3m telescope in
Chile, whose aim was to determine the source color.  We
will make use of these observations only for that purpose, i.e., not for the
modeling.  On the other hand, OGLE-2019-BLG-0249 was the object of intensive
ground-based observations.  These require special handling,
as described in Section~\ref{sec:survey+followup-event}.
To the best of our knowledge, there were
no other ground-based follow-up observations of any of these events.

The KMT and OGLE data were reduced using difference image analysis 
\citep{tomaney96,alard98},
as implemented by each group, i.e., \citet{albrow09} and
\citet{wozniak2000}, respectively.  

\section{{Light Curve Analysis}
\label{sec:anal}}

\subsection{{Preamble}
\label{sec:anal-preamble}}

With one exception that is explicitly noted below, we reproduce here
Section~3.1 of \citet{2018subprime}, which describes
the common features of the light-curve analysis.  We do so
(rather than simply referencing that paper) to provide easy access to
the formulae and variable names used throughout this paper.
The reader who is interested in more details should consult
Section~3.1 of \citet{2018prime}.  Readers who are already familiar
with these previous works can skip this section, after first reviewing
the paragraph containing Equation~(\ref{eqn:probmu2}), below.

All of the events can be initially approximated by 1L1S models, which
are specified by three \citet{pac86}
parameters, $(t_0,u_0,t_\e)$, i.e., the time of lens-source closest
approach, the impact parameter in units of $\theta_\e$ and the Einstein
timescale,
\begin{equation}
t_\e = {\theta_\e\over\mu_\rel}; \qquad
\theta_\e = \sqrt{\kappa M\pi_\rel}; \qquad
\kappa\equiv {4\,G\over c^2\,\au} \simeq 8.14\,{\mas\over M_\odot},
\label{eqn:tedef}
\end{equation}
where $M$ is the lens mass, $\pi_\rel$ and $\bmu_\rel$ are the
lens-source relative parallax and proper-motion, respectively,
and $\mu_\rel \equiv |\bmu_\rel|$.  The notation ``$n$L$m$S'' means $n$ lenses
and $m$ sources.   In addition to these 3 non-linear parameters, there are
2 flux parameters, $(f_S,f_B)$, that are required for each observatory,
representing the source flux and the blended flux.

We then search for ``static'' 2L1S solutions, which generally require 4
additional parameters $(s,q,\alpha,\rho)$, i.e., the planet-host separation
in units of $\theta_\e$, the planet-host mass ratio, the angle of the
source trajectory relative to the binary axis, and the angular source
size normalized to $\theta_\e$, i.e., $\rho=\theta_*/\theta_\e$.

We first conduct a grid search with $(s,q)$ held fixed at a grid of values
and the remaining 5 parameters allowed to vary in a Monte Carlo Markov chain
(MCMC). After we identify one or more local minima, we refine these by
allowing all 7 parameters to vary.

We often make use of the   heuristic analysis introduced by \citet{kb190253}
and modified by \citet{kb211391} based on further investigation in 
\citet{2018prime}.
If a brief anomaly at $t_{\rm anom}$
is treated as due to the source crossing the planet-host axis,
then one can estimate two relevant parameters
\begin{equation}
s^\dagger_\pm = {\sqrt{4 + u_{\rm anom}^2}\pm u_{\rm anom}\over 2}; \quad
\tan\alpha ={u_0\over\tau_{\rm anom}},
\label{eqn:heuristic}
\end{equation}
where $u_{\rm anom}^2= \tau_{\rm anom}^2 + u_0^2$ and
$\tau_{\rm anom} = (t_{\rm anom}-t_0)/t_\e$.
Usually,
$s^\dagger_+>1$ corresponds to anomalous bumps and
$s^\dagger_-<1$ corresponds to anomalous dips.   
This formalism predicts that if there are two degenerate solutions, $s_{\pm}$,
then they both have the same $\alpha$ and that there exists a $\Delta\ln s$
such that
\begin{equation}
s_\pm = s^\dagger_{\rm pred}\exp(\pm \Delta \ln s),
\label{eqn:heuristic2}
\end{equation}
where $\alpha$ and $s^\dagger$ are given by Equation~(\ref{eqn:heuristic}).
To test this prediction in individual cases, we can compare the purely
empirical quantity $s^\dagger\equiv \sqrt{s_+ s_-}$ with prediction
from Equation~(\ref{eqn:heuristic}), which we always label with a subscript,
i.e., either $s^\dagger_+$ or $s^\dagger_-$.  This formalism can also be used
to find ``missing solutions'' that have been missed in the grid search,
as was done, e.g., for the case of KMT-2021-BLG-1391 \citep{kb211391}.

For cases in which the anomaly is a dip, the mass ratio $q$ can be estimated,
\begin{equation}
q = \biggl({\Delta t_{\rm dip}\over 4\, t_\e}\biggr)^2
{s^\dagger\over |u_0|}|\sin^3\alpha|,
\label{eqn:qeval}
\end{equation}
where $\Delta t_{\rm dip}$ is the full duration of the dip.  
In some cases, we investigate whether the microlens parallax vector,
\begin{equation}
\bpi_\e\equiv {\pi_\rel\over \theta_\e}\,{\bmu_\rel\over\mu_\rel}
\label{eqn:piedef}
\end{equation}
can be constrained by the data.  When both $\pi_\e$ and $\theta_\e$ are
measured, they can be combined to yield,
\begin{equation}
M = {\theta_\e\over\kappa\pi_\e}; \qquad
D_L = {\au\over \theta_\e\pi_\e + \pi_S},
\label{eqn:mpirel}
\end{equation}
where $D_L$ is the distance to the lens and
$\pi_S$ is the parallax of the source.

To model the parallax effects due to Earth's orbital motion, we add
two parameters $(\pi_{\e,N},\pi_{\e,E})$, which are the components of
$\bpi_\e$ in equatorial coordinates.  We also add (at least initially)
two parameters $\bgamma =[(ds/dt)/s,d\alpha/dt]$, where
$s\bgamma$ are the first derivatives of projected lens orbital position 
at $t_0$, i.e., parallel and perpendicular to the projected separation of the
planet at that time, respectively.
In order to eliminate unphysical solutions, we impose a  constraint
on the ratio of the transverse kinetic to potential energy,
\begin{equation}
\beta \equiv \bigg|{\rm KE\over PE}\bigg|
= {\kappa M_\odot {\rm yr}^2\over 8\pi^2}\,{\pi_\e\over\theta_\e}\gamma^2
\biggr({s\over \pi_\e + \pi_S/\theta_\e}\biggr)^3 < 0.8 .
\label{eqn:betadef}
\end{equation}
It often happens that $\bgamma$ is neither significantly constrained
nor significantly correlated with $\bpi_\e$.  In these cases, we suppress
these two degrees of freedom.

Particularly if there are no sharp caustic-crossing features in the light curve,
2L1S events can be mimicked by 1L2S events.  Where relevant, we test for
such solutions by adding at least 3 parameters 
$(t_{0,2},u_{0,2},q_F)$ to the 1L1S models.
These are the time of closest approach and impact parameter of the
second source and the ratio of the second to the first source flux
in the $I$-band. 
If either lens-source approach can be interpreted
as exhibiting finite source effects, then we must add one or two further
parameters, i.e., $\rho_1$ and/or $\rho_2$.  And, if the two sources
are projected closely enough on the sky, one must also consider
source orbital motion.

In a few cases, we make kinematic arguments that solutions are unlikely
because their inferred proper motions $\mu_\rel$ are too small.   If planetary
events (or, more generally, anomalous events with planet-like signatures)
traced the overall population of microlensing events, then the fraction with
proper motions less than a given $\mu_\rel\ll \sigma_\mu$ would be, 
\begin{equation}
p(\leq\mu_\rel) ={(\mu_\rel/\sigma_\mu)^3\over 6\sqrt{\pi}}
\rightarrow 4\times 10^{-3}\biggl({\mu_\rel\over1\,\masyr}\biggr)^3
\qquad {\rm (old)},
\label{eqn:probmu}
\end{equation}
where (following \citealt{gould21})
the bulge proper motions are approximated as an isotropic Gaussian
with dispersion $\sigma_\mu = 2.9\,\masyr$.

However, subsequent to the work of \citet{2018prime} and \citet{2018subprime},
\citet{masada} showed that the proper-motion distribution of observed
planetary microlensing events scales
$\propto d\mu\,\mu^\nu\exp(-(\mu/2\sigma_\mu)^2)$
where $\sigma_\mu=3.06\pm 0.29\,\masyr$ and $\nu=1.02\pm0.29$.  Hence,
in place of Equation~(\ref{eqn:probmu}), we adopt
\begin{equation}
p(\leq\mu_\rel) ={(\mu_\rel/2\sigma_\mu)^{\nu+1}\over [(\nu+1)/2]!}
\rightarrow {\mu_\rel^2\over 4\sigma_\mu^2}  
\rightarrow 2.8\times 10^{-2}\biggl({\mu_\rel\over1\,\masyr}\biggr)^2,
\label{eqn:probmu2}
\end{equation}
where we have evaluated at $\sigma_\mu=3.0\,\masyr$ and $\nu=1$.
For example, $p(\leq 0.5\,\masyr) = 0.7\%$ and
$p(\leq 0.1\,\masyr) = 0.03\%$.

\subsection{{KMT-2019-BLG-0298}% ob190445
  \label{sec:anal-kb190298}}
\begin{figure}
\plotone{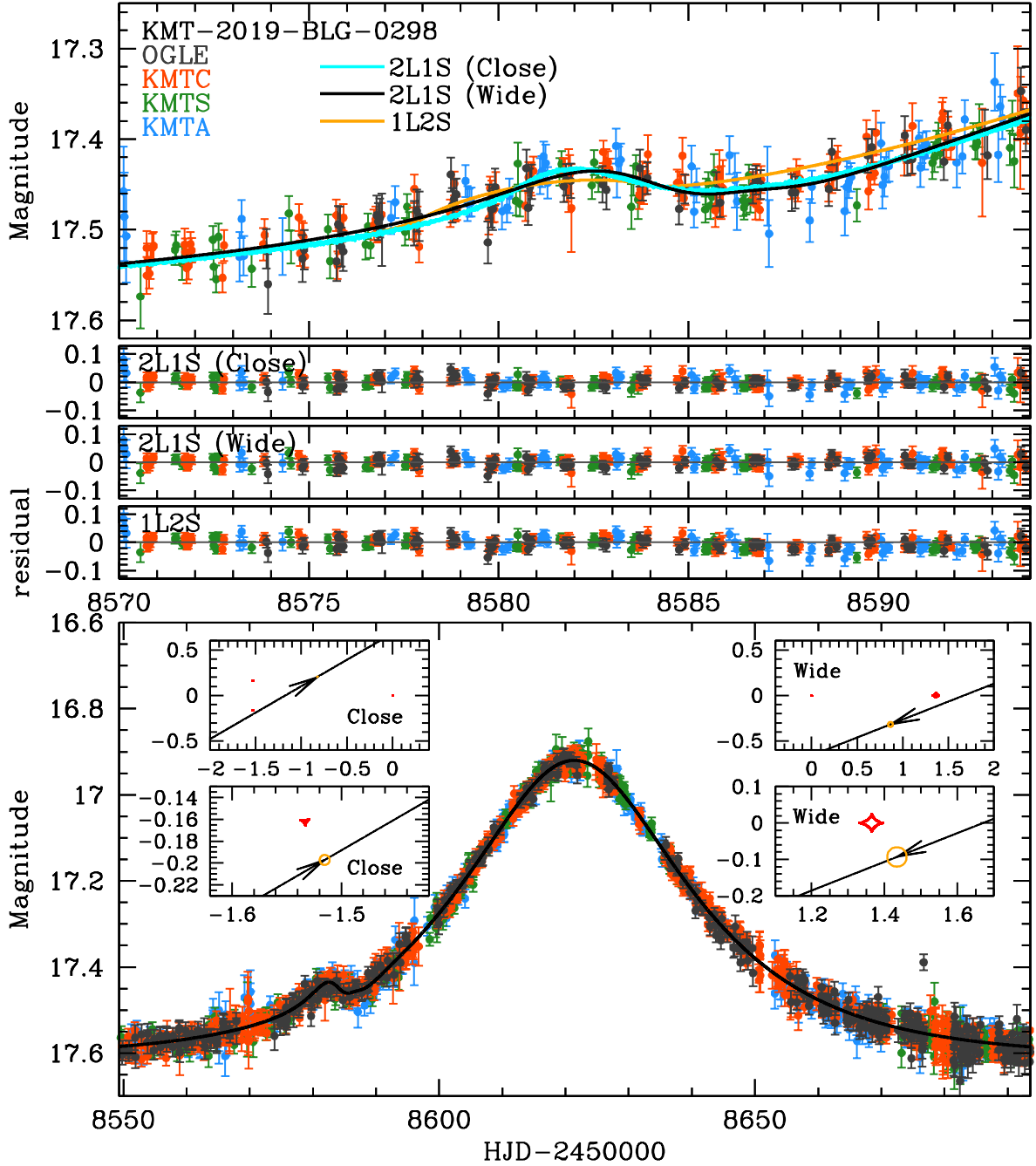}
\caption{Data (color-coded by observatory) together with the predictions
and residuals for the two planetary models of KMT-2019-BLG-0298 specified in 
Table~\ref{tab:kb0298parms},
plus the 1L2S model.
The caustic topologies are shown in the insets for both the close and
wide geometries, but the wide geometry is decisively favored by
$\Delta\chi^2=19$.  The 1L2S model is excluded at $\Delta\chi^2=125$.
}
\label{fig:0298lc}
\end{figure}

% gaia -1.6962 0.4689 -4.200 0.638 -5.908 0.411  1.250 ruwe
Figure~\ref{fig:0298lc} shows a low-amplitude 
$(\Delta I\simeq ~0.5)$ microlensing
event, peaking at $t_0=8621.4$ and punctuated by a smooth bump at 
$t_{\rm anom}\simeq 8582.$, i.e., $-39.4$ days before peak.  Assuming that the
source is unblended (as is reasonable for such a bright source), the remaining
\citet{pac86} parameters are $u_0=0.60$ and $t_\e=28\,$days.  Then
$\tau_{\rm anom} = -1.41$ and $u_{\rm anom} = 1.53$.  Because the source is bright
(so, large), while the caustics are likely to be small (because $u_{\rm anom}>1$),
we consider that the bump could be due to either a major-image or minor-image
perturbation.  For these,
Equation~(\ref{eqn:heuristic}) predicts $s^\dagger_+=2.02$ and $\alpha=158^\circ$,
and $s^\dagger_-=0.49$ and $\alpha=338^\circ$, respectively.

The grid search returns two solutions, whose refinements are shown
in Table~\ref{tab:kb0298parms}.  The wide solution is substantially preferred by
$\Delta\chi^2=19$.  For this solution, the heuristic prediction of $\alpha$
is confirmed, while the fit value of $s=s_{\rm outer}=1.89$ indicates that there
could be another solution near $s_{\rm inner}=(s^\dagger_+)^2/s_{\rm outer} = 2.16$.
As a matter
of due diligence, we seed an MCMC with this value and indeed find a local
minimum at $s_{\rm inner}=2.22$.  However, this solution is ruled out at
$\Delta\chi^2=205$, which confirms its failure to be detected in the grid
search.  The reasons that the degeneracy is decisively broken in this case
is that the inner/outer degeneracy is most severe for angles
$\alpha=\pm90^\circ$ \citep{zhang22}, whereas in this case
$\alpha=(90+68)^\circ$. 

Although, there is no signature of finite-source effects in the light curve
(i.e., all values $\rho<0.046$ are consistent at $1\,\sigma$), the absence
of a signal actually places significant constraints: $\rho<0.061$ at $2\,\sigma$
and $\rho<0.077$ at $3\,\sigma$.  That is, sufficiently larger sources
would be impacted by the caustic.  See the inset in Figure~\ref{fig:0298lc}.
Hence, when we carry out a Bayesian analysis
in Section~\ref{sec:phys-kb190298},
we will ultimately incorporate a $\rho$-envelope function
to represent this constraint.  For the present, however, we simply note
that, in light of the source-radius estimate $\theta_*=6.1\,\muas$ derived
in Section~\ref{sec:cmd-kb190298}, the $1\,\sigma$ range corresponds to
Einstein radii,
$\theta_\e > 0.13\,\mas$ and lens-source relative proper motions,
$\mu_\rel > 1.7\,\masyr$.  These values imply that while there is no
compelling reason to believe that $\theta_\e$ is large, it could be
relatively large, e.g., $\theta_\e\sim 0.75\,\mas$ and still be consistent
with the typical range of microlensing proper motions, $\mu_\rel\la 10\,\masyr$.
If so, this would favor a nearby lens and thus a potentially large
(hence, measurable) microlens parallax, $\pi_\e$.

\begin{figure}
\epsscale{0.8}
\plotone{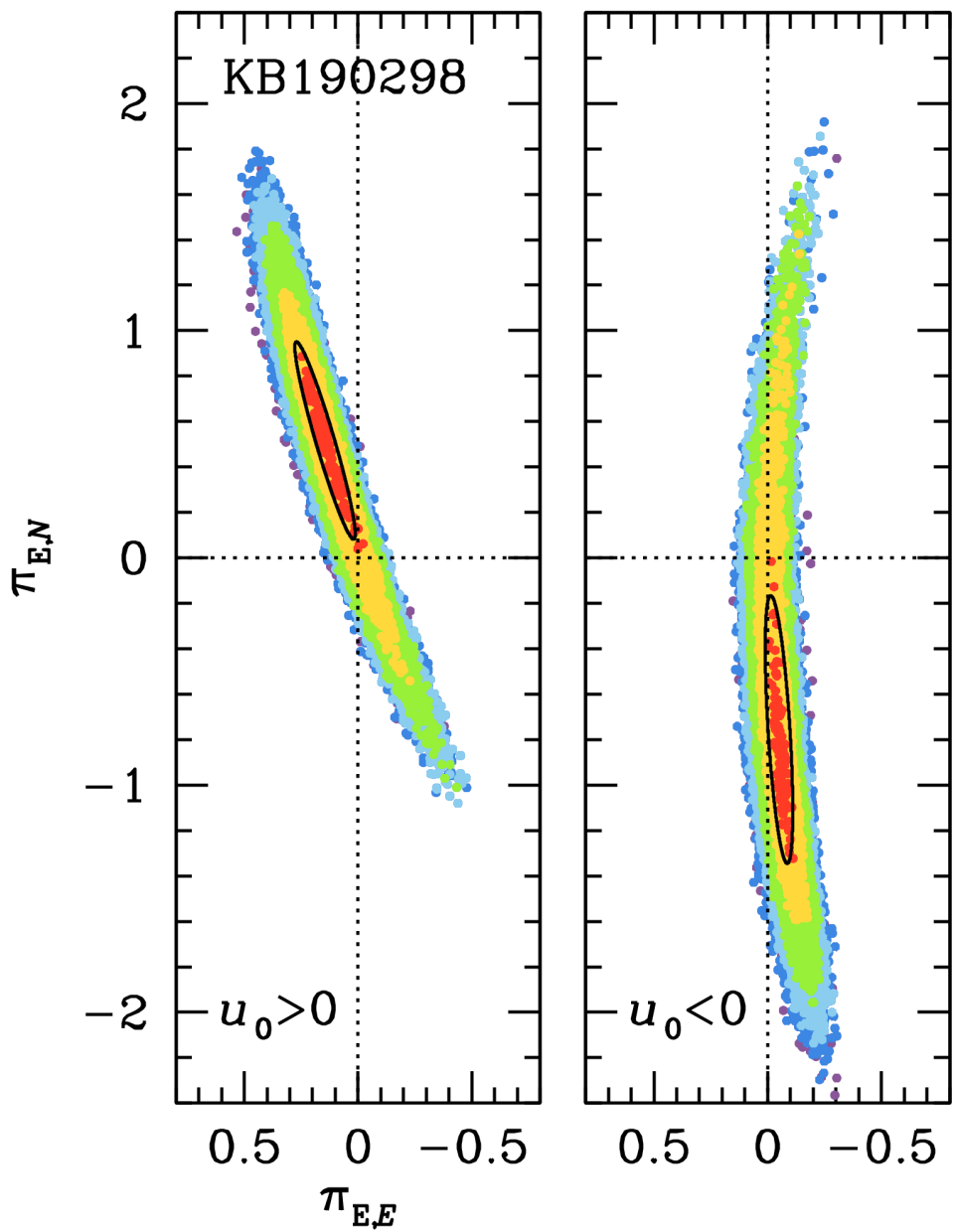}
  \caption{Scatter plot on the $\bpi_\e$ plane derived from the MCMC of the two
parallax models of KMT-2019-BLG-0298 presented in Table~\ref{tab:kb0298parms},
    %~\ref{tab:0298par},
color coded (red, yellow, green, cyan, blue) for $\Delta\chi^2 <(1,4,9,16,25)$.
The contours are effectively 1-D, with widths
$\sigma(\pi_{\e,\parallel})\equiv \sigma_\parallel\sim 0.039$ and 0.046. The black
contours show the mean and covariances ($\Delta\chi^2=1$)
that are used in Section~\ref{sec:phys-kb190298}.
}
\label{fig:0298par}
\end{figure}

Therefore, as a matter of due diligence, and despite the relatively
short Einstein timescale, $t_\e=28\,$day, we undertake a parallax analysis.
The results, given in Table~\ref{tab:kb0298parms},
are consistent with $\bpi_\e=0$, and in this sense the parallax is
``undetected''.  Nevertheless, as shown in Figure~\ref{fig:0298par},
the fit does place
1-dimensional (1-D) constraints on $\bpi_\e$, and we will incorporate these
into the Bayesian analysis in Section~\ref{sec:phys-kb190298}.
However, because $\bpi_\e$
is poorly constrained in the orthogonal direction, the free fit allows
for $\pi_\e$ values that would be very unlikely in a posterior fit.  These
would unphysically broaden the errors in $q$ and other parameters.  We note
that (except for $u_0$), all the parameter values from the parallax fits
are consistent with those from the standard fits at $1\,\sigma$.  Hence,
we will finally report $(s,q)$ and the physical parameters that are derived
from them based on the standard fit of Table~\ref{tab:kb0298parms}.

Because the anomaly is a smooth, featureless bump, we must
also consider the possibility that it is due to an extra source, i.e., 1L2S,
rather than an extra lens.  However, we find that such models are
excluded by $\Delta\chi^2 = 125$.

\subsection{{KMT-2019-BLG-1216} %ob191033
\label{sec:anal-kb191216}}
\begin{figure}
\plotone{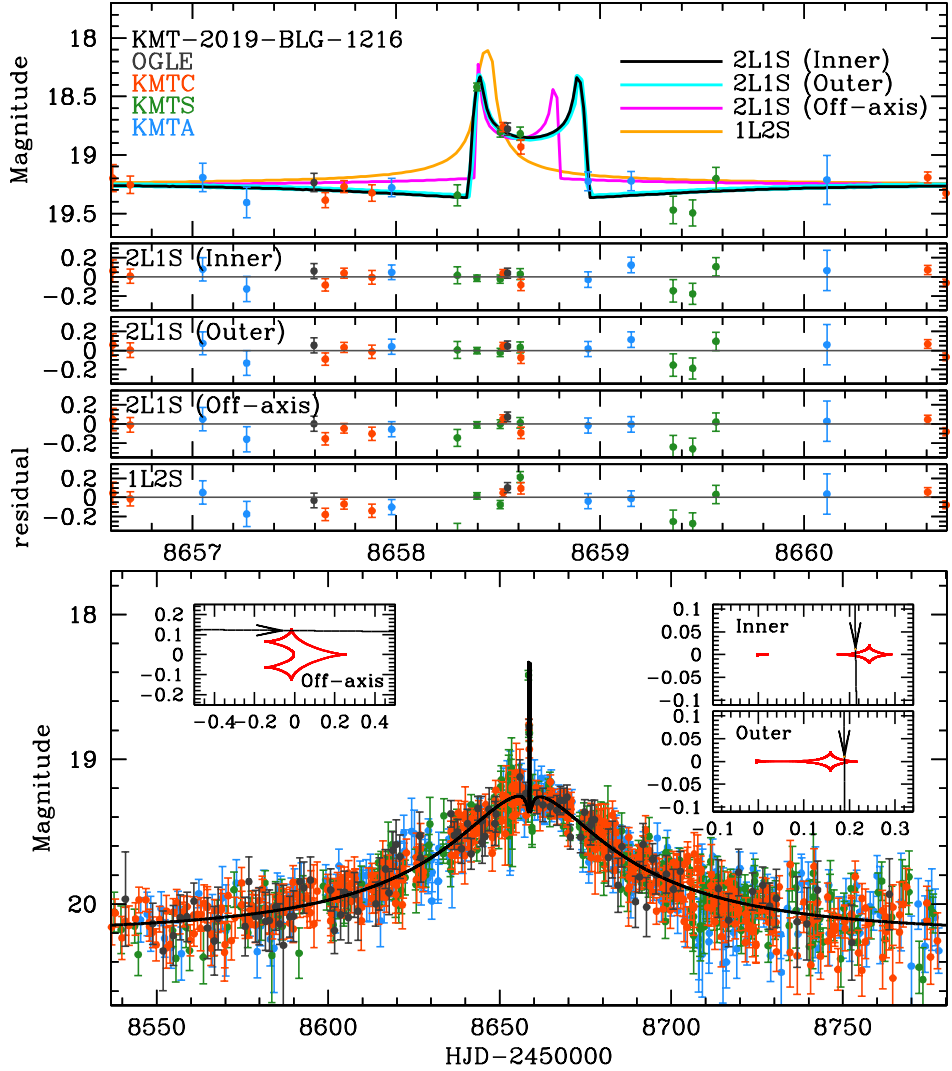}
\caption{Data (color-coded by observatory) together with the predictions
and residuals for the three planetary models of KMT-2019-BLG-1216 specified in 
Table~\ref{tab:kb1216parms},
plus the 1L2S model.
The caustic topologies are shown in the insets for the inner, outer, and
off-axis models.  The first two are perfectly degenerate but the last is
decisively disfavored at $\Delta\chi^2=24$. The 1L2S model is excluded
at $\Delta\chi^2=57$.
}
\label{fig:1216lc}
\end{figure}

%high 8658.396 - 8658.612  %low 8658.300 - 8658.941
Figure~\ref{fig:1216lc} shows a low-amplitude, generally smooth 
microlensing event, peaking at $t_0=8658.44$, except for 6 elevated points
(from 3 observatories) over an interval of 5.2 hours.  The maximum
extent of the deviation, defined by the two limiting points that lie
on the 1L1S curve, is 15.4 hours and is centered at
$t_{\rm anom}\simeq 8658.62$, i.e., just $+0.18$ days after peak.  A 1L1S
fit to the unperturbed parts of the light curve yields
$u_0=0.18$ and $t_\e=90\,$days.  Then,
$\tau_{\rm anom} = +0.002$ and $u_{\rm anom} = 0.18$.  The first anomalous point
is about 0.4 mag brighter than the others and is almost 1 mag brighter than
the point that precedes it by 2.3 hours.  Therefore, the anomaly is very
likely to be a caustic entrance that is followed by a caustic trough, but it is
difficult to make a further assessment by eye.  Within the heuristic
framework of an on-axis (or near-axis) anomaly, Equation~(\ref{eqn:heuristic})
predicts $s^\dagger_+=1.094$ and $\alpha=89.4^\circ$.

The grid search returns three solutions, whose refinements are shown
in Table~\ref{tab:kb1216parms}  and are illustrated in Figure~\ref{fig:1216lc}.
Two of these are a classical
inner/outer degeneracy \citep{gaudi97}, in which (as is often the case,
\citealt{ob190960}), the outer-solution caustic has a resonant topology,
with the source intersecting its ``planetary wing''.  These two solutions have
nearly identical values of $\alpha=89.6^\circ$, which are in excellent agreement
with the heuristic prediction, and $s^\dagger=\sqrt{s_{\rm inner}s_{\rm outer}}=1.094$,
which is also in near perfect agreement.

However, there is also a third solution, which has a fully resonant topology,
and in which the source intersects an off-axis cusp.  As illustrated by
Figure~\ref{fig:1216lc},
the degeneracy of the inner/outer solutions is intrinsic,
while the off-axis solution degeneracy is only possible because of a lack of
data during the latter part of the caustic perturbation.  Nevertheless,
because the off-axis solution is disfavored by $\Delta\chi^2=24$, we reject it.

As a matter of due diligence, we check for 1L2S solutions, but find that these
are rejected by $\Delta\chi^2= 57$.

\begin{figure}
\plotone{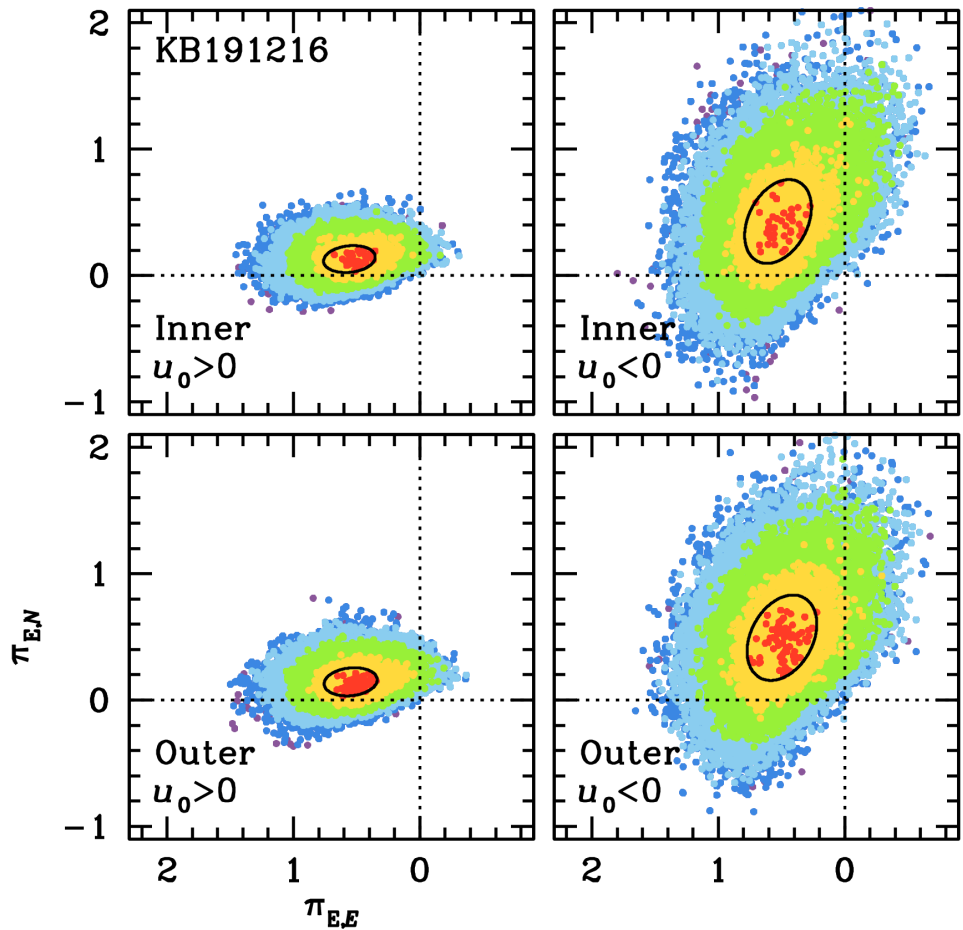}
  \caption{Scatter plot on the $\bpi_\e$ plane derived from the MCMC of the four
    parallax models of KMT-2019-BLG-1216 presented in
    Table~\ref{tab:kb1216parallax},
color coded (red, yellow, green, cyan, blue) for $\Delta\chi^2 <(1,4,9,16,25)$.
The black contours show the mean and covariances ($\Delta\chi^2=1$)
that are used in Section~\ref{sec:phys-kb191216}.
}
\label{fig:1216par}
\end{figure}

While $\rho$ is not well measured, there is weak $\chi^2$ minimum at
$\rho\sim 6\times 10^{-4}$ and a secure $3\,\sigma$ upper limit of
$\rho< 11\times 10^{-4}$.  In Section~\ref{sec:cmd-kb191216}, we will show
that $\theta_*\simeq 0.40\,\muas$.  Hence, these $\rho$ values correspond to
$\theta_\e\sim 0.67\,\mas$ and $\theta_\e> 0.36\,\mas$.  Thus, it is at
least plausible that $\theta_\e$ is relatively large, which would be
consistent with a nearby lens and so a relatively large (hence, measurable)
microlens parallax $\pi_\e=\pi_\rel/\theta_\e$.  Therefore, despite the faintness
of the source, we attempt a parallax analysis.  The results are shown in
Table~\ref{tab:kb1216parallax} and illustrated in Figure~\ref{fig:1216par}.

We will approach this $\bpi_\e$ measurement cautiously.  While there is
no reason to doubt this measurement based on the modeling, the best-fit
values of $\pi_\e$ are relatively large, and the improvement is only
$\Delta\chi^2=10.5$ for 4 degrees of freedom (dof).  Even assuming
Gaussian statistics, this has false alarm probability of
$p = (1 + \Delta\chi^2/2)\exp(-\Delta\chi^2/2)=3\%$.

Our orientation toward such a measurement depends on our prior expectation
on the magnitude of $\pi_\e$.  For a typical microlensing event, the
expected value is much closer to zero, and the fraction of events with
such large $\pi_\e$ values is small.  In such conditions, a $p=3\%$
measurement cannot be considered compelling: in addition to the
relatively high false-alarm probability, the large parallax could be due
to systematics.  However, KMT-2019-BLG-1216 is far from typical: it has
an exceptionally large $t_\e$ and there is evidence for a possibly large
$\theta_\e$.  Therefore, we will begin the Bayesian analysis in
Section~\ref{sec:phys-kb191216} by examining the posterior distributions
in the absence of the $\bpi_\e$ constraint before deciding whether to
incorporate it.

\subsection{{KMT-2019-BLG-2783} %ogle BLG509.29 2081.27 690.92 chip gap
\label{sec:anal-kb192783}}
\begin{figure}
\plotone{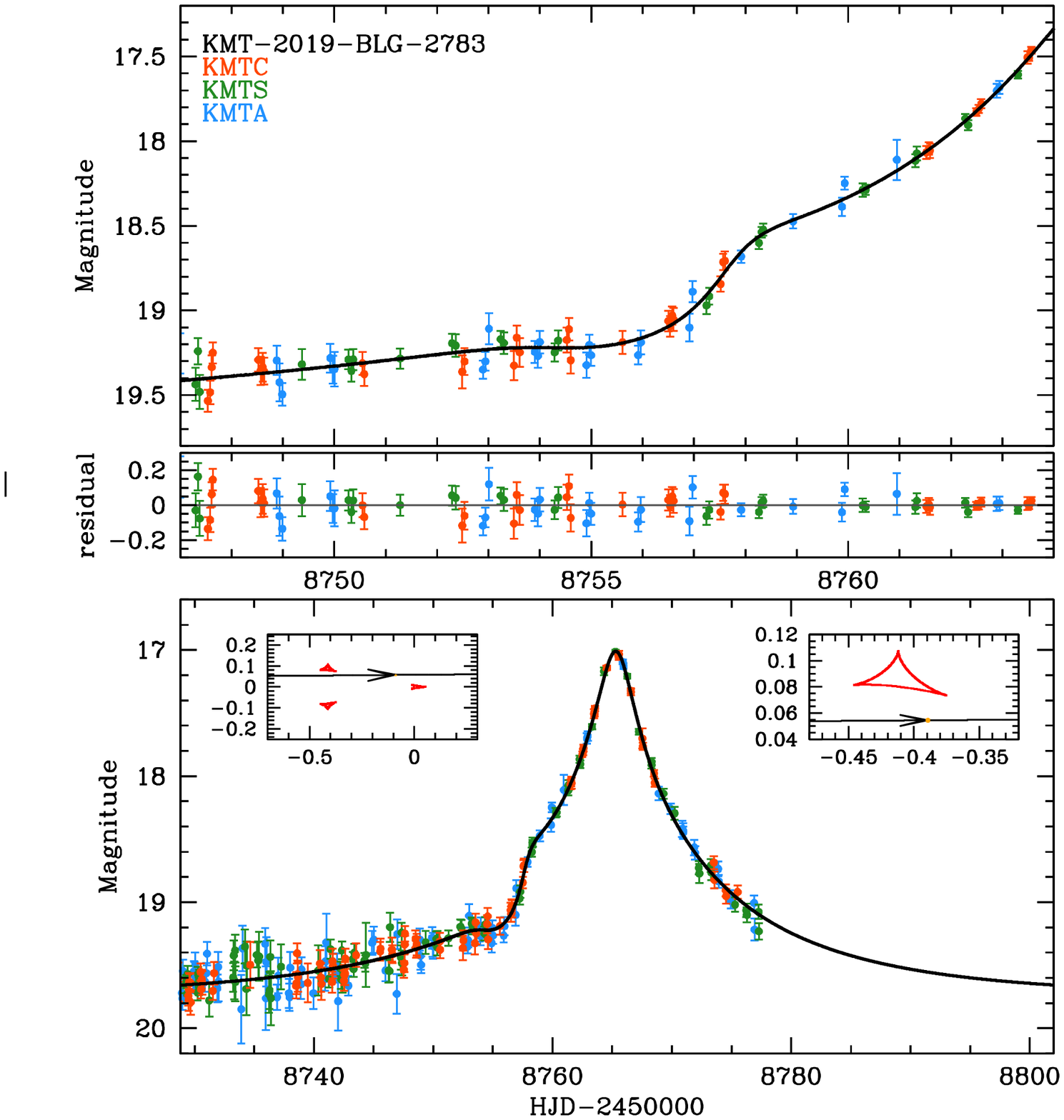}
\caption{Data (color-coded by observatory) together with the prediction
and residuals for the model of KMT-2019-BLG-2783 specified in 
Table~\ref{tab:kb2783parms}.
The caustic topology is shown in the insets.  
}
\label{fig:2783lc}
\end{figure}

Figure~\ref{fig:2783lc} shows a smooth, somewhat complex, perturbation on the
rising wing of a microlensing event that peaks at $t_0=8764.4$, i.e.,
close to the end of the season.  When this anomaly is excised, a 1L1S fit yields
$u_0=0.06$ and $t_\e=24\,$days.  The anomaly is characterized by a dip at
$t_{\rm anom,dip}\simeq 8756.3$,
followed by a bump at $t_{\rm anom,bump}\simeq 8758.0$.
If the dip is regarded as the driving feature, then
$\tau_{\rm anom,dip} = -0.34$ and $u_{\rm anom} = 0.34$, $s^\dagger_-=0.84$ and
$\alpha_-=350^\circ$, while if the bump is regarded as the driving feature, then
$\tau_{\rm anom,bump} = -0.27$ and $u_{\rm anom} = 0.27$, $s^\dagger_+=1.14$ and
$\alpha_+=168^\circ$.  In either case, if the heuristic prediction is correct,
then the non-driving feature (bump or dip, respectively) would have to be
naturally explained by the resulting geometry.

In fact, the grid search returns only a single solution whose
refinement is shown in Table~\ref{tab:kb2783parms}.
The $s^\dagger_-$ prediction is
qualitatively confirmed, while the $\alpha$ prediction is off by $\sim
10^\circ$.  The $s^\dagger_-$ inaccuracy derives from the difficulty
in judging the exact position of the dip in the presence of the bump.  As can
be seen from Figure~\ref{fig:2783lc}, the error in the $\alpha$ estimate is
due to the generic problem that minor-image caustics lie off-axis,
which is exacerbated by the fact that $q$ is large, implying that the
separation of the caustics is also large \citep{han06}.  As
anticipated in the previous paragraph, the bump is then naturally
explained by the fact that the source passes close to a cusp as it
exits the trough between the two minor-image caustics.  Indeed, there
is also a bump before the dip, which is much weaker because the source
passes much farther from the cusp.  This bump is hardly noticeable in
the data because of larger error bars, but it can be discerned in the
model.

The source passes about 0.015 from the cusp, which creates a $3\,\sigma$
limit, $\rho<0.01$.  This is of relatively little interest because,
given the estimate $\theta_*=0.39\,\muas$ that is derived in
Section~\ref{sec:cmd-kb192783},
it corresponds to a limit $\mu_\rel > 0.6\,\masyr$, which excludes only
a small fraction of parameter space.  Nevertheless, we will include
the constraints on $\rho$ via an envelope function when we carry out
the Bayesian analysis in Section~\ref{sec:phys-kb192783}.

Due to the brevity of the event, the faintness of the source, as well
as the absence of any data more than 10 days after $t_0$, we do not
attempt a parallax analysis.

\subsection{{OGLE-2019-BLG-0249} %kb192688
  \label{sec:anal-ob190249}}
\begin{figure}
\plotone{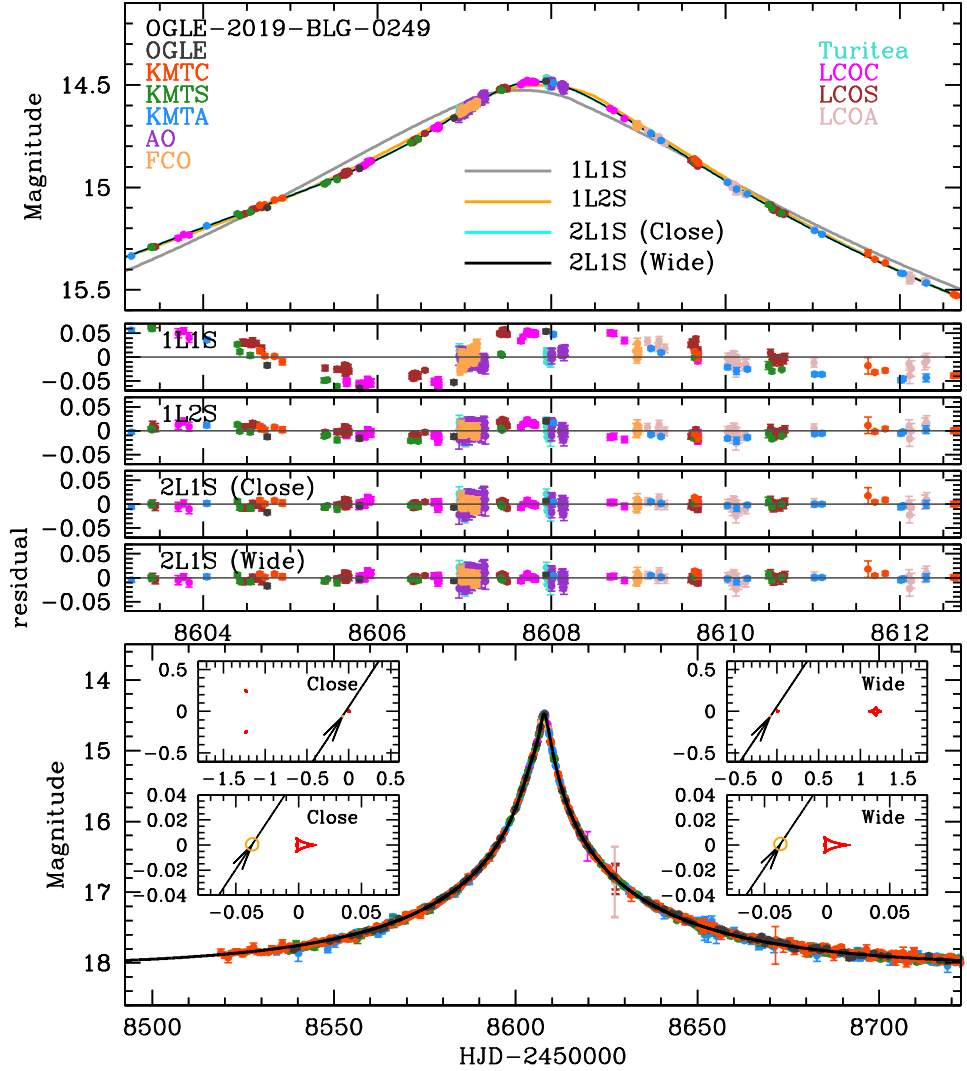}
\caption{Data (color-coded by observatory) together with the predictions
and residuals for the two planetary models of OGLE-2019-BLG-0249 specified in 
Table~\ref{tab:ob0249followup} with
the caustic topologies shown as insets.  Also shown are the 1L1S and
1L2S models.
}
\label{fig:0249lc}
\end{figure}

%tan alpha = -1.51 = u0/tau_anom, u0= 0.0305, tE=76,
%t_anom = tE*tau_anom = tE*u0/tan alpha = -1.53 day
Figure~\ref{fig:0249lc} shows a long microlensing event of a relatively
bright source that, in the absence of model light curves, might be
taken for a 1L1S event.  However, the residuals to the 1L1S model clearly
show a dip at $t_{\rm anom}=8606.0$, i.e., $\Delta t_{\rm anom}\sim-1.5\,$day
before the peak at $t_0\simeq 8607.5$.  Because the source is bright, it is
plausible
to guess that it might be unblended.  This turns out to be not precisely
the case, but proceeding on this assumption, $u_0 = 0.04$ and $t_\e=58\,$day.
Hence, $\tau_{\rm anom}=0.026$, $u_{\rm anom}=0.048$,
$s^\dagger_+=0.976$ and $\alpha=303^\circ$.  Table~\ref{tab:ob0249parms} shows
that the $\alpha$ prediction is accurate to high precision, but
$s^\dagger=\sqrt{s_{\rm close}s_{\rm wide}}=0.985$ is slightly off.  The reason
is that the source flux is only about 76\% of the baseline flux.  This
does not affect $\alpha$, which can be written in terms of the invariant
$t_\eff \equiv u_0 t_\e$ (for high-magnification events, \citealt{mb11293}), as
$\tan\alpha = t_\eff/\Delta t_{\rm anom}$.  However, it does affect $s^\dagger$,
with the 24\% blending fraction driving $s^\dagger$ about 24\% closer to
unity\footnote{That is, in the limit $u_{\rm anom}\ll 1$,
$s^\dagger_\pm \rightarrow 1\pm u_{\rm anom}/2$, while
$u_{\rm anom}^2 = u_0^2 + \tau_{\rm anom}^2
\rightarrow (t_\eff^2 + (\Delta t_{\rm anom}^2))/t_\e^2$.  Hence,
$s^\dagger_\pm \rightarrow 1\pm \eta/t_\e$, where
$\eta\equiv 0.5\sqrt{t_\eff^2 + (\Delta t_{\rm anom})^2}$, in which the first
term is an invariant and the second is a direct observable.
}.

\subsubsection{{A Survey+Followup Event} 
  \label{sec:survey+followup-event}}

In addition to survey data from OGLE and KMT, OGLE-2019-BLG-0249 was
intensively observed by many follow-up observatories
(see Figure~\ref{fig:0249lc}), in part because it was a {\it Spitzer}
target and in part because it was a moderately-high magnification event
($A_\max=33$) in a low-cadence field (see Table~\ref{tab:names}).
The current AnomalyFinder series of papers includes the analysis of
events only if they are (1) identified as anomalous by the AnomalyFinder
algorithm, which is applied to KMT data alone, and (2) have a plausible
planetary solution based on survey data alone.  This ``publication grade''
analysis then (3) lays the basis for deciding whether the planet is ultimately
included in the complete AnomalyFinder sample.  From the standpoint of
(2) and (3), it is therefore essential to ask how the event would have
been evaluated in the absence of followup data.

Nevertheless, if this evaluation determines that the planet should be
in the AnomalyFinder papers and/or if it is included in the final sample,
the followup data may be used to improve the characterization of the
planet.

OGLE-2019-BLG-0249 is the first planet with extensive followup data to be
included in this series, following 7 previous papers containing a total of
%I zang: 1: II hwan 6; III goul: 10; IV wang: 1; V jung 9; VI zang 3; VII zang 7
37 planets and ``possible planets''.  There have, of course, been other
published planets that had extensive followup data
and that will ultimately enter the
AnomalyFinder sample.  For example, the planetary anomaly in
OGLE-2019-BLG-0960 was originally discovered in followup data, and therefore
\citet{ob190960} carefully assessed that this planet could be adequately
characterized based on survey data alone.  Another relevant example is
OGLE-2016-BLG-1195, for which the MOA group obtained intensive
data over peak (including the anomaly) by using their survey telescope in
followup mode \citep{ob161195a}.  In principle, one should assess whether
this anomaly would have been adequately characterized had MOA observed at
its normal cadence.  However, as a practical matter, this is unnecessary
because the KMT and {\it Spitzer} groups showed that this planet could be
adequately characterized based on an independent survey-only data set
\citep{ob161195b}.

\subsubsection{{Survey-Only Analysis} 
\label{sec:survey-only}}

Thus, we began by analyzing the survey data alone.  These results have already
been reported above in Table~\ref{tab:ob0249parms}, when we compared them to the
heuristic predictions.  We note that before making these fits, we removed
the KMTC points during the two days $|{\rm HJD}^\prime - 8706|<1$, due to
saturation and/or significant nonlinearity of this very bright target.
We also checked that if these excluded points were re-introduced, which
we do not advocate, the parameters were affected by $\la 2\,\sigma$.

In addition to the two planetary solutions shown in
Table~\ref{tab:ob0249parms}, there are two
local minima derived from the grid search that, when refined, have
binary-star mass ratios, i.e., $q\sim 0.15$ and $q\sim 0.25$, with
source trajectories passing roughly parallel to a side of a
\citet{cr1,cr2} caustic (not shown).  This is a common form of planet/binary
degeneracy for dip-type anomalies \citep{han08}.
However, in the present case,
these binary solutions are rejected by $\Delta\chi^2=64$.
See Table~\ref{tab:ob0249parms}, 

As a matter of due diligence, we also fit the data to 1L2S models.
These usually give poor fits to dip-type anomalies, but there can be
exceptions.  However, in this case, we find that 1L2S is ruled out
by $\Delta\chi^2=475$.

\subsubsection{{Followup Data} 
  \label{sec:followup-data}}

The followup observations were all, directly or indirectly, initiated
in response to an alert that this event would be monitored by {\it Spitzer}.
Although the {\it Spitzer} observations themselves could not begin until
9 July (due to telescope-pointing restrictions), i.e., 66 days after $t_0$,
OGLE-2019-BLG-0249 was chosen by the {\it Spitzer} team on 29 April
(6 days before $t_0$)
in order to ``claim'' any planets that were discovered (which would also
ultimately require that the microlens parallax be measured at sufficient
precision).  See the protocols of \citet{yee15}.

On 30 April, the Tsinghua Microlensing Group, working with the {\it Spitzer}
team, initiated observations on three 1-meter telescopes from the Las Cumbres
Observatory at the same locations as the KMT telescopes, which we designate
in parallel as LCOC, LCOS, and LCOA, using an SDSS $i$ filter.  

%UT 07:59
Based on these observations, combined with ongoing survey observations by
OGLE and KMT, these teams  noted that the event was probably anomalous
and, on this basis, alerted the microlensing community by email.
Because this alert was
triggered by an anomaly, such observations can be used only to characterize
the planet, but not to ``claim'' its detection according to the {\it Spitzer}
protocols \citep{yee15}. (However, from Section~\ref{sec:survey-only}, we can
see that this issue has subsequently become moot.)
Four observatories in the
Microlensing Follow Up Network ($\mu$FUN), which is composed mainly
of small telescopes, responded to this alert, i.e., the
(Auckland, Farm Cove, Kumeu, Turitea) observatories, respectively, in
(Auckland, Pakuranga, Auckland, Palmerston) New Zealand, with respectively,
(0.41, 0.36, 0.41, 0.36) meter mirrors, and respectively,
($R$, white, $R$, $R$) filters.

We found that the Kumeu observations were not of sufficient quality to
include them in the analysis.  In addition, there were two other observatories,
both in Chile, i.e., the Danish 1.5 meter and the SMARTS 1.3 meter,
that began observations on May 10 and 11, respectively, i.e., 5--6 days
after $t_0$.  We do not include these observations because they were
taken too late to help constrain any of the event parameters.

\subsubsection{{Survey+Followup Analysis} 
  \label{sec:survey+followup-anal}}

Table~\ref{tab:ob0249followup}
shows the parameters after incorporating the followup data
into the fit.  The values of $q$ increase by about 10\%, corresponding
to $\sim 2\,\sigma$, which is not surprising given that the additional
data are concentrated on the anomaly.  The changes in $s$ are similar.
From the comparisons of Tables~\ref{tab:ob0249parms} and
\ref{tab:ob0249followup}, the most puzzling (and potentially
most consequential) change is that $\rho$ drops by a factor $\sim 2$
without much change in the error bar. We investigate this and find that
these three parameters are tightly correlated, which is very plausible
given that they are all derived from the same short feature in the
light curve, so that the three parameter changes are all expressions
of the same additions to the data set.

\subsubsection{{Parallax Analysis} 
  \label{sec:parallax-ob190249}}

\begin{figure}
\plotone{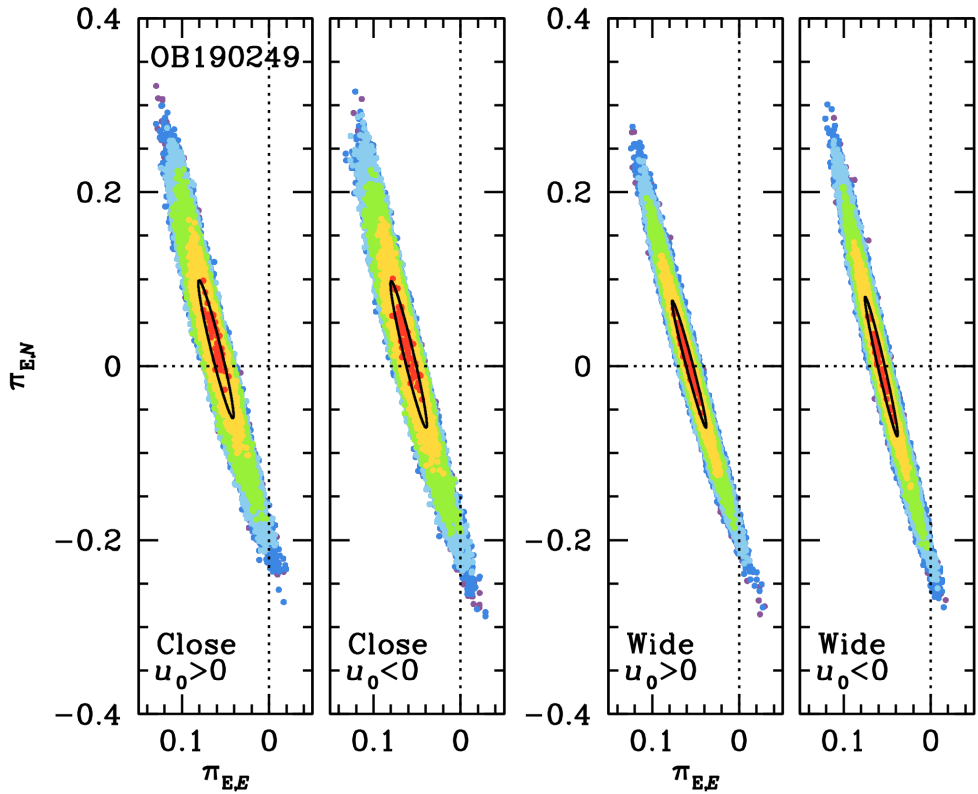}
  \caption{Scatter plot on the $\bpi_\e$ plane derived from the MCMC of the four
    parallax models of OGLE-2019-BLG-0249 presented in
    Table~\ref{tab:ob0249followup},
color coded (red, yellow, green, cyan, blue) for $\Delta\chi^2 <(1,4,9,16,25)$.
The contours are effectively 1-D, with typical
widths $\sigma_\parallel\sim 0.0080$ and 0.0065, for the close and
wide solutions respectively.  The black contours show the mean and covariances
that are used in Section~\ref{sec:phys-ob190249}.
}
\label{fig:0249par}
\end{figure}

Because the event is long ($t_\e\sim 75\,$day) and reaches
relatively high magnification ($A_\max\sim 30$), and because the source is
relatively bright ($I_S\sim 18$), it is plausible that substantial parallax
information can be extracted.  We therefore add four parameters, i.e., $\bpi_\e$
and $\bgamma$, and report the results in Table~\ref{tab:ob0249followup}.
A scatter plot of
the MCMC on the $\bpi_\e$ plane is shown in Figure~\ref{fig:0249par}
for each of the
four solutions.  As in the case of KMT-2019-BLG-0298, the contours are
essentially 1-D, with axis ratios $\sim 10$.  However, contrary to that
case, even the long axes of the error ellipses are relatively small,
$\sigma_\perp\sim 0.08$, which is comparable to the offsets of the
1-D contours from the origin.  Hence, the argument given in
Section~\ref{sec:anal-kb190298} for adopting the standard-model
parameters (but incorporating the $\bpi_\e$ constraints) does not apply,
and we therefore use the full parallax solutions
from Table~\ref{tab:ob0249followup} when we carry out the Bayesian analysis
in Section~\ref{sec:phys-ob190249}.

\subsection{{OGLE-2019-BLG-0679} %kb192688
\label{sec:anal-ob190679}}
\begin{figure}
\plotone{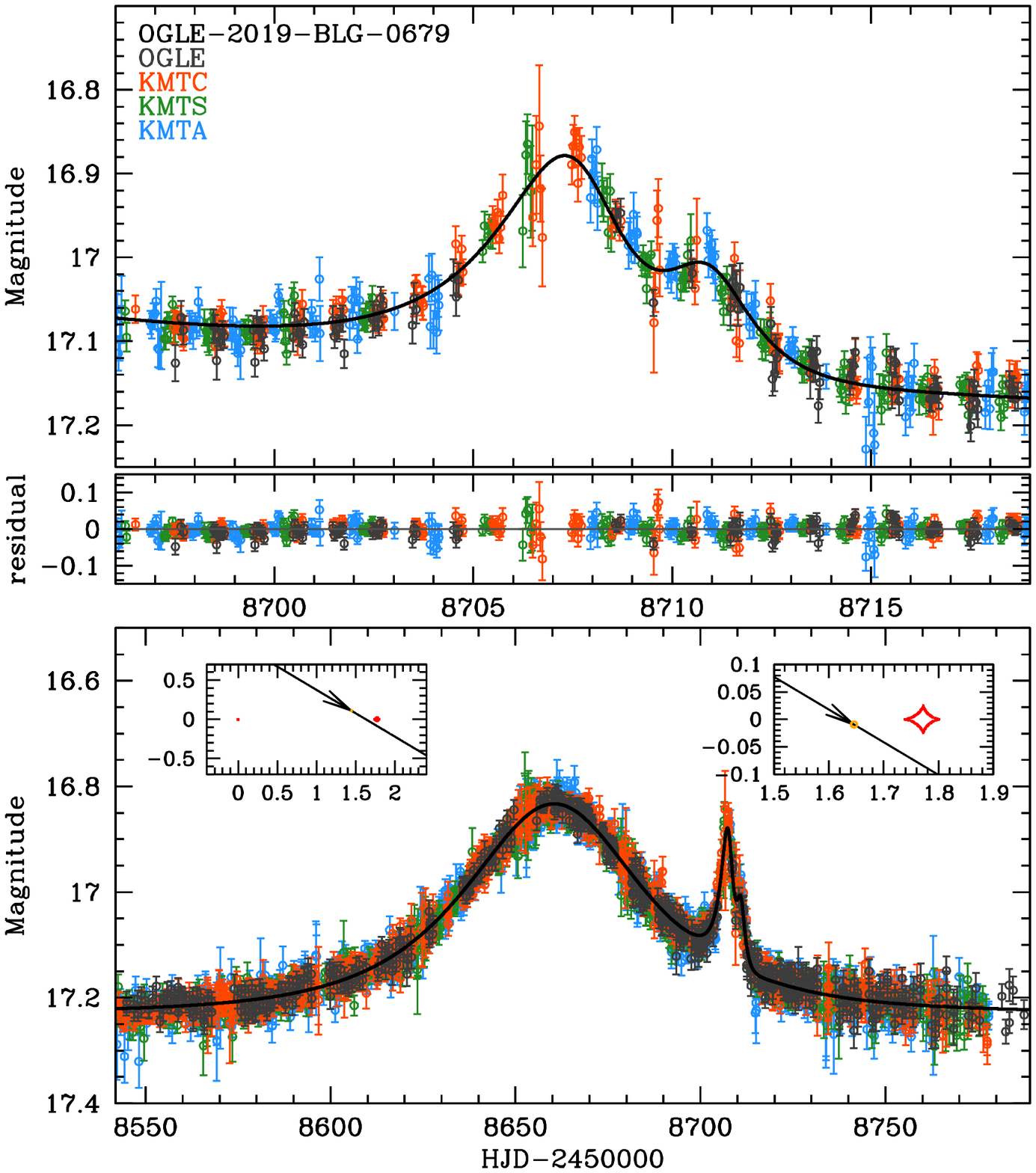}
\caption{Data (color-coded by observatory) together with the prediction
and residuals for the model of OGLE-2019-BLG-0679 specified in 
Table~\ref{tab:ob0679parms}.
The caustic topology is shown in the insets.
}
\label{fig:0679lc}
\end{figure}

Figure~\ref{fig:0679lc} shows a roughly 10-day bump, which peaks at
$t_{\rm anom}\simeq 8707.6$ and is itself punctuated by a shorter 2-day bump
on its falling wing, all on the 
falling wing of a microlensing event that peaks at $t_0=8660.7$.
When this anomaly is excised, a 1L1S fit (assuming no blending, as
is plausible for such a bright source) yields
%dt_anom = 8707.6-8660.7=46.9
$u_0=0.87$ and $t_\e=31\,$days.  Hence, $\tau_{\rm anom}=1.51$, $u_{\rm anom}=1.74$,
$s^\dagger_+=2.20$ and $\alpha=30^\circ$.

The grid search returns only one solution, whose refinement is shown in
Table~\ref{tab:ob0679parms}.  The value of $\alpha$ is in good agreement with
the heuristic prediction, while the fitted value of $s_{\rm inner}=2.22$
is in ``surprising'' agreement with $s^\dagger_+$, given that the anomaly
does not appear to be caustic crossing.  Figure~\ref{fig:0679lc} shows that
the solution has an ``inner'' topology.  In fact, if we had used the
fit values for $u_0$ and $t_\e$ (as opposed to those assuming no blending),
we would have derived $s^\dagger = 2.13$, which would suggest
that there might be another solution at
$s_{\rm outer}=(s^\dagger_+)^2/s_{\rm inner}=2.04$.  However, it is clear from the
caustic topology in Figure~\ref{fig:0679lc} that the peak of the bump is
due to the source passing the on-axis cusp and the shorter, post-peak bump
is due to passage of the off-axis cusp.  Hence, in a hypothetical ``outer''
solution, this extra bump would occur before the peak of the main bump.  Thus,
there is no degeneracy.

Although $\rho$ is not measured, the constraints on $\rho$ are of some
interest.  That is, we will show in Section~\ref{sec:cmd-ob190679}
that $\theta_*\sim 7.0\,\muas$, so the $3\,\sigma$ limit, $\rho<0.03$,
rules out $\mu_\rel<2.8\,\masyr$, which is a reasonably well populated
part of parameter space.  We will therefore incorporate the
$\rho$ envelope function when we carry out the Bayesian analysis in
Section~\ref{sec:phys-ob190679}.

Because the source is relatively bright ($I_S\sim 17.3$) and the anomaly
is long after the peak and
has two features that are separated by 4 days, we attempt a parallax
analysis.  That is, while in many cases, the change in source trajectory
induced by parallax could be compensated (or mimicked) by lens orbital
motion, this is much more difficult when the model must accommodate
additional light-curve features.  See, e.g., \citet{angould}.

\begin{figure}
\plotone{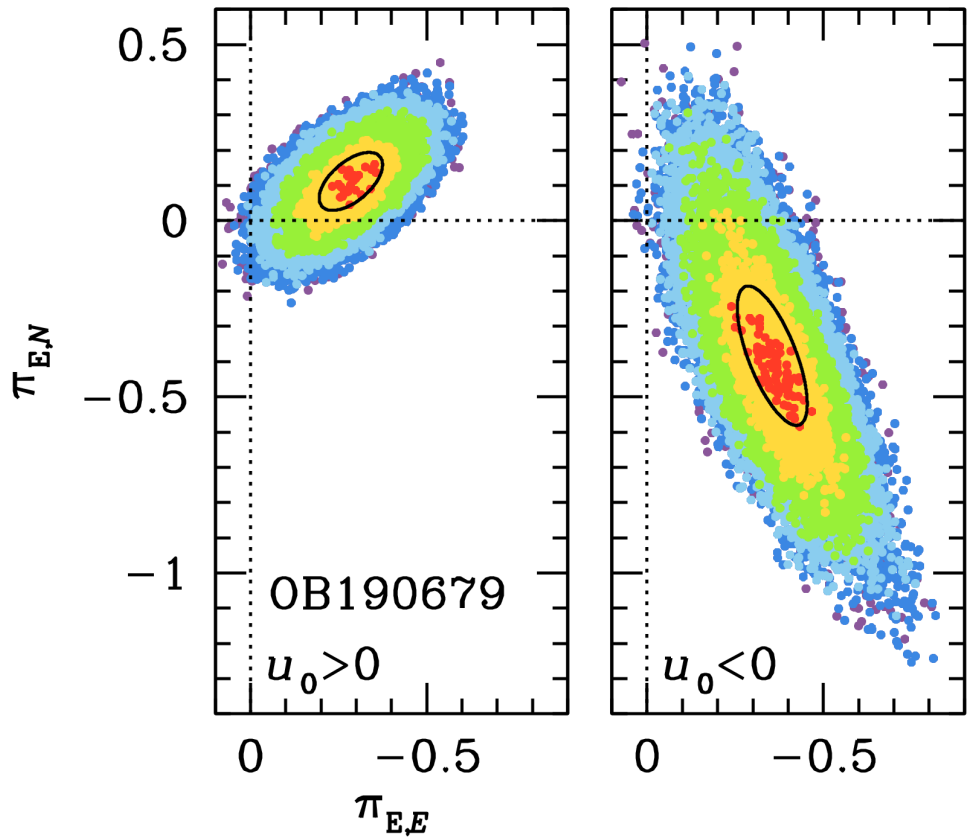}
  \caption{Scatter plot on the $\bpi_\e$ plane derived from the MCMC of the two
parallax models of OGLE-2019-BLG-0679 presented in Table~\ref{tab:ob0679parms},
color coded (red, yellow, green, cyan, blue) for $\Delta\chi^2 <(1,4,9,16,25)$.
The black contours show the mean and covariances
that are used in Section~\ref{sec:phys-ob190679}.
}
\label{fig:0679par}
\end{figure}

The results are shown in Table~\ref{tab:ob0679parms} and illustrated in the
$\bpi_\e$ scatter plot from the MCMC in Figure~\ref{fig:0679par}.
Including parallax and orbital motion improves the fit $\Delta\chi^2=29$.
Nevertheless, as we explain in some detail in Section~\ref{sec:phys-ob190679},
we will adopt the standard-model parameters for purposes of this paper.
However, we document the details of the parallax fit here in anticipation that
they will be useful when the ground-based and space-based parallax fits are
later integrated.  While we do not know what the space-based parallax fits
will reveal, we do note that preliminary reduction of the {\it Spitzer}
data shows a fall of $\sim 20$ flux units over 37 days, which should be enough
to strongly constrain the parallax.  In brief, when quoting parameters from
this paper, only the ``Standard'' column in Table~\ref{tab:ob0679parms} should
be used.

\subsection{{OGLE-2019-BLG-0344} %kb190149
  \label{sec:anal-ob190344}}

\begin{figure}
\plotone{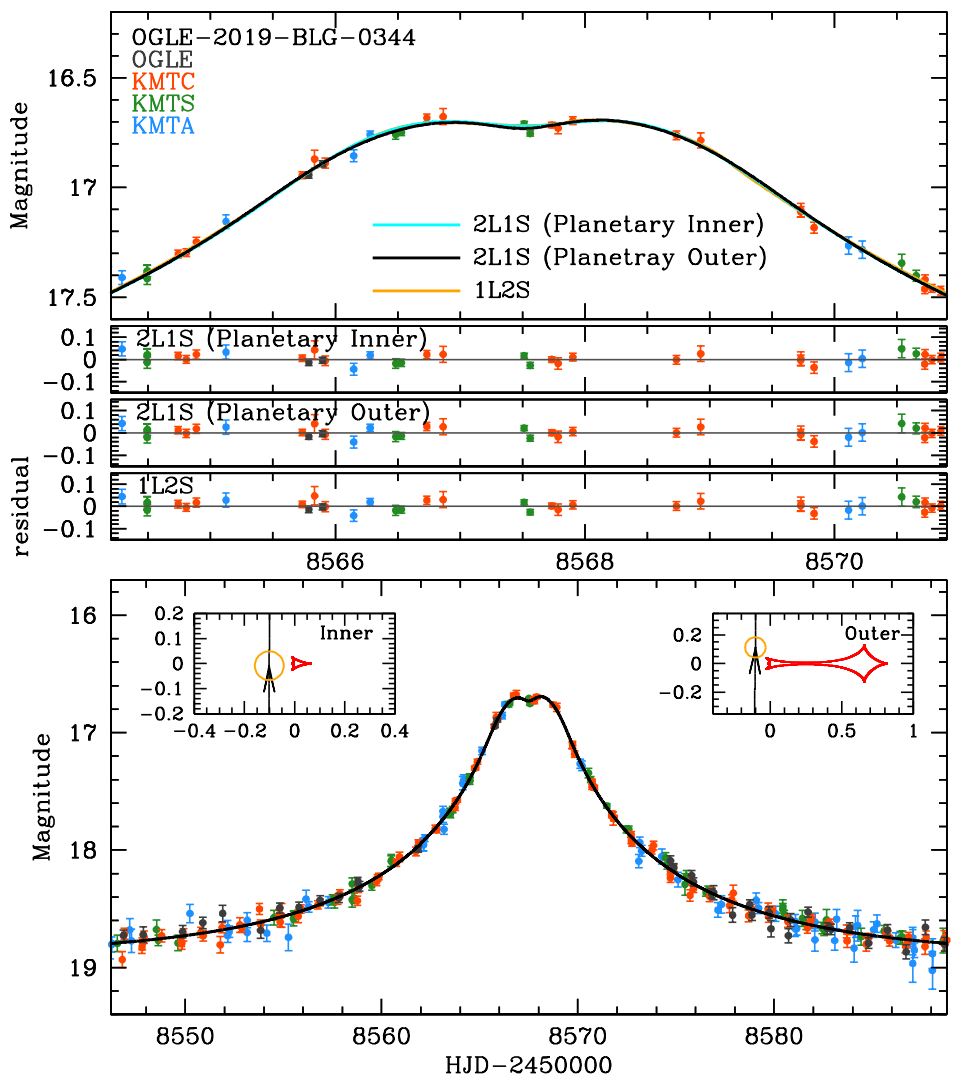} % kb190149/ob190344
\caption{Data (color-coded by observatory) together with the predictions
and residuals for the two planetary models of OGLE-2019-BLG-0344 specified in 
Table~\ref{tab:ob0344parms}
as well as the 1L2S model specified in Table~\ref{tab:ob0344-1l2s}.
All three models can explain the data equally well.  In addition, there
are 6 other non-planetary 2L1S models, most of which also can explain
the data equally well.  See Figure~\ref{fig:0344bin}.  Hence, this event
cannot be cataloged as ``planetary''.
}
\label{fig:0344lc}
\end{figure}

Figure~\ref{fig:0344lc} shows a moderately-high magnification 
microlensing event, peaking at $t_0=8567.53$ and punctuated by a short dip
that is almost exactly at peak.  A 1L1S fit to the data (with the anomaly
excluded) yields $u_0=0.10$ and $t_\e=14\,$days.  Hence,
$\tau_{\rm anom}=0$, $u_{\rm anom}=0.1$, $s^\dagger_-= 0.95$, and $\alpha=270^\circ$.

A grid search does indeed return two planetary solutions whose refinements
are shown in Table~\ref{tab:ob0344parms}
and that are in good agreement with these predictions,
i.e., $s^\dagger=\sqrt{s_{\rm inner}s_{\rm outer}}=0.95$, and $\alpha=270^\circ$.
However, it also returns six other solutions.  Before discussing these,
we first note that the planetary solutions are somewhat suspicious in that
they have relatively large values of $\rho\simeq 0.06$.  We will show
in Section~\ref{sec:cmd-ob190344} that $\theta_*\simeq 1.03\,\muas$.
If these solutions are correct, they would therefore imply $\theta_\e=17\,\muas$
and $\mu_\rel=0.44\,\masyr$.  The first of these falls in the category
of ``exciting if true'', while the second has a relatively implausible
$p=0.5\%$ probability according to Equation~(\ref{eqn:probmu2}).  Therefore,
we also show for comparison the solutions with $\rho=0$, which are disfavored
by $\Delta\chi^2=8$.

\begin{figure}
\plotone{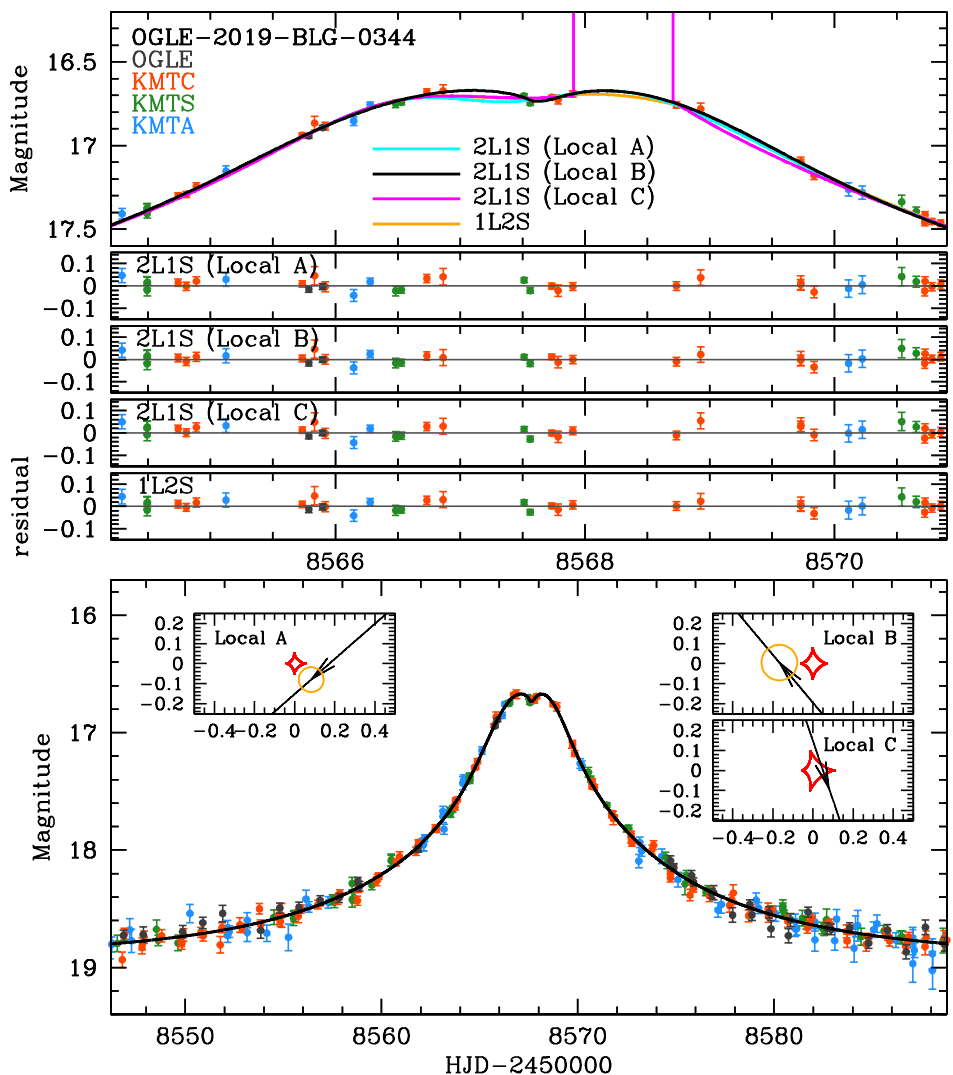} % kb190149ob190344
\caption{Data (color-coded by observatory) together with the predictions
and residuals for the three close-binary models of OGLE-2019-BLG-0344
specified in 
Table~\ref{tab:ob0344bin},
as well as the 1L2S model.
Two of the three close-binary models can explain the data about as well
as the planetary models.  In addition, there are three wide-binary models
that are not shown.  These viable binary-lens models provide an additional
reason that this event cannot be cataloged as ``planetary''.
}
\label{fig:0344bin}
\end{figure}

The six other solutions come in three pairs, which each approximately
obey the close/wide degeneracy \citep{dominik99}.  We label these
pairs (A,D), (B,E), and (C,F).  The close solutions are given in
Table~\ref{tab:ob0344bin}
and illustrated in Figure~\ref{fig:0344bin}.
One of these also has an implausibly
large $\rho$, so we show the $\rho=0$ solutions in all cases.  The bottom
line is that if we consider the free $\rho$ case, then Local B is preferred
over either planetary solution by $\Delta\chi^2=6$, while if we consider the
$\rho=0$ case, then Local A is within $\Delta\chi^2<1$ of 
either planetary solution.  Hence, there is no reason to believe that the
companion is a planet rather than another star.  To avoid clutter, we do
not present a table or figure for the three wide solutions, but the situation
is qualitatively similar.

Finally, we investigate 1L2S models, which are shown in
Table~\ref{tab:ob0344-1l2s}.
In this case, the apparent ``dip'' is the result of two sources of nearly
equal brightness successively passing the lens, with nearly equal impact
parameters and with an interval of 2.0 days.  The values of $\rho_1$ and
$\rho_2$ are each poorly measured, and if we were to take them at face
value, then the two stars would nearly overlap in projection.  Hence,
we also consider the $\rho_1=\rho_2=0$ case.  This has the best $\chi^2$
for any of the $\rho=0$ cases.

We conclude that the lens-source system could be either 1L2S or 2L1S and,
if the latter, the lens could equally well be planetary or binary in
nature.  Hence, we strongly counsel against classifying
this event as ``planetary.''

\subsection{{KMT-2019-BLG-0304}
\label{sec:anal-kb190304}}

\begin{figure}
\plotone{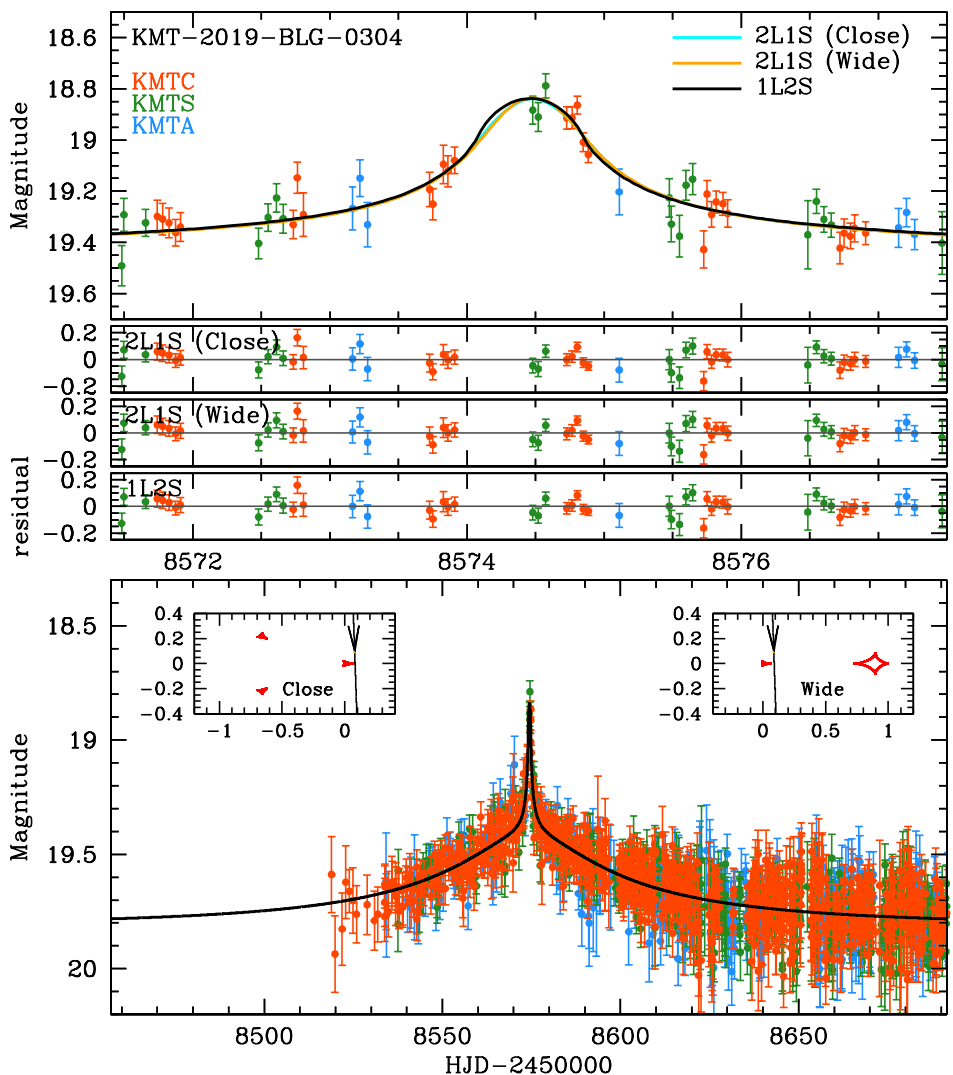}
\caption{Data (color-coded by observatory) together with the predictions
and residuals for the two planetary models of KMT-2019-BLG-0304 specified in 
Table~\ref{tab:kb0304parms},
as well as the 1L2S model specified in Table~\ref{tab:kb0304-1l2s}.
All three models can explain the data equally well.  Hence, this event
cannot be cataloged as ``planetary''.
}
\label{fig:0304lc}
\end{figure}

In many ways, KMT-2019-BLG-0304 is very similar to OGLE-2019-BLG-344
(Section~\ref{sec:anal-ob190344}), except that the anomaly near peak
is a bump rather than a dip.
Figure~\ref{fig:0304lc} shows a moderately-high magnification 
microlensing event, peaking at $t_0=8574.0$ and punctuated by a short bump
at $t_{\rm anom}=8574.5$, i.e. just $\Delta t_{\rm anom}= 0.5\,$day after peak.
The source is extremely faint, $I_S\sim 23$, which implies
(e.g., \citealt{mb11293}) that in 1L1S and 2L1S fits, the parameter
combinations $t_\eff \equiv u_0 t_\e$, $t_* \equiv \rho t_\e$,  and
$t_q \equiv q t_\e$, will be much better determined than $(u_0,t_\e,\rho,q)$.
For the 1L1S fit, we find $t_\eff=13.7\,$day.   This implies
$\alpha=\tan^{-1}(t_\eff/\Delta t_{\rm anom})= 88^\circ$, which is independent of
$t_\e$.  On the other hand, the prediction for $s^\dagger_+$ does depend on
$t_\e$.  Noting that $u_{\rm anom}= u_0/\sin\alpha \simeq u_0$, this can
be written as
\begin{equation}
s^\dagger_+ = {1\over 2}
\biggl(\sqrt{4 + {u_0^2\over\sin^2\alpha}} + {u_0\over\sin\alpha}\biggr)
\rightarrow 1 + {u_0\over 2} + {u_0^2\over 8}. 
\label{eqn:sdaggerspec}
\end{equation}
Adopting $t_\e=165\,$day as a fiducial value, this implies $u_0=0.083$, and
thus, $s^\dagger_+ = 1.04$.

A grid search does indeed return two planetary solutions whose refinements
are shown in Table~\ref{tab:kb0304parms},
which are in good agreement with these predictions,
i.e., $s^\dagger=\sqrt{s_{\rm inner}s_{\rm outer}}=1.05$, and $\alpha=88^\circ$.
In contrast to the case of KMT-2019-BLG-0304, there are no other
2L1S solutions.  However, as in that case, there is a competitive 1L2S
model, whose parameters are given in Table~\ref{tab:kb0304-1l2s}.

At present, there is no way to distinguish between these two solutions.
The ``free $\rho$'' 1L2S solution does predict an unusually low proper motion,
$\mu_\rel\sim 0.2\,\masyr$.  However, as shown in Table~\ref{tab:kb0304-1l2s},
the 1L2S solution remains competitive even when we impose $\rho=0$.
In principle, the solutions could be distinguished by measuring the colors
of the two sources: because the secondary source is $\sim 3.7$ mag fainter
than the primary, it should be substantially redder.  However, the event is
heavily extincted, $A_I\sim 4.4$, so that even the primary source does not
yield a good color measurement from the entire event.  Hence, measurement of
the color of the secondary source, likely 5 mag fainter in the $V$ band, is
completely hopeless.  Therefore, we strongly counsel against including
this event as planetary.

We note that the 2L1S and 1L2S models do predict very different $t_\e$ and
therefore (because $f_S t_\e$ is an invariant), different source fluxes.
Hence, it is conceivable that these could be distinguished by measuring
the source flux from future adaptive optics (AO) observations on
next-generation extremely large telescopes (ELTs).  However, we only
mention this possibility and do not pursue it in the present context.

\section{{Source Properties}
\label{sec:cmd}}

As in Section~\ref{sec:anal-preamble}, above, we begin by reproducing
(with slight modification) the
preamble to Section~4 of \citet{2018subprime}.  Again, this is done
for the convenience of the reader.  Readers who are familiar with
\citet{2018subprime} may skip this preamble.

If $\rho$ can be measured from the light curve, then 
one can use standard techniques \citep{ob03262} to determine the
angular source radius, $\theta_*$ and so infer
$\theta_\e$ and $\mu_\rel$:
\begin{equation}
\theta_\e = {\theta_*\over \rho}; \qquad
\mu_\rel = {\theta_\e\over t_\e}.
\label{eqn:mu-thetae}
\end{equation}
However, in contrast to the majority of published by-eye discoveries
(but similarly to most of new AnomalyFinder discoveries reported
in \citealt{ob191053,af2,logqlt-4,kb190253,2018prime,2018subprime}),
most of the planetary
events reported in this paper have only upper limits on $\rho$,
and these limits are mostly not very constraining.  
As discussed by \citet{2018prime}, in these cases,
$\theta_*$ determinations are not likely to be of much use, either now
or in the future.  Nevertheless, the source color and magnitude measurement
that are required inputs for these determinations may be of use in the
interpretation of future high-resolution observations, either by space
telescopes or AO on large ground-based telescopes
\citep{masada}.
Hence, like \citet{2018prime}, we calculate $\theta_*$ in all cases.
% poor rho: kb190298, kb191216, kb192783, kb190304

Our general approach is
to obtain pyDIA \citep{pydia} reductions of KMT data at 
one (or possibly several) observatory/field combinations.  These
yield the microlensing light curve and field-star photometry on the
same system.  We then determine the source color by regression of the
$V$-band light curve on the $I$-band light curve.  For the $I$-band
source magnitudes, we adopt the values and errors from the parameter
tables in Section~\ref{sec:anal} after aligning the reporting system
(e.g., OGLE-IV or KMT pySIS) to the pyDIA system via regression of
the $I$-band light curves.  While 
\citet{2018prime} were able to calibrate the KMT pyDIA color-magnitude
diagrams (CMDs) using published
field star photometry from OGLE-III \citep{oiiicat} or OGLE-II 
\citep{oiicat1,oiicat2,oiicat3}, only 3 of the 7 subprime-field events in
this paper are covered by these catalogs.  Hence, for the remaining 4,
we work directly in the KMTC pyDIA magnitude system.  Because the
$\theta_*$ measurements depend only on photometry relative to the clump,
they are unaffected by calibration.  In the current context, calibration
is only needed to interpret limits on lens light.  Where relevant,
we carry out an alternative approach to calibration.

We then follow the standard method of \citet{ob03262}.  We adopt the
intrinsic color of the clump $(V-I)_{0,\rm cl}= 1.06$ from \citet{bensby13}
and its intrinsic magnitude from Table~1 of \citet{nataf13}.
We obtain 
$[(V-I),I]_{\rm S,0} = [(V-I),I]_{\rm S} + [(V-I),I]_{\rm cl,0} - [(V-I),I]_{\rm cl}$.
We convert from $V/I$ to $V/K$ using the $VIK$ color-color relations of
\citet{bb88} and then derive $\theta_*$ using the
relations of \citet{kervella04a,kervella04b} for giant and dwarf sources,
respectively.  After propagating errors, we
add 5\% in quadrature to account for errors induced by the overall method.
These calculations are shown in Table~\ref{tab:cmd}.
Where there are multiple
solutions, only the one with the lowest $\chi^2$ is shown.  However,
the values of $\theta_*$ can be inferred for the other solutions by noting
the corresponding values of $I_S$ in the event-parameter tables and using
$\theta_*\propto 10^{-I_S/5}$.  In any case, these are usually the same within
the quoted error bars.

Where relevant, we report the astrometric 
offset of the source from the baseline object.

Comments on individual events follow, where we also note any
deviations from the above procedures.

\subsection{{KMT-2019-BLG-0298}% ob190445
\label{sec:cmd-kb190298}}

The positions of the source and clump centroid are shown in blue and red
respectively in Figure~\ref{fig:allcmd1} together with the background
of neighboring field stars.  The blended light is consistent with zero,
so it is not represented in the CMD.  The source position (derived from
difference imaging) is offset from the baseline object by 24 mas, which
is consistent with measurement error.

On the other hand, the $1\,\sigma$ error on the blended flux is about 7\%
of the source flux, which would correspond to $I_B\sim 20.4$.  According to
the KMT website\footnote{This site uses the $A_K$ map of \citet{gonzalez12}
and assume $A_I= 7\,A_K$.}, there
are $A_I\sim 2.9\,$mag of extinction toward this line of sight, while the mean
distance modulus of the bar is 14.30 \citep{nataf13}.  Hence, a bulge lens
star that saturated this $1\,\sigma$ limit would have $M_I\sim 3.2$,
implying that no useful limit can be placed on flux from the lens.

While the normalized source size is not well measured, it is constrained
to be $\rho<0.075$ at $3\,\sigma$.  From the values of $\theta_*=6.12\,\mas$
and $t_\e=27.7\,$day in Tables~\ref{tab:cmd} and \ref{tab:kb0298parms},
we therefore obtain
$\theta_\e > 0.082\,\mas$ and $\mu_\rel > 1.1\,\masyr$.  As can be seen
from Equation~(\ref{eqn:probmu2}), this is only marginally constraining.
Nevertheless, we will use the $\rho$ envelope function in
Section~\ref{sec:phys-kb190298} to constrain the Bayesian analysis.

% gaia -1.6962 0.4689 -4.200 0.638 -5.908 0.411  1.250 ruwe
Finally, we note that {\it Gaia} \citep{gaia16,gaia18}
reports a source proper motion
\begin{equation}
\bmu_S(N,E) = (-5.91 \pm 0.41, -4.20\pm 0.64)\,\masyr \qquad ({\rm Gaia}).
\label{eqn:gaia0298}
\end{equation}
There are two reasons for mild caution regarding this result.
First, the same solution yields a $3.6\,\sigma$ negative parallax,
$\pi_S = -1.696\pm 0.469\,\mas$.  Second, the Gaia RUWE parameter is 1.25.
It is extremely unlikely that the large negative parallax is due to normal
statistical fluctuations if one interprets the error bars naively.
\citet{2018subprime} showed that ``high'' RUWE numbers are indicative
of spurious source proper motions in microlensing events.  While, these
``high'' values were all above 1.7, i.e., far above the RUWE value of 1.25
for KMT-2019-BLG-0298, it is still the case that this RUWE value is somewhat
above average.  Noting that \citet{rybizki22} found that {\it Gaia} errors
in microlensing fields are typically underestimated by a factor of two,
we accept the estimate of Equation~(\ref{eqn:gaia0298}), but we double the
error bars.  With this revision, the negative parallax becomes $<2\,\sigma$.

We note that the source is a typical bulge clump giant, both from its
proper motion and its position on the CMD.

\subsection{{KMT-2019-BLG-1216} %ob191033
  \label{sec:cmd-kb191216}}

The positions of the source and clump centroid are shown in blue and red
respectively in Figure~\ref{fig:allcmd2}, while the blended light is shown
in green.

Our procedures differ substantially from most other events in this paper.
First, we do not obtain a reliable source color from regression because
the $V$-band signal is too weak.  Therefore, to determine the source position
on the CMD, it is unnecessary to make use of the pyDIA reductions.  Instead, we
go directly from the OGLE-IV value and error shown in
Table~\ref{tab:kb1216parms} to the calibrated OGLE-III system by finding the
$I$-band offset between OGLE-III and OGLE-IV from
comparison stars.  We then find the offset relative to the clump
(Table~\ref{tab:cmd}) and infer from this offset  the $(V-I)_{S,0}$ intrinsic
color using the {\it Hubble Space Telescope (HST)} CMD from Baade's Window
\citep{holtzman98}.

To find $I_B$, we subtract this source flux from the flux of the baseline
object in the OGLE-III catalog, $I_{\rm base}=20.15$.  Unfortunately, there
is no color measurement for this object in the OGLE-III catalog.  Therefore,
to estimate its color, we first identify its counterpart
in the KMTC pyDIA catalog.  After transforming the photometry to the
OGLE-III system, we find agreement for $I_{\rm base}$ within 0.03 mag.
Therefore, we transform the pyDIA $(V-I)_{\rm base}$ into the OGLE-III
system and then proceed to find $(V-I)_B$ in the usual way.

The baseline object is offset from the source by 120 mas, which means
it cannot be the lens, and in fact cannot be dominated by the lens.
If there were no errors in the estimates of $I_{\rm base}$ and $I_S$,
this would imply that the blend flux would place a very conservative upper limit
on the lens flux.  In fact, $I_S$ has a 0.25 mag error from the modeling,
although this has only a small effect on $I_B$ because $>70\%$ of the
baseline light comes from the blend.  The error in the DoPhot \citep{dophot}
photometry of the baseline object is of greater concern.  While it is
encouraging that OGLE-III and KMTC pyDIA agree closely on this measurement,
both could be affected by the mottled background of these crowded fields
\citep{mb03037}.  Therefore, to be truly conservative, we place a
limit on the lens flux of twice the inferred blend flux, i.e.,
$I_L > I_B - 0.75 = 19.76$.

When calculating $\theta_*$, we take account of the correlation between
$I_{S,0}$ and $(V-I)_{S,0}$ in the above-described color-magnitude-relation
method.  That is, at each possible offset (i.e., taking account of the
0.25 mag error in $I_S$) we allow for a 0.1 mag spread in $(V-I)_{S,0}$,
centered on the value for that $I_S$.  Then we consider the ensemble
of all such estimates within the quoted error of $I_S$.

\subsection{{KMT-2019-BLG-2783} 
\label{sec:cmd-kb192783}}

The positions of the source and clump centroid are shown in blue and red
respectively in Figure~\ref{fig:allcmd2}.  After transforming the source
flux to the OGLE-III system and comparing to the OGLE-III baseline
object, we find that the blended light is consistent with zero.
From the pyDIA analysis, we find that the source is offset from the
baseline object by only 14 mas, which is consistent with zero within
the measurement errors.

We find that all values of $\rho<(8,12,15)\times 10^{-3}$ are consistent at
$(1,2,3)\,\sigma$.  Given the value $\theta_*=0.39\,\mas$ from
Table~\ref{tab:cmd} and $t_\e=23.6\,$day from Table~\ref{tab:kb2783parms},
these values correspond to $\mu_\rel > (0.75,0.50,0.40 )\,\masyr$.  These are
virtually unconstraining according to Equation~(\ref{eqn:probmu2}).
Nevertheless, we will include a $\rho$-envelope function in the Bayesian
analysis of Section~\ref{sec:phys-kb192783}.

Based on the absence of blended light, we set the limit on lens flux
at half the source flux, i.e., $I_L>I_S +0.75=20.93$.

\subsection{{OGLE-2019-BLG-0249} %kb190109
\label{sec:cmd-ob190249}}

The positions of the source and clump centroid are shown in blue and red
respectively in Figure~\ref{fig:allcmd1}.  The blended light (green) cannot
be determined from the KMT pyDIA analysis because there is no
true ``baseline'' during 2019.  Rather, we find the blended flux
from OGLE-IV and transform to the pyDIA system, $I_B = 19.75\pm 0.05$.
While the source color is determined with high precision from this
very bright event, the error in the blend color is very large,
$(V-I)_B=2.70\pm 0.22$.  Nevertheless, the color plays no significant
role because, as we will show, the blend is unlikely to be
related to the event.

Using a special pyDIA reduction, with a late-season-based template,
we find that the source position (derived from difference images)
is offset from the baseline object by 53 mas.  Taking account
of the fact that these late-season images are still magnified by
$A\sim 1.06$,  this separation should be corrected to
$\Delta\theta_{S,\rm base} = 56\,\mas$.  This implies that the separation
of the source (and so, lens) from the blend is
$\Delta\theta_{S,B} = (1+f_S/f_B)\Delta\theta_{S,\rm base}=240\,\mas$.
Hence, on the one hand, it cannot be the lens and is very unlikely
to be a companion to either the source or the lens.  On the other hand,
it lies well within the point-spread function (PSF), so it is
undetectable in seeing-limited images.
% xbase = (xb*fb + 0*fs)/(fs+fb) ; xb =

When $\bpi_\e$ and $\bgamma$ are included in the fits,
there are well defined minima in $\rho$ for the wide solutions, but less so
for the close solutions.  See Figure~\ref{fig:rho}.  Hence, we will use
$\rho$-envelope functions in the Bayesian analysis of
Section~\ref{sec:phys-ob190249} in all cases.

We set the limit on lens flux
as that of the blend flux, i.e., $I_L>
I_{B,\rm Cousins} \simeq I_{B,\rm pyDIA} - 0.1 = 19.65$, where we have
estimated (based on other events) a 0.1 mag offset between the
pyDIA and standard systems.  Note that while
it is true that the blend flux could be underestimated due the
mottled background of crowded bulge fields, it is also the
case that the lens flux can comprise no more than $3/4$ of all
the blended flux: otherwise the remaining light would be so far
from the source as to be separately resolved.

Finally, we adopt the {\it Gaia} proper motion measurement,
\begin{equation}
\bmu_S(N,E) = (-5.29 \pm 0.39, -3.92\pm 0.77)\,\masyr \qquad ({\rm Gaia}),
\label{eqn:gaia0249}
\end{equation}
noting that it has a RUWE value of 1.00.

\subsection{{OGLE-2019-BLG-0679} %kb192688
\label{sec:cmd-ob190679}}

\begin{figure}
\plotone{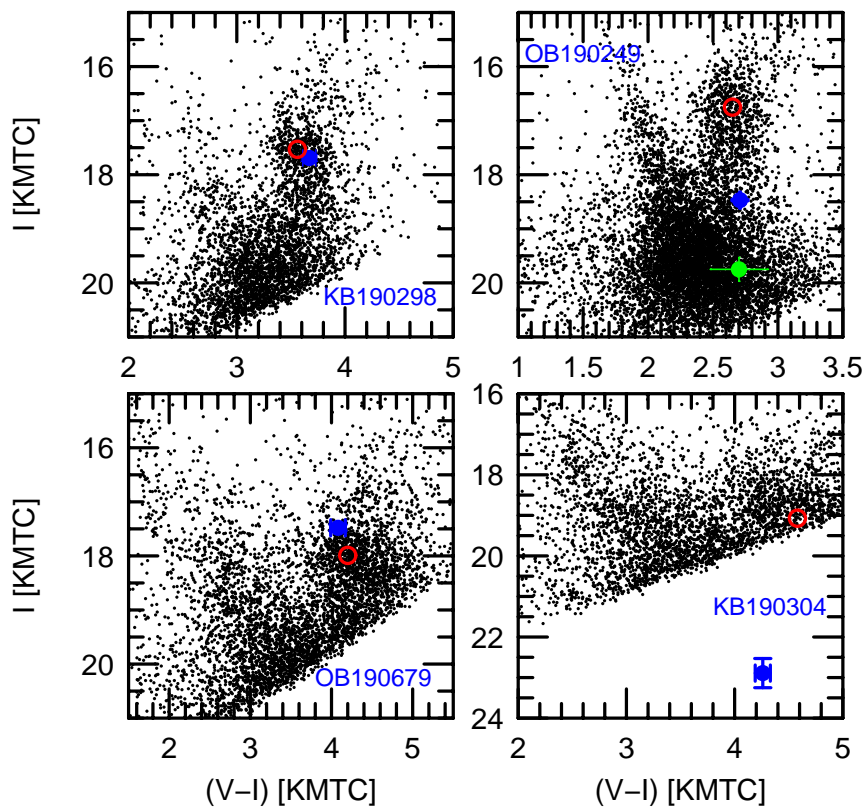}
\caption{Color-magnitude diagrams for 4 of the 7 events analyzed in
this paper, each identified by an abbreviation, e.g.,  KB190298 for
KMT-2019-BLG-0298.  The centroid of the red clump and the lens position
are always shown in red and blue, respectively.  Where relevant, the
blended light is shown in green.  When there are multiple solutions,
we show only the source and blend for the lowest-$\chi^2$ solution.
}
\label{fig:allcmd1}
\end{figure}

\begin{figure}
\plotone{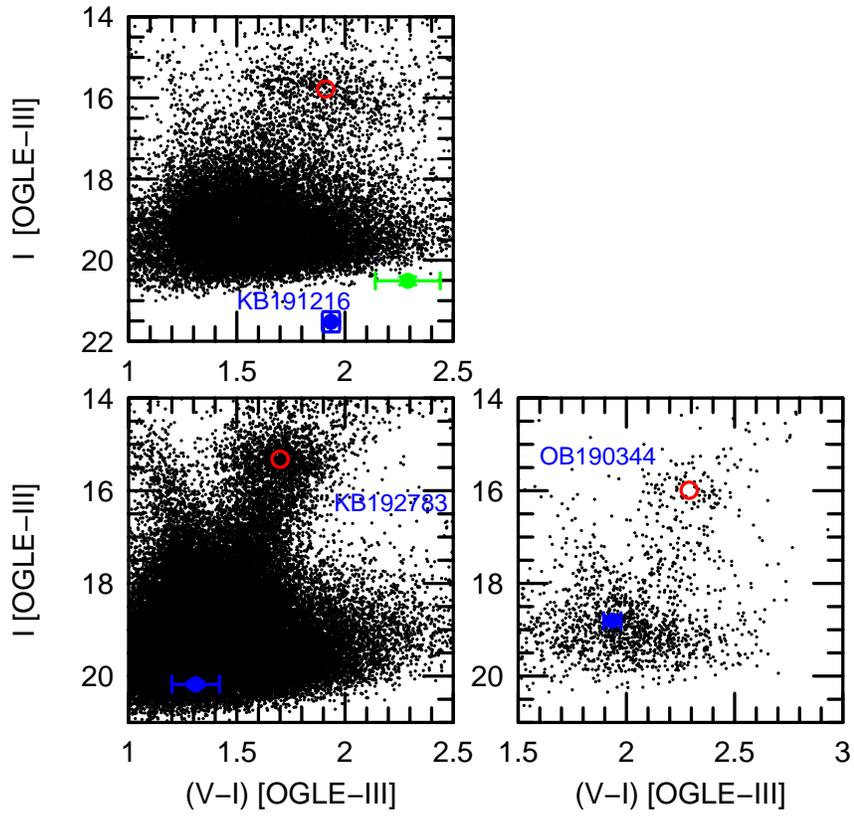}
\caption{Color-magnitude diagrams for the remaining 3 events analyzed in
this paper.  See caption of Figure~\ref{fig:allcmd1}.
}
\label{fig:allcmd2}
\end{figure}

\begin{figure}
%\plotone{rho-envelope.eps}
\plotone{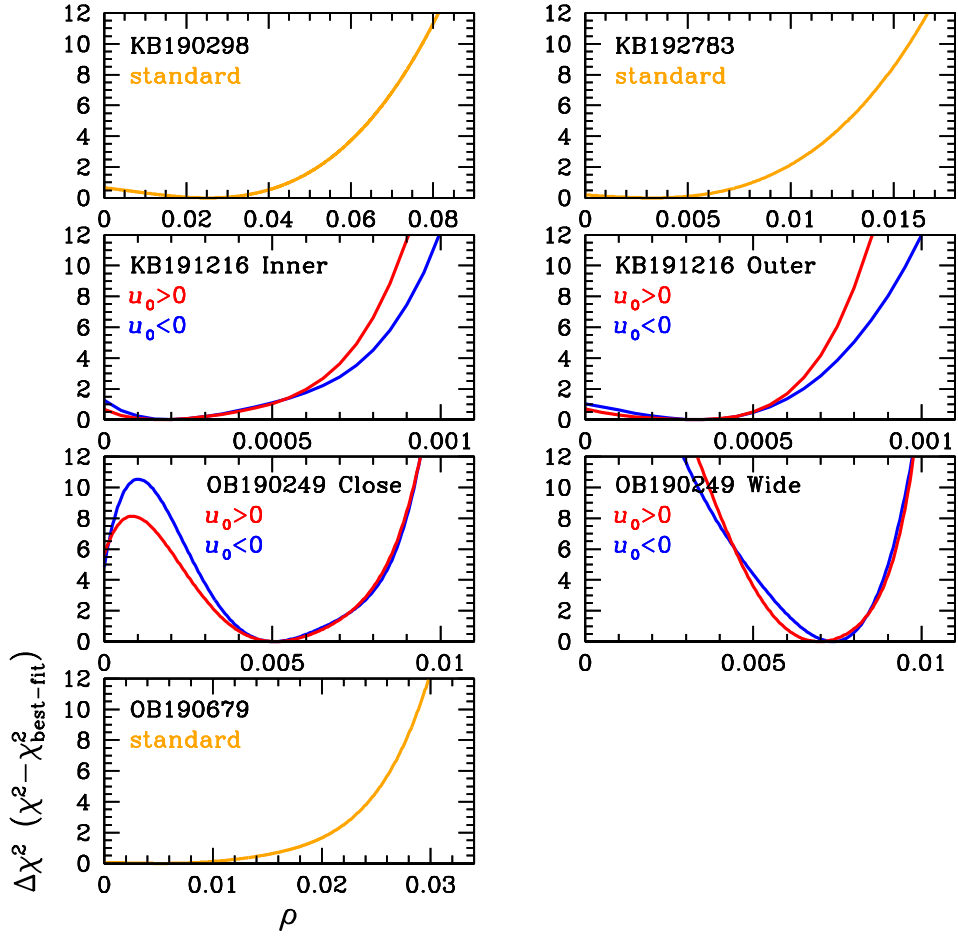}
  \caption{Envelope functions for $\chi^2$ versus $\rho$ for all 5 planetary
    events in this paper.
}
\label{fig:rho}
\end{figure}

The positions of the source and clump centroid are shown in blue and red
respectively in Figure~\ref{fig:allcmd1}.  According to the better ($u_0<0$)
solution in Table~\ref{tab:ob0679parms}, blended light is detected at
$2.5\,\sigma$.  For Gaussian statistics, this would have a low false-alarm
probability, $p\sim 0.7\%$.  Nevertheless, as we now discuss, we treat this
detection cautiously.

The key point is that the offset between the source and the baseline
object is only 17 mas, which is consistent with zero
within the measurement error.  This would
naturally be explained if there were no blended light or, as a practical matter,
much less than is recorded in Table~\ref{tab:ob0679parms}.  In principle, it
might also be explained by the blend being associated with the event, either
the lens or a companion to the lens or the source.  However, this possibility is
itself somewhat problematic.  That is, this is a heavily extincted
field, $A_I=3.8$, so if this blend is behind most of the dust, then
$I_{B,0}\sim 15.7$.  Hence, if it is in the bulge (e.g., as a companion
to the source or as part of the lens system) then it is a giant, i.e.,
$M_I\sim 1.3$.  And to be an unevolved main-sequence lens (or companion
to the lens), it would have to be at $D_L\la 2\,\kpc$.  Of course, this
is not impossible, but it is far from typical.

Secondly, there is much experience showing that microlensing photometry
does not obey Gaussian statistics, so the $p<1\%$ false-alarm probability
cannot be taken at face value.  The combination of this reduced confidence
with the low prior probability for so much blended light so close to the
lens is what makes us cautious about this interpretation.  While we
adopt the source flux as measured by the fit (i.e., less than the baseline
flux), we do not claim to have detected blended light, and therefore
we do not show an estimate of the blend in Figure~\ref{fig:allcmd1}.

For both the standard and parallax fits, $\rho$ is poorly constrained.
Hence, we will apply the $\rho$-envelope function
%s, as derived from the parallax solutions,
in the Bayesian analysis of
Section~\ref{sec:phys-ob190679}. See Figure~\ref{fig:rho}.

The color measurement, which is tabulated in Table~\ref{tab:cmd}
and illustrated in Figure~\ref{fig:allcmd1}, presented some difficulties
because the event is low amplitude and suffers heavy extinction.
Both factors contribute to low flux variation in the $V$ band.
Our usual approach, based on regression of the
magnified event in KMTC data, yields $(V-I)_S =4.01\pm 0.12$.
By comparison, the color of baseline object, which has more than twice
the flux of the difference object even at the peak of the event,
has an identical central value but substantially smaller error,
$(V-I)_{\rm base} =4.01\pm 0.06$.  This coincidence of color values would be
a natural consequence of the zero-blending hypothesis, but the errors are
too large to draw any strong conclusions.  Note that both central values
place the source $\Delta(V-I) = -0.19$ blueward of the clump.

We made two further efforts to clarify the situation.  First, we made
independent reductions of KMTS data.  These produced a similar color,
but with an error bar that was more than twice as large.  Combining
KMTC and KMTS yields $(V-I)_S =4.01\pm 0.11$, which is not a significant
improvement.

Second we found the offset from the clump in $(I-H)$,
making use of ANDICAM $H$-band data for the light curve and VVV data
for the baseline-object and field-star $H$-band photometry.
And we compare these to
the offsets in $(V-I)$ using the color-color relations of \citet{bb88}.
For the baseline object, we find $\Delta(I-H)_{\rm base} = -0.12\pm 0.02$,
corresponding to $\Delta(V-I)_{\rm base} = -0.10\pm 0.02$, which is marginally
consistent with the direct $V/I$ measurement at $1.4\,\sigma$.

On the other hand, the regression of the $H$-band light curve leads to
$\Delta(I-H)_S = +0.17\pm 0.06$.  This is inconsistent at $4.6\,\sigma$
with the color offset of baseline object.
In principle, $\Delta(I-H)$ need not be the
same for the source and the baseline because the baseline can have
a contribution from blended light of a different color.  However, to explain
such a large offset from just 15\% of the $I$-band light would require
an extraordinarily red blend.  Considering, in particular, that the blend
lies just 1.3 mag below the clump, we consider this to be very unlikely.

In the face of this somewhat contradictory evidence, we adopt
$\Delta(V-I)_S = -0.12\pm 0.09$.  That is, first, given that the baseline
light is dominated by the source, it provides the best first guidance to
the source color.  We then adopt a compromise value between the $V/I$
and $I/H$ determinations that is consistent with both at $\sim 1\,\sigma$.
This value is also well within the $1\,\sigma$ interval of the source-color
determination in $V/I$.  For the error bar, we adopt the offset between
these two baseline-object determinations, in recognition of the fact that
they disagree by more than $1\,\sigma$.  We consider that the
$I/H$ determination of the source color is most likely spurious.

Although unsatisfying, any errors in our adopted resolution of this issue
do not have significant implications for the results reported in this paper.
The source color (as well as the degree of blending) only impact the
$\theta_*$ determination, and only at $\la 15\%$.  This would be of some
concern if we had a precise $\rho$ measurement, in which case it would
impact $\theta_\e$ at the same level.  However, we basically have only
an upper limit on $\rho$, and this fact completely dominates the uncertainty
in $\theta_\e$.  We have presented a thorough documentation of this issue
mainly for reference, in case it becomes relevant to the interpretation
of {\it Spitzer} data.  That is, when {\it Spitzer} data do not cover the
peak of the light curve (as appears to be the case for OGLE-2019-BLG-0679),
the parallax measurement can sometimes be substantially improved if the
{\it Spitzer} source flux is independently constrained via a
ground-{\it Spitzer} color-color relation together with a ground-based
color measurement.  Hence, a thorough understanding of potential uncertainties
in the latter can be of direct relevance.

We do not attempt to place any limit on the lens light.  At the $2\,\sigma$
level, $I_B>18.8$, which (assuming the blend lies behind most of the dust)
corresponds to $I_{B,0} > 15.0$, and so is not constraining.

Finally, we note that {\it Gaia} reports a proper motion measurement
\begin{equation}
\bmu_S(N,E) = (-6.04 \pm 0.28, -6.35\pm 0.51)\,\masyr \qquad ({\rm Gaia\ DR3}).
\label{eqn:gaia0679}
\end{equation}
However, it also reports a RUWE number, 1.75.  Based on a systematic
investigation of {\it Gaia} proper motions of microlensed sources,
\citet{2018subprime} concluded that such high-RUWE measurements were often
spurious or, at least, suspicious.  In the present case, caution is further
indicated by the fact that the {\it Gaia} DR2 measurement,
$\bmu_S(N,E) = (-6.98 \pm 0.79, -3.42\pm 1.12)\,\masyr$, is inconsistent
with the DR3 measurement, even though they are based mostly on the same data.

These discrepancies lead us to make our own independent measurement
of $\bmu_S$ based on almost 10 years of OGLE-IV data, which yields,
\begin{equation}
\bmu_S(N,E) = (-5.32 \pm 0.37, -8.36\pm 0.16)\,\masyr \qquad ({\rm OGLE-IV}).
\label{eqn:oiv0679}
\end{equation}
This measurement is strongly inconsistent with the {\it Gaia} DR3 measurement,
casting further doubt upon the latter.  However, as we discuss in
Section~\ref{sec:phys-ob190679}, the OGLE-IV measurement is,
similar to {\it Gaia} DR3, in significant tension with other information
about the event.

Therefore, we do not incorporate any $\bmu_S$
measurement in the Bayesian analysis of
Section~\ref{sec:phys-ob190679}.  Nevertheless, as we will discuss in that
section, the various estimates of $\bmu_S$ raise enough concerns about the
microlensing $\bpi_\e$ measurement as to convince us to report the
``Standard'' (i.e., non-parallax) solution for our final values.

\subsection{{OGLE-2019-BLG-0344} %kb190149
\label{sec:cmd-ob190344}}

As discussed in Section~\ref{sec:anal-ob190344}, this event has
planetary solutions, but it cannot be claimed as a planet.  Hence, the
CMD analysis is presented solely for completeness.  Because the
source is consistent with being unblended in the planetary fits, we
simply adopt the parameters of the OGLE-III baseline object as those
of the microlensed source.  These are shown as a blue circle in
Figure~\ref{fig:allcmd2}, while the source centroid is shown as a red circle.

\subsection{{KMT-2019-BLG-0304}
\label{sec:cmd-kb190304}}

Due to heavy extinction, the red clump on the CMD is partially truncated
by the $V$-band threshold.  Therefore, we determine the height of the
clump in the $I$-band by matching the pyDIA to the
VVV catalog \citep{vvv-survey,vvvcat} and then determining the
$(V-I)$ color from the portion of the red clump that survives truncation.  This
is shown as a red circle in Figure~\ref{fig:allcmd1}.
We then determine the offset
from the clump in the $I$ band (Table~\ref{tab:cmd}) and then apply the
{\it HST} color-magnitude relation, as in Section~\ref{sec:cmd-kb191216}.
Note that the color of the red clump centroid plays no role in this
calculation, and it is shown in Figure~\ref{fig:allcmd1} only to maintain
a consistent presentation with other events.

\section{{Physical Parameters}
\label{sec:phys}}

To make Bayesian estimates of the lens properties, we follow the same
procedures as described in Section~5 of \citet{2018prime}.  We refer the
reader to that work for details.  Below, we repeat the text from
Section~5 of \citet{2018subprime} for the reader's convenience.

In Table~\ref{tab:physall},
we present the resulting Bayesian estimates
of the host mass $M_{\rm host}$, the planet mass $M_{\rm planet}$, 
the distance to the lens system $D_L$, and the planet-host projected
separation $a_\perp$.  For the three of the five
events, there are two or more competing solutions.  For these cases
(following \citealt{2018prime}),
we show the results of the Bayesian analysis for each solution separately,
and we then show the ``adopted'' values below these.  For $M_{\rm host}$,
$M_{\rm planet}$, and $D_L$, these are simply the weighted averages of the
separate solutions, where the weights are the product of the two
factors at the right side of each row.  The first factor is simply
the total weight from the Bayesian analysis.  The second is 
$\exp(-\Delta\chi^2/2)$ where $\Delta\chi^2$ is the $\chi^2$ difference
relative to the best solution.  For $a_\perp$,
we follow a similar approach provided that either the individual solutions
are strongly overlapping or that one solution is strongly dominant.  
If neither condition were met, we would enter ``bi-modal'' instead.
However, in practice, this condition is met for all 3 events for which
there is potentially an issue.  Note that in all cases (including those
with only one solution), we have provided symmetrized error bars in the
``adopted'' solution, for simplicity of cataloging.  The reader interested in
recovering the asymmetric error bars can do so from the table.

We present Bayesian analyses for 5 of the 7 events, but not for
OGLE-2019-BLG-0344 and KMT-2019-BLG-0304, for which
we cannot distinguish between competing interpretations of the event.
See Sections~\ref{sec:anal-ob190344} and \ref{sec:anal-kb190304}.
Figures~\ref{fig:bayes1} and \ref{fig:bayes2}
show histograms
for $M_{\rm host}$ and $D_L$ for these 5 events.

\subsection{{KMT-2019-BLG-0298}% ob190445
  \label{sec:phys-kb190298}}

As discussed in Section~\ref{sec:anal-kb190298}, we accept the event parameters
from the standard (7-parameter) solution in Table~\ref{tab:kb0298parms}, but
incorporate the $\bpi_\e$ constraints from the parallax-plus-orbital-motion
solution.  Again, the reason for this is that the $\bpi_\e$ constraints are
essentially 1-D, so the parallax MCMC explores regions of very high $|\bpi_\e|$,
which would be highly suppressed after incorporating Galactic priors.

In the Bayesian analysis, there are four constraints, i.e., on
$t_\e$, $\bmu_S$, $\rho$, and $\bpi_\e$. The first is $t_\e=27.71\pm0.62\,$day
from Table~\ref{tab:kb0298parms}.  The second is
$\bmu_S(N,E) = (-5.91\pm0.82,-4.20\pm 1.28)\,\masyr$ from
Section~\ref{sec:cmd-kb190298}.  The third is given by
$\exp(-\Delta\chi^2(\rho)/2)$, where $\Delta\chi^2(\rho)$ is the envelope
function that is shown in Figure~\ref{fig:rho}.  For the fourth, we represent
the $\bpi_\e$ scatter plots shown in Figure~\ref{fig:0298par}
as Gaussian ellipses
(also illustrated in this figure) with
means and covariance matrices derived from the MCMC.  These have central values
and error bars similar to those shown in Table~\ref{tab:kb0298parms}
(based on medians) and with correlation coefficients 0.95 and 0.60 for
the $u_0>0$ and $u_0<0$ solutions, respectively.  They are highly linear
structures with minor axes 
$\sigma_\parallel = (0.039,0.046)$ 
and axis ratios of
$\sigma_\perp/\sigma_\parallel = (11.0,12.3)$ for the respective cases.

% 0.9466 0.0392 0.4313 11.0168  u0+
% 0.6036 0.0458 0.5633 12.2971  u0-

\begin{figure}
\plotone{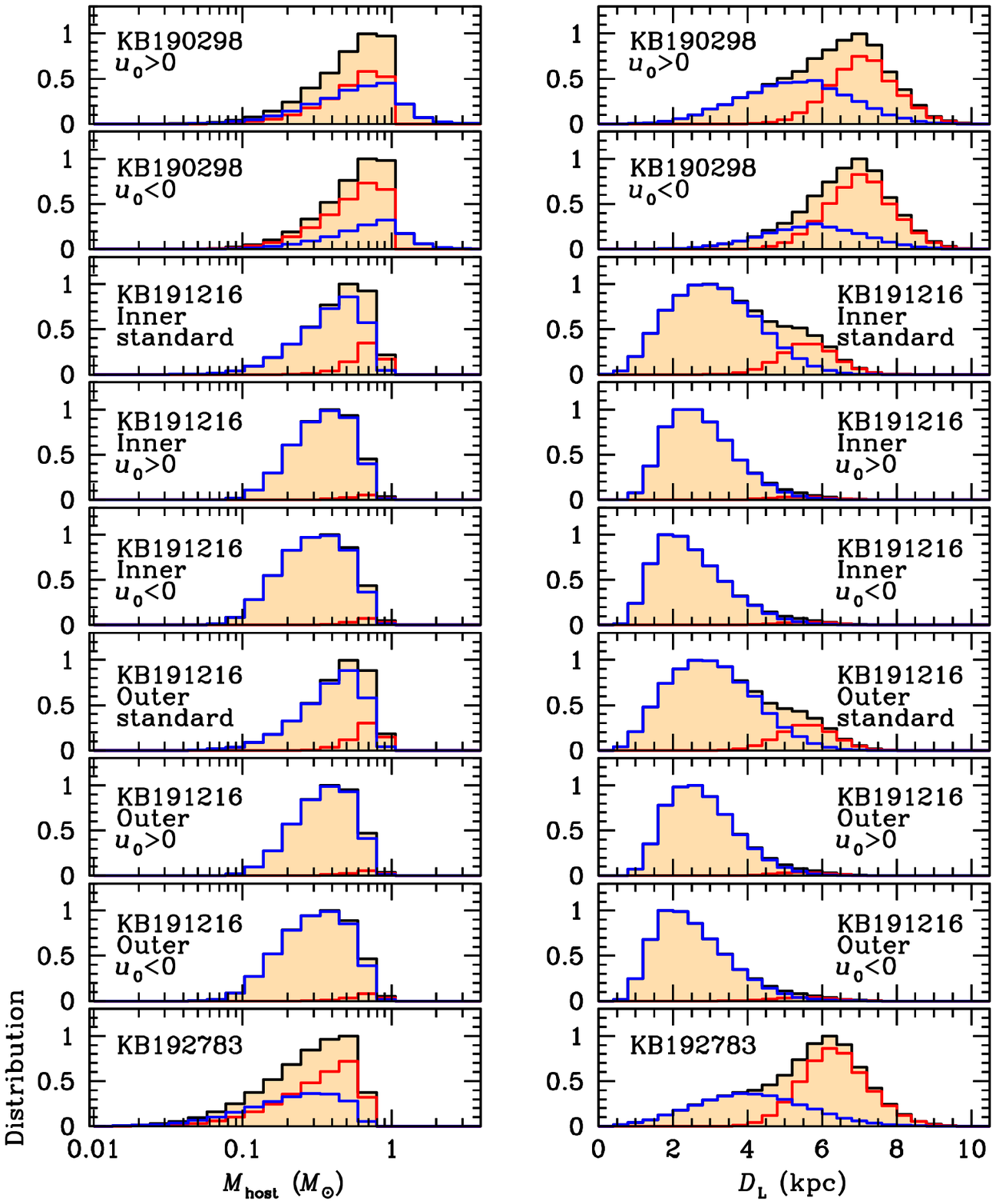}
\caption{Histograms of the host mass (left) and lens distance (right)
for 3 of the 5 unambiguously planetary events, as derived from the 
Bayesian analysis.  Disk (blue) and bulge (red) distributions are shown
separately, with their total shown in black.
}
\label{fig:bayes1}
\end{figure}

The Bayesian estimates (Table~\ref{tab:physall} and Figure~\ref{fig:bayes1})
favor $M\sim 0.7\,M_\odot$
hosts that are
in or near the bulge, i.e., small $\pi_\rel$.  This preference is due to the
$\bpi_\e$ constraint, which is, effectively, a 1-D structure passing
through the origin.  Hence, for randomly oriented $\bmu_\rel$ (so $\bpi_\e$),
the fraction of surviving simulated events scales $\propto\pi_\e^{-1}$, while
the very weak constraints on $\rho$ (so $\mu_\rel$), imply that typical
$\mu_\rel \sim 5\,\masyr$ are favored, so $\theta_\e\sim 0.4\,\mas$.
Low $\pi_\e$ then drives $\pi_\rel=\pi_\e\theta_\e$ to low values, and
it drives $M=\theta_\e/\kappa\pi_\e$ to the higher range of the available
mass function.  Nevertheless, the fact that the $(u_0>0)$ solution for $\bpi_\e$
closely tracks the direction of Galactic rotation (i.e., $\sim 30^\circ$ north
through east), combined with the fact that the source is measured to
be moving at $\mu_S\sim 7.3\,\masyr$ at $\sim -145^\circ$ (north through east,
i.e., almost anti-rotation),
permits disk hosts with very small $D_L$.  See Figure~\ref{fig:bayes1}.

\subsection{{KMT-2019-BLG-1216} %ob191033
\label{sec:phys-kb191216}}

As discussed in Section~\ref{sec:anal-kb191216}, we adopt a cautious
attitude toward incorporating the $\bpi_\e$ measurement.  That is,
given the relatively high value of $\pi_\e\sim 0.6$,
the $p=3\%$ false-alarm probability of this measurement would be too
high to accept it for typical microlensing events, for which $\pi_\e$ is
generally much closer to zero.  Therefore, we begin the Bayesian analysis using
the standard (7-parameter) solution in Table~\ref{tab:kb1216parms}.  There are
then three constraints, i.e., on $t_\e$ (from Table~\ref{tab:kb1216parms}),
on $\rho$ (from the envelope function in Figure~\ref{fig:rho}), and on the
lens flux, $I_L>19.76$ from Section~\ref{sec:phys-kb191216}.
The results are shown in Table~\ref{tab:physall}
and illustrated in Figure~\ref{fig:bayes1}.

The results favor nearby lenses $D_L\sim 3.5\,\kpc$, corresponding to
$\pi_\rel \sim 0.17\,\mas$.  The reason is that while $\rho$ is not measured,
it is constrained at, e.g., $2\,\sigma$ to be $\rho\la 8.5\times 10^{-4}$,
corresponding to $\theta_\e>0.47\,\mas$.
Because this is a long event, this threshold
corresponds to $\mu_\rel>1.9\,\mas$, which is moderately low.
Hence, somewhat bigger $\theta_\e$ are favored by Galactic kinematics,
e.g., $\theta_\e\sim 0.8\,\mas$, which would also correspond to the weak
minimum of the $\rho$-envelope function.  Considering the ``effective top''
of the mass function $M\la 1\,M_\odot$, these values respectively imply
$\pi_\rel \ga 0.03\,\mas$ and $\pi_\rel \ga 0.08\,\mas$.  For $M\sim 0.5\,M_\odot$,
i.e., closer to the peak of the mass function, these values are doubled.  Hence,
nearby lenses are strongly favored, while a broad range of masses is permitted.

The Bayesian results do not give any reason to be suspicious of the $\bpi_\e$
measurement.  The main takeaway from Figure~\ref{fig:bayes1}
is that despite the powerful
Galactic priors favoring bulge lenses (e.g., \citealt{mb09387}),
which tend to ``override'' the $\rho$
constraint, disk lenses are strongly favored.  It is notable that
the direction of $\bpi_\e$, for ($u_0 <0$), is consistent with that
of Galactic rotation at $1\,\sigma$.  While the central value of this
solution, $\pi_\e=0.7\pm 0.2$, is substantially higher than would be naively
indicated by the Bayesian analysis, the error is large.  Therefore
we incorporate this result.

We find that the main effect of incorporating the $\bpi_\e$ measurement is
to effectively eliminate the bulge and near-bulge lenses, which
(as explained above) were previously allowed due to the Galactic priors
``overriding'' the $\rho$ constraint.

\subsection{{KMT-2019-BLG-2783} 
  \label{sec:phys-kb192783}}
There is only one solution, upon which there are three constraints, i.e.,
on $t_\e=23.6\,$day (from Table~\ref{tab:kb2783parms}),
on $\rho$ (from the envelope function in Figure~\ref{fig:rho}), and on
the lens flux, $I_L>20.93$ from Section~\ref{sec:cmd-kb192783}.  However,
given that the $2\,\sigma$ limit, $\rho<0.011$, corresponds to
$\mu_\rel > 0.5\,\masyr$, the $\rho$ constraint effectively plays no role.
On the other hand, the lens-flux constraint, combined with the low extinction
($A_I\sim 0.73$, see Table~\ref{tab:cmd}) eliminates solar-type lenses
even in the bulge, and then progressively eliminates increasingly
less massive stars for increasingly nearby disk lenses.  The net
result is that the Bayesian results are compatible with a very broad
range of distances, but a mass distribution that is sharply curtailed
at the high end.  See Figure~\ref{fig:bayes1}.

\subsection{{OGLE-2019-BLG-0249} %kb190109
  \label{sec:phys-ob190249}}
There are four solutions (two parallax solutions for each of the close and wide
topologies), on which there are five constraints, i.e., on $t_\e$, $\rho$,
$I_L$, $\bmu_S$ and $\bpi_\e$.  The first comes from
Table~\ref{tab:ob0249followup}, the second from the $\rho$-envelope functions
discussed in Section~\ref{sec:cmd-ob190249} and shown in
Figure~\ref{fig:rho}, the third is $I_L>19.65$ (from
Section~\ref{sec:cmd-ob190249}), and the fourth is from {\it Gaia}
(Equation~(\ref{eqn:gaia0249})).  Finally, we characterize the
$\bpi_\e$ constraints as 2-D Gaussian distributions, whose
$\Delta\chi^2=1$ contours are shown as black ellipses in
Figure~\ref{fig:0249par}.  These have central values and error bars similar
to those shown in Table~\ref{tab:ob0249followup}
(based on medians) and with correlation coefficients
(0.91,0.92,0.95,0.94) for
the (close, $u_0>0$; close, $u_0<0$; wide, $u_0>0$; wide, $u_0<0$)
solutions, respectively.  They are highly linear
structures with minor axes 
$\sigma_\parallel = (0.0080,0.0079,0.0064,0.0067)$ 
and axis ratios of
$\sigma_\perp/\sigma_\parallel = (9.9,11.0,12.0,12.1)$
for the respective cases.

The result is that the host is very well constrained to be an upper
main-sequence star that is in or near the bulge.  See Table~\ref{tab:physall}
and Figure~\ref{fig:bayes2}.  The reason that
these constraints are much tighter than for any other event analyzed
in this paper is that, while neither $\theta_\e$ nor (the vector)
$\bpi_\e$ is well measured, both $\theta_\e$ and (the scalar) $\pi_\e$
are reasonably well constrained.

In the case of $\theta_\e$, the $\rho$-envelope functions have relatively
broad, but nonetheless well-defined, minima.  It is true that these
functions turn over for $\rho\la 0.001$ for the close solution.  However,
these values typically result in masses $M\ga 5\,M_\odot$, and so are excluded
by the mass function.  Regarding $\bpi_\e$, despite the high axis ratios
mentioned above, the (scalar) $\pi_\e$ are well constrained (and to very similar
values) in the four cases because the lines from the origin that are
perpendicular to these linear structures all pass though the $1\,\sigma$
contour, with $\sigma_\perp\simeq \pi_{\e,\rm best}$.

%close+ 0.911, 0.0080, 0.0786,  9.86
%close- 0.922, 0.0079, 0.0866, 10.96
%wide+  0.948, 0.0064, 0.0762, 11.96
%wide-  0.937, 0.0067, 0.0815, 12.08

\subsection{{OGLE-2019-BLG-0679} %kb192688
  \label{sec:phys-ob190679}}

As foreshadowed in Sections~\ref{sec:anal-ob190679} and \ref{sec:cmd-ob190679},
we ultimately decided to report final results (for this paper) based on the
``Standard'' solution of Table~\ref{tab:ob0679parms}.  We detail our reasons
for this decision at the end of this subsection.

Hence, there is only one solution on which there are
two constraints , i.e., on $t_\e$, and $\rho$.
The first comes from Table~\ref{tab:ob0679parms}, and the second comes from the
$\rho$-envelope function discussed in Section~\ref{sec:cmd-ob190679} and
shown in Figure~\ref{fig:rho}.  

\begin{figure}
\plotone{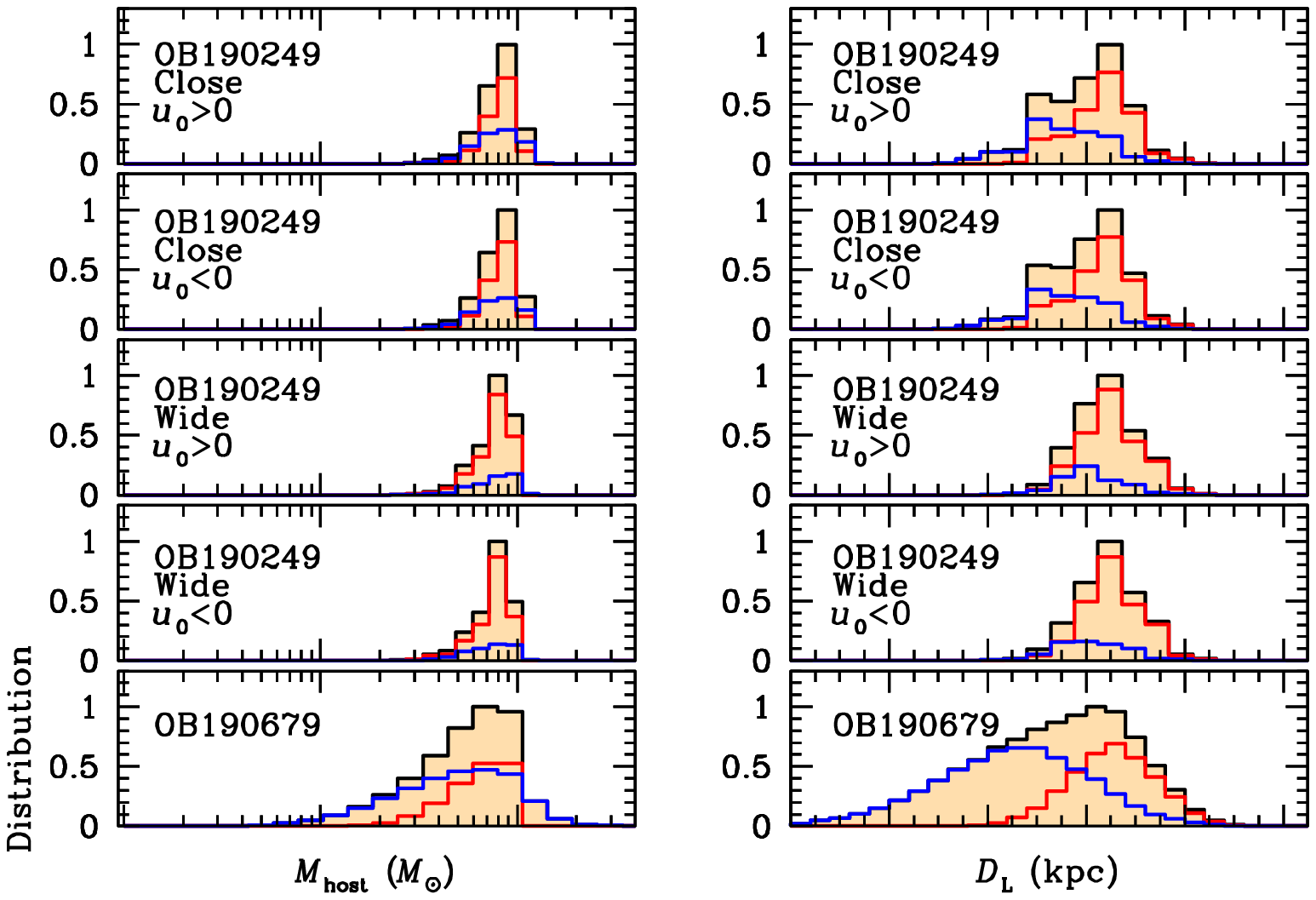}
\caption{Histograms of the host mass (left) and lens distance (right)
for the remaining 2 unambiguously planetary events, as derived from the 
Bayesian analysis.  Disk (blue) and bulge (red) distributions are shown
separately, with their total shown in black.
}
\label{fig:bayes2}
\end{figure}

Table~\ref{tab:physall} and Figure~\ref{fig:bayes2} show that the posterior
distributions of both mass and distance
are very broad, and the lens system can almost equally well reside in the bulge
or disk.  This is a consequence of the fact that the only measured constraint
is $t_\e$, while $\theta_\e$ effectively has only a lower limit.

We made the decision to adopt the ``Standard'' solution as follows.
We first carried out Bayesian analyses for both the ``Standard'' and
``Parallax'' solutions to understand how they differ not only with
respect to the parameters that we normally report (in Table~\ref{tab:physall})
and display (in Figure~\ref{fig:bayes2}) but also for the source
proper motion, $\bmu_S$, which is normally considered a nuisance parameter.
Before continuing, we note that including the parallax measurement somewhat
reduced the estimates of the host mass and distance but left broad
distributions for both.

\begin{figure}
\plotone{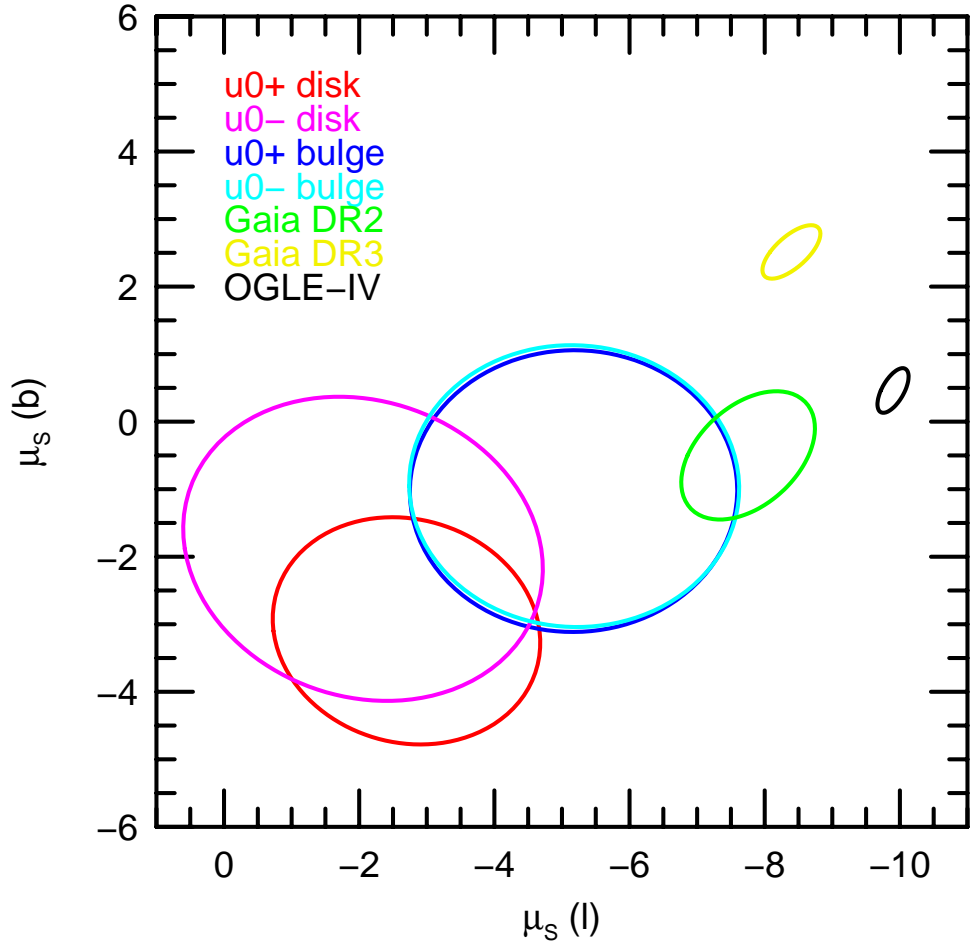}
\caption{Comparison of the posterior distributions of $\bmu_S$ for
OGLE-2019-BLG-0679, for four classes of model (as indicated in the
legend) to the {\it Gaia} DR2, {\it Gaia} DR3, and OGLE-IV
measurements of the same quantity.  In particular, {\it Gaia} DR3
and OGLE-IV are both 
in strong tension with all the other error ellipses, including
{\it Gaia} DR2.
}
\label{fig:ell}
\end{figure}

We found that, regardless of which solution ($u_0 > 0$ or $u_0 < 0$)
was correct, and regardless of whether the host was assumed to be
in the disk or the
bulge, both the {\it Gaia} DR3 and OGLE-IV measurements of $\bmu_S$ were
inconsistent at $\ga 2\,\sigma$
with the posterior $\bmu_S$ distributions.  See Figure~\ref{fig:ell}.
These tensions can be understood by considering the example of disk
lenses in the $u_0>0$ solution (red), for which $\pi_{\e,b} \simeq 0.3\pm 0.1$.
Because $\bmu_\rel$ and $\bpi_\e$ have the same direction, this implies that
$\mu_{\rel,b}$ should also be positive\footnote{Actually, what is directly
relevant is $\mu_{\rel,\hel,b}$, but the difference, which is relatively small,
is ignored here in the interest of simplicity.}.  As the prior distributions
of $\bmu_{\hel,S}$ and $\bmu_{\hel,L}$ are basically symmetric in $b$,
while $\bmu_{\rel,\hel}= \bmu_{\hel,L}-\bmu_{\hel,S}$, the posterior distributions
are driven to positive and negative values for the lenses and sources,
respectively.

This ``conflict'' may well have a perfectly reasonable explanation.
As discussed in Section~\ref{sec:cmd-ob190679}, the {\it Gaia} DR3
measurement may simply be wrong, as signaled both by its high RUWE number
and its strong disagreement with both {\it Gaia} DR2 and OGLE-IV.
See Figure~\ref{fig:ell}. 
Similarly, the OGLE-IV measurement may be wrong.
Alternatively, it may be that either {\it Gaia} DR3 or OGLE-IV 
is correct (or basically correct), that
the host lies in the bulge (blue and cyan ellipses), and that the event
characteristics are $2\,\sigma$ outliers.  In addition, it could be that
the $\bpi_\e$ measurement suffers from unrecognized systematics.
Given that any of these three explanations is possible in principle,
that they can lead to very different results,
and that the matter will probably be resolved within a year by the
{\it Spitzer} parallax measurement, the most prudent course is to defer
judgment until the {\it Spitzer} data can be properly evaluated.  This
course will also minimize the possibility that confusion will propagate
though the literature.

% + -0.548, 0.0536, 0.0997, 1.86
% -  0.683, 0.0710, 0.2256, 3.18

\subsection{{OGLE-2019-BLG-0344} %kb190149
  \label{sec:phys-ob190344}}

Because there is no compelling reason to believe that the planetary solution
is correct, we do not present a Bayesian analysis.

\subsection{{KMT-2019-BLG-0304}
\label{sec:phys-kb190304}}

Because there is no compelling reason to believe that the planetary solution
is correct, we do not present a Bayesian analysis.

\section{{Discussion}
\label{sec:discussion}}

We have analyzed here all 5 of the previously unpublished planets found by the
KMT AnomalyFinder algorithm toward the 21 KMT subprime fields.  We also
analyzed the two events that have nonplanetary solutions, but are consistent
with planetary interpretations.  Such events are rarely published, but
they are a standard feature of the AnomalyFinder series because they can
be important for understanding the statistical properties of the sample as
a whole.  A total of 5 such events were previously published from the 2018
season \citep{2018prime,2018subprime}.

\subsection{{Summary of All 2019 Subprime AnomalyFinder Planets}
\label{sec:2019summary}}

Table~\ref{tab:all2019events} shows these 5 planets and 2 ``possible planets''
in the context of the ensemble of all such 2019 subprime AnomalyFinder events.
The horizontal line distinguishes between objects that we judge as likely
to enter the final statistical sample and those that we do not.  Note that
among the latter, OGLE-2019-BLG-1470 is definitely planetary in nature, but
it has a factor $\sim 3$ discrete degeneracy in its mass ratio, $q$. 
On the other hand, KMT-2019-BLG-0414 has an alternate, orbiting
binary-source (xallarap) solution that is disfavored by only $\Delta\chi^2=4$,
and so it cannot be claimed as a planet.

Thus, of the 11 events (containing 12 planets) that are ``above the line'',
almost half are published here.  This is the main accomplishment of the
present work.  Among these 5 planets, none is truly exceptional in its
own right, although OGLE-2019-BLG-0679 has a relatively large normalized
projected separation, $s=2.18$.  Indeed, among the 53 previously published
(or summarized) AnomalyFinder planets from 2018 and 2019
\citep{2018prime,2018subprime,af2,logqlt-4}, only one had a larger
separation, i.e., OGLE-2018-BLG-0383, with $s=2.45$ \citep{ob180383}.

\subsection{{2018+2019 Planets: 4 Discrete Characterizations}
\label{sec:2018+2019summary}}

Because the AnomalyFinder planets for the 2019 prime fields \citep{af2},
as well as all of the 2018 fields \citep{2018prime,2018subprime}, have
previously been published (or summarized), our
work permits several types of comparison between different seasons,
different classes of planets, and different methods and conditions of
discovery.  At the highest level we can compare the 2018 and 2019 seasons
in terms of number of planets found by field type (prime versus subprime),
method of discovery (by-eye versus AnomalyFinder), source trajectory
(caustic crossing or not caustic crossing), and type of perturbation
(major image, minor image, or central caustic).
Table~\ref{tab:sum} presents these comparisons and summaries.

\subsubsection{{Statistical Consistency of 2018 and 2019}
\label{sec:2018+2019consistency}}

The first point is that the 2018 and 2019 seasons are consistent
with respect to all of these breakdowns.  For example, there were 33 and
25 total detections, respectively, i.e., a difference of $8\pm \sqrt{58}$
according to Poisson statistics.  Of all the various comparisons that
one could make among the various subcategories, the most ``discrepant'' is 
in the difference between the fraction of events identified by the
AnomalyFinder, 52\% versus 37\%, i.e., a difference of  $15\% \pm 13\%$,
according to binomial statistics.
Similarly, 
the fraction of planets found via major versus minor image perturbations:
55\% versus 42\%, i.e., a difference of $13\% \pm 14\%$.

Combining
five tests, i.e., the Poisson test of total detections and the 4 binomial
tests of Table~\ref{tab:cmd}, we find $\chi^2=3.37$ for 5 dof.

\subsubsection{{AnomalyFinder Yielded 40\% of All Detections}
\label{sec:AFvsByEye}}

Given that the two seasons are statistically consistent, we should
ask what can be learned from their combined statistics.  In particular,
with 58 planets, this is a factor more than 2.5 times larger than any
other homogeneously detected planetary microlensing
sample \citep{suzuki16}.  Perhaps the most
striking feature of Table~\ref{tab:sum} is that 23 of the 58 planets
(40\%) were initially identified by AnomalyFinder, despite the fact that
KMT's publicly available data (the same as are input to AnomalyFinder)
had previously been systematically searched by several
experienced modelers.  This may indicate the difficulty of by-eye searches
in the era of massive microlensing data sets.  It also shows that samples
derived from by-eye searches alone are not even approximately complete.

At the same time, AnomalyFinder has not replaced by-eye searches:
the two actually work hand-in-hand.  AnomalyFinder typically identifies
of order 250 candidates (after human review of a much larger 
candidate list) that
each requires detailed investigation to various levels.  The first step
in these massive reviews is to consult the summaries of systematic by-eye
investigations, particularly those of C.~Han, thereby reducing the number
that require new or additional investigations by a factor 3--5.  The by-eye
searches also serve as a check on the AnomalyFinder completeness.  In fact,
for 2021, we deliberately accelerated the by-eye searches with three new
``mass production'' papers \citep{kb211391,mp2021-2,mp2021-3}, as well
as many other papers on individual planets (see \citealt{mp2021-2} for a list),
so that about 18 planets that are suitable for statistical studies were
identified and prepared for publication prior to running the AnomalyFinder
algorithm.

In contrast to the other three statistical indicators that are discussed
below, the AnomalyFinder fraction of planets depends on a human factor.
For example, when the 2016-2017 data are analyzed, the fraction could go
down simply because there has been more time to apply the by-eye approach.
And, going forward, the rate could go down because humans have learned
more about planetary signatures based on the results from 2018-2019.  On
the other hand, the rate could go up if humans become less diligent, knowing
that the planets will ``eventually'' be found anyway.

\subsubsection{{50\%$\pm$7\% of Planets Have Caustic Crossings}
\label{sec:caustic-crossings}}

\citet{zhu14} predicted that for about half of the planets detected in a
KMTNet-like survey, the source would cross a caustic.  These crossings
are important because they allow the normalized source radius, $\rho$,
to be measured, which in turn enables measurement of $\theta_\e$ and
$\mu_\rel$.  In addition to helping to characterize the planet, these
measurements allow one to predict when the source and lens will be
sufficiently separated to resolve them using AO on large telescopes,
which can lead to measurements
of the host and planet masses and the system distance.  The 2018-2019
AnomalyFinder statistical sample confirms this prediction of \citet{zhu14}
at relatively high statistical precision.

We note that a large minority of planetary events that do not have caustic
crossing nevertheless yield good $\rho$ measurements because the planet
is detected when the source passes over a magnification``ridge'' that extends
from the tip of a cusp.  See, for example, OGLE-2016-BLG-1195
\citep{ob161195a,ob161195b}.  \citet{masada} showed, based on a larger 
(but inhomogeneous) sample of 102 planetary events, which substantially
overlaps the current one, that about 2/3 yield $\rho$ measurements, even
though only about 1/2 have caustic crossings.

Of the five planetary events analyzed in the present work, only one
(KMT-2019-BLG-1216) has a caustic crossing.  Yet, due to inadequate data
over the caustic, $\rho$ is not well measured.  This problem is likely
to be much more common in subprime fields, particularly those that (like 
KMT-2019-BLG-1216) have cadences of $\Gamma=0.4\,{\rm hr}^{-1}$.
None of the 4 planetary events that lacked caustic crossings yielded
precise $\rho$ measurements, although for OGLE-2019-BLG-0249,
$\rho$ was reasonably well constrained.

\subsubsection{{50\%$\pm$7\% of Major/Minor Image Perturbations Are Major}
\label{sec:major-minor}}

It has long been known that for microlensing events with high, or even moderate,
sensitivity to planets, the $(\log s,\log q)$ sensitivity diagrams are
nearly symmetric about zero in $\log s$.  One aspect of this symmetry
is understood at a very deep level, while another aspect remains,
to the best of our knowledge, completely unexplored.

\citet{griest98} showed that, for low $q$, there is a deep symmetry in the
lens equation for $s\leftrightarrow s^{-1}$ in the immediate neighborhood
of the host (or, more accurately, the ``center of magnification'').
For example, the very first planet to exhibit such a degeneracy,
OGLE-2005-BLG-071 \citep{ob05071}, has nearly identical $\chi^2$ for the
two solutions \citep{ob05071b}.
Hence, because the magnification pattern is nearly identical for the
two cases, a given source trajectory will generate very similar light curves,
and hence nearly equal detectabilities.   As a result, all published
sensitivity diagrams for high-magnification events (whose planet sensitivity
is completely dominated by the source passage close to the center of
magnification), are nearly perfectly symmetric.  See, for example,
OGLE-2007-BLG-050 \citep{ob07050} and OGLE-2008-BLG-279 \citep{ob08279}.

By contrast, for source trajectories that pass closer to the planetary
caustics than to the central caustics,  the magnification structures,
and hence the resulting light-curve morphologies, are completely different.
Major images generally have much larger caustics that are flanked
by narrow magnification ridges, while minor images have smaller caustic pairs
that are threaded by broad magnification troughs.  Because of these two very
different morphologies, one might expect the symmetry in the sensitivity
profiles to break down.

%To our knowledge, there has been no systematic investigation of whether these two very different morphologies lead to symmetric sensitivity profiles, and in the case  that they do not, to what extent this impacts the overall symmetry in $\log s$ for the sensitivity of a microlensing survey taken as a whole.

In the very first systematic study of such sensitivity, \citet{gaudi02}
presented $(\log s,\log q)$ plots for 43 microlensing events.
Despite the fact that they span a very broad range of peak
magnifications, many of these events display rough symmetry in their
sensitivity profiles.  However, in detail, many individual events also
have an asymmetry in the minimum detectable $\log q$ for $\pm\log s$,
with more sensitivity for $\log s < 0$. On the other hand, their Figure 13,
which combines the sensitivities of these 43 events, shows a slight
deviation from symmetry toward positive $\log s$.  However, they do not comment
upon either effect.

%Being from a very broad range of peak magnifications, these are not strictly symmetric as they are for typical high-magnification events.  Their Figure~13, which combines the sensitivities of these 43 events, shows a slight deviation from symmetry toward positive $\log s$, but they do not comment upon this small effect.

\begin{figure}
\plotone{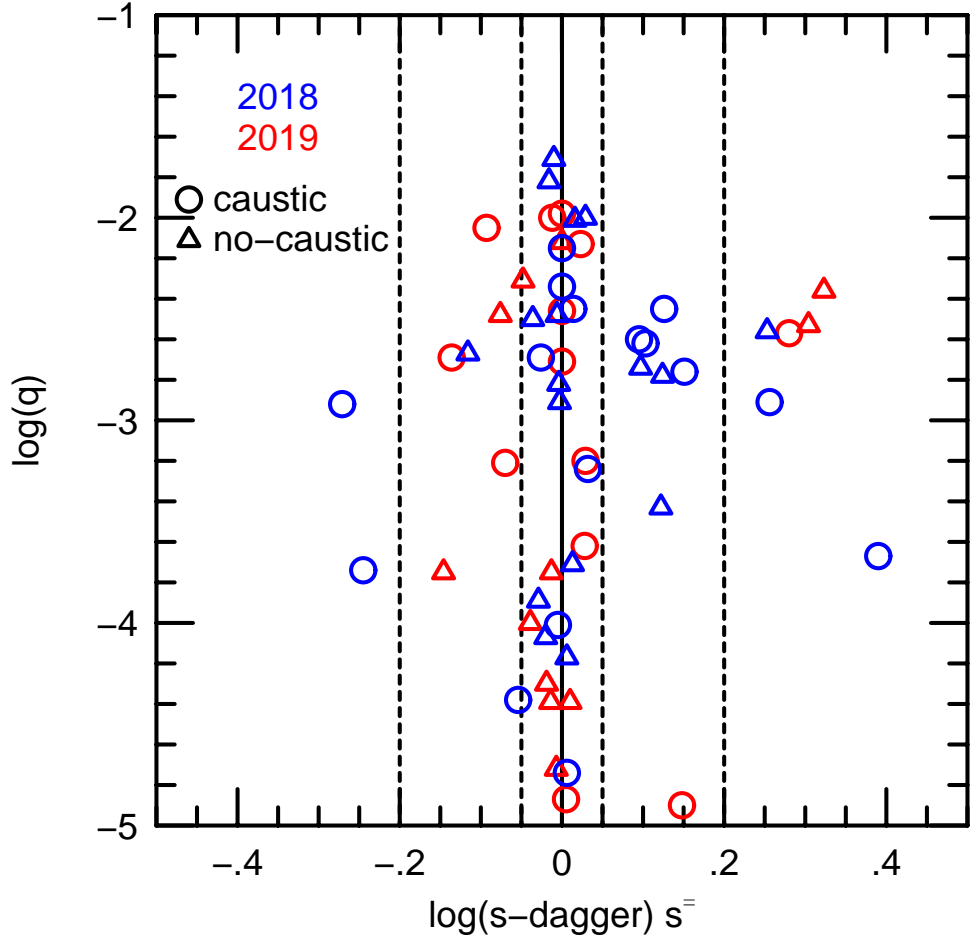}
\caption{Scatter plot of $\log s^\dagger$ versus $\log q$ for 58 AnomalyFinder
planets from 2018 and 2019.  Here,
$s^\dagger_\pm=(\sqrt{u_{\rm anom} + 4}\pm u_{\rm anom})/2$, $u_{\rm anom}$ is the
lens-source separation normalized to $\theta_\e$ at the time of the anomaly,
and the ``$\pm$'' refers to major and minor image perturbations, respectively.
A majority of detections have $|\log s^\dagger|<0.05$ (inner dashed lines),
for which the same light-curve morphology can almost equally be generated
by positive or negative $\log s$, including with large absolute values.
Hence, there is no correspondence between the signs of $\log s^\dagger$ and
$\log s$ in this regime.  By contrast, for $|s^\dagger|> 0.2$
(outer dashed lines), the anomalies are generally associated with the
planetary caustics, so that $\log s^\dagger$ and $\log s$ have the same sign.
In this regime there is suggestive evidence (6 versus 2 detections) that
there is more overall senstivity to wide separation planets.
}
\label{fig:sdagger}
\end{figure}

Here, we investigate detections in the 2018-2019 AnomalyFinder sample
from the standpoint of image perturbations rather than planet-host separation.
As will become clear, these represent orthogonal perspectives.
Figure~\ref{fig:sdagger} shows a scatter plot of $\log q$ versus
$\log s^\dagger$, for which positive and negative values correspond to
major-image and minor-image perturbations, respectively.
See Equation~(\ref{eqn:heuristic}).

There are three notable features.  First, a majority (35/58) of the
planets lie within $|\log s^\dagger|<0.05$.  In this regime, there is essentially
no correlation between the signs of $\log s^\dagger$ and $\log s$ because
either light-curve morphology can almost equally be generated by $s>1$ and
$s<1$ lens geometries.  The fact that a majority of detections lie in this
narrow zone simply reflects the well-known fact that planet sensitivity 
is higher for relatively high \citep{gouldloeb,abe13} and very
high \citep{griest98} magnification events.  Note that, in this regime,
$u_{\rm anom}\simeq (\ln 100)|\log s^\dagger|$, so $u_{\rm anom}< 0.23$, i.e.,
$A_{\rm anom} > 4.4$.
Nevertheless, it is still of interest that the
detections are about equally distributed between positive and negative
values in this inner zone.  That is, the light-curve morphologies are
generally very different for positive and negative $\log s^\dagger$
(perturbations of the major and minor images), but apparently this
leads to very similar planet sensitivities.  This question could be
investigated to much higher precision based on already existing (or future)
planet-sensitivity studies by subdividing the simulations according to $\alpha$
into those with perturbations of the major or minor image.
 
% Eight cases
% kb180030 -2.56 1.58 1.23 Jung+2022 planetary caustic
% ob180596 -3.74 0.51 1.19 Jung+2019 planetary caustic
% ob180383 -3.67 2.45 2.04 Wang+2022 planetary caustic
% ob180932 -2.92 0.54 1.33 Goul+2022 planetary caustic
% ob180567 -2.91 1.81 1.25 Jung+2021 planetary caustic
% kb190298 -2.53 1.85 1.52 Jung+2023 planetary caustic
% ob190679 -2.36 2.18 1.63 Jung+2023 planetary caustic
% ob191180 -2.57 1/87 1/38 Chun+2023 planetary caustic
The second notable feature is that for $|\log s^\dagger|> 0.2$
(outer dashed lines), there are substantially more (6 versus 2) major-image
than minor-image planets.  In this regime, the anomalies are generally
closely associated with the planetary caustics (and this is so for all
8 cases from Figure~\ref{fig:sdagger}).  Moreover, both of the minor-image
perturbations are caustic-crossing, whereas this is the case for
only half of the major-image perturbations.  Because of small-number
statistics, no strong conclusions can be drawn from either of these
two comparisons.  However, both conform to our naive impression that
for planets that are far from the Einstein ring, it should be easier
to detect the isolated bump due to a wide-separation planet than the
weak dip of a close-separation planet, unless the source actually interacts
with one of the two small caustics.  Again this issue can be more precisely
explored from detailed simulations than from current planet samples, due
to small-number statistics.

The third notable feature is that in the intermediate region, i.e.,
the transition between the central-caustic and the planetary-caustic regimes,
there are about an equal number (7 versus 8) of major-image and
minor-image
perturbations. This suggests that in this regime, the substantially different
light-curve morphologies lead to about equal sensitivity.  This is again
deserving of systematic study via simulations.  One might also note
that all but one of the major-image perturbations in this regime
are from 2018,
while all but two 
of the minor-image perturbations are from 2019,  However, as
we cannot imagine any physical cause for this near dichotomy, we ascribe
it to the random ``noticeable effects'' that one often discovers when
viewing scatter plots.

\subsubsection{{55\%$\pm$7\% of Detections Are From Prime Fields}
\label{sec:prime-subprime}}

KMT devotes approximately half\footnote{In KMT's nominal schedule,
exactly half of the time is devoted to prime fields.  However, during
2016-2019, the schedule alternated between this nominal schedule and
an alternate one, according to the need to support {\it Spitzer}
microlensing \citep{yee15}.  During these alternate times, KMTC
kept to the nominal schedule, while KMTS and KMTA devoted 5/8 of their
time to the prime fields. Given the better weather at KMTC, the overall
fraction of time devoted to prime fields during these alternate times
was about 57\%.}
% 3/2 * 1/3 + 3/4 * 1/6 = 1/2 + 1/8 = 5/8
% 1/2 * 0.8 + 5/8 * 2 * 0.65 = 0.4 + 0.81 = 1.21 versus
% 1/2 * 0.8 + 3/8 * 2 * 0.65 = 0.4 + 0.49 = 0.89 121/210 ~ 4/7
of its observing time to the 6 prime fields and the other half to the 21
subprime fields.  While there are many considerations that go into this
division, such as sensitivity to the Galactic distribution of planets,
probing planets in a broad range of mass ratios, and probing other types
of dim or dark objects like black holes, one consideration is certainly
``return of planets on observing-time investment''.  Prior to the start
of KMT's commissioning observations in 2015, \citet{henderson14} had
already shown that there would be diminishing returns from concentrating
all observations on the ``most productive'' fields.  This understanding,
as well as the experience of OGLE, which pioneered a multi-tiered
observing approach, contributed to KMT adopting this strategy.  Thus,
it is of some interest that the planet return is in fact approximately
proportional to the invested observing time.

\subsection{{6-D Distribution}
\label{sec:6D}}
\begin{figure}
\plotone{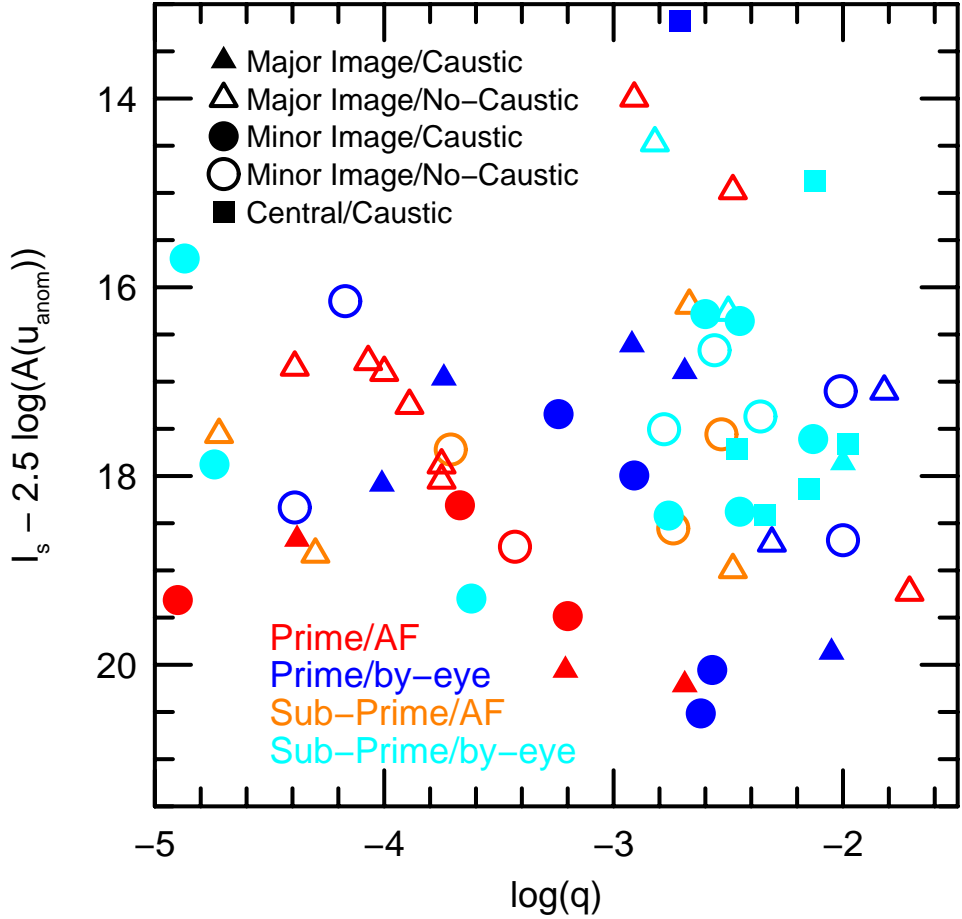}
\caption{6-dimensional scatter plot of 58 planets from 2018--2019.
(1) Abscissa: log mass ratio.
(2) Ordinate: source magnitude of unperturbed event at time of anomaly.
The remaining four dimensions are shown in the legend and are the same
as in Figure~14 from \citet{2018subprime}.
}
\label{fig:6d}
\end{figure}

In Figure~\ref{fig:6d}, we show a six-dimensional (6-D) representation of the
58 planets from 2018 and 2019 AnomalyFinder searches that are discussed
in this section.  It is an update to Figure~14 from \citet{2018subprime} which
included the subset of 33 planets from that paper.  To recapitulate their
description, it includes 2 continuous dimensions (given by the axes)
and 4 discrete dimensions that are represented by colors and point types.
The abscissa and ordinate are $\log q$ and 
$I_{S,\rm anom}\equiv I_S - 2.5\log[A(u_{\rm anom})]$,
with the latter being the source brightness in the unperturbed event
at the time of the anomaly.  The description of the symbols is identical
to those of \citet{2018subprime}, and they are also given in the legend.

Previously, \citet{2018subprime} had noted a ``paucity of by-eye detections
of non-caustic-crossing events (open bluish symbols) at low-$q$:
i.e., 1 out of 5 for $\log q<-3$ compared to
7 out of 12 for $\log q>-3$''.  This trend is strongly confirmed by
the larger sample: 2 out of 12 for $\log q<-3$ compared to
9 out of 16 for $\log q>-3$.  They also noted that 14 out of their
16 caustic-crossing planets were discovered by eye and that the
remaining two were both in prime fields and at low $\log q <-3$.
They suggested this was ``a regime where machines may do better
than people because the relatively weak signals of low-$q$ events
are spread out over a greater number of data points.''  In the
2019 sample, almost equal numbers of caustic-crossing planets were found by
each method, so that total now is 23 out of 29 caustic-crossing planets,
i.e., still heavily favoring
by-eye detections.  Moreover, the other trend is strongly confirmed:
now out of 6 AnomalyFinder caustic-crossing planets, none are from
subprime fields, and only one had $\log q>-3$.  This strengthens the evidence
for the \citet{2018subprime} conjecture that machines excel in
the high-cadence, low-$q$ regime for caustic-crossing planets.

One feature of this diagram noted by \citet{2018subprime} that is not
confirmed is the apparent threshold of AnomalyFinder detection at
$I_{S,\rm anom}= 18.75$.  There had been only one major
exception (OGLE-2018-BLG-0962), which has $I_{S,\rm anom}= 20.4$.
While there are still no detections fainter than this (previous) outlier,
The 1.5 magnitudes, $18.75<I_{S,\rm anom} \la 20.25$, are now ``filled in''
with a total of 10 planets.  Thus, $I_{S,\rm anom} \sim 20.25$ now appears
to be the detection floor.  This will be tested as additional seasons are
analyzed in this series.

\acknowledgments
This research has made use of the KMTNet system operated by the Korea Astronomy and Space Science Institute (KASI) 
at three host sites of CTIO in Chile, SAAO in South Africa, and SSO in Australia. 
Data transfer from the host site to KASI was supported by the Korea Research Environment Open NETwork (KREONET). 
%%This research has made use of the KMTNet system operated by the Korea
%%Astronomy and Space Science Institute (KASI) and the data were obtained at
%%three host sites of CTIO in Chile, SAAO in South Africa, and SSO in
%%Australia.
%
This research was supported by the Korea Astronomy and Space Science Institute 
under the R\&D program (Project No. 2023-1-832-03) supervised by the Ministry of Science and ICT.
W.Zang acknowledges the support from the Harvard-Smithsonian Center for Astrophysics through the CfA Fellowship.
Work by C.H. was supported by the grants of National Research Foundation 
of Korea (2020R1A4A2002885 and 2019R1A2C2085965).
J.C.Y. acknowledges support from US NSF Grant No. AST-2108414.
Y.S. acknowledges support from BSF Grant No. 2020740.
W.Zang, H.Y., S.M., and W.Zhu acknowledge support by the National Science Foundation of China (Grant No. 12133005).
W.Zhu acknowledges the science research grants from the China Manned Space Project with No.\ CMS-CSST-2021-A11.
R.Poleski was supported by Polish National Agency for Academic Exchange grant ``Polish Return 2019''.
This research uses data obtained through the Telescope Access Program (TAP), which has been funded by the TAP member institutes. 
The authors acknowledge the Tsinghua Astrophysics High-Performance Computing platform at Tsinghua University for providing computational and data storage resources that have contributed to the research results reported within this paper.

 \begin{deluxetable}{llrrrrr}
 \tablecolumns{7} \tablewidth{0pc}
 \tablecaption{\textsc{Event Names, Cadences, Alerts, and Locations}}
 \tablehead{\colhead{Name} & 
\colhead{$\Gamma\,({\rm hr}^{-1})$} &
\colhead{Alert Date} &
\colhead{RA$_{\rm J2000}$} &
\colhead{Dec$_{\rm J2000}$} &
\colhead{$l$} &
\colhead{$b$} }
%\hline
 \startdata
KMT-2019-BLG-0298 & 1.0 & 05 Apr 2019 & 17:39:30.72 & $-27$:38:17.30  & $+0.40$  & $+1.83$  \\ % no O-III or O-II pydia2/KMTC15/cmd.mon
OGLE-2019-BLG-0445& 0.4 \\
\hline
KMT-2019-BLG-1216 & 0.4 & 11 Jun 2019 & 17:53:55.35 & $-35$:08:11.90 & $-4.43$ & $-4.69$ \\  % BLG119.5 3 1433.68 1307.42 oiii/cmd.mon
OGLE-2019-BLG-1033& 0.2 \\
\hline
KMT-2019-BLG-2783 & 1.0 & Post Season & 17:57:10.06  & $-33$:47:18.67 & $-2.92$ & $-4.59$ \\  % BLG142.1 3 1877.57 2595.42  also BLG142.2 oiii/cmd.mon
\hline
OGLE-2019-BLG-0249& 0.1 & 09 Mar 2019 & 17:41:36.84 & $-34$:42:06.30
& $-5.35$ & $-2.30$ \\ % no O-III or O-II A_I = 2.10 nataf = 14.62 pydia/KMTC37/cmd.mon
KMT-2019-BLG-0109 & 0.4 \\ %gaia 0.1054  0.5908 -3.923 0.768 -5.286 0.387 0.999 19.53 1.84
\hline
OGLE-2019-BLG-0679& 0.1 & 05 May 2019 & 17:42:57.70 & $-27$:46:22.37 & $+0.69$ & $+1.11$ \\ % no O-III or O-II A_I = 3.80 nataf = 14.41 pydia/KMTC18/cmd.mon
KMT-2019-BLG-2688 & 1.0 \\ %gaia 0.1586 0.3856 -6.353 -6.036 0.279 1.747
\hline
OGLE-2019-BLG-0344& 0.2 & 20 Mar 2019 & 17:23:52.38 & $-29$:32:48.59  & $-3.08$  & $+3.66$  \\ % BLG331.8 3 1267.11 1282.09 oiii/cmd.mon
KMT-2019-BLG-0149 & 0.4 \\
\hline
KMT-2019-BLG-0304 & 1.0 & 05 Apr 2019 & 17:41:18.70 & $-32$:31:47.82  & $-3.54$  & $-1.10$ \\ % no O-III or O-II pydia/KMTC17/cmd.mon
%\hline
%KMT-2019-BLG-0967 & 1.0 & 27 May 2019 & 17:35:53.47 & $-27$:57:52.31 & $-0.31$ & $+2.33$ \\
\hline
%kb192688 ob190679 spitzer good signal ogle lead?
%kb190109 ob190249 spitzer some signal mao group?
%kb192084 ob191352 fspl diff  YKJ: weak amom dchi2=40 (unrenorm); sys@8631.
%KB191860 ob191239 0.045 must be included in review
%kb192802 ob191443 kuang+ in prep cannot be in AF
%kb192814 ob191470 published Kuang+
 \enddata
% \tablecomments{a: Manually alerted by MOA, based on OGLE alert}
 \label{tab:names}
 \end{deluxetable}

\begin{deluxetable}{lr|rrrrr}
%%\tabletypesize{\scriptsize}
\tabletypesize{\footnotesize}
\tablecaption{Standard \& Parallax 2L1S Models for KMT-2019-BLG-0298}
\tablewidth{0pt}
\tablehead{
\multicolumn{1}{l}{Parameters} &
%\multicolumn{1}{c}{Close} &
%\multicolumn{3}{c}{Wide} \\
%\multicolumn{1}{c}{} &
\multicolumn{1}{c}{Close} &
\multicolumn{1}{c}{Wide} &
\multicolumn{1}{c}{Wide} &
\multicolumn{1}{c}{Wide} \\
\multicolumn{1}{c}{} &
\multicolumn{1}{c}{Standard} &
\multicolumn{1}{c}{Standard} &
\multicolumn{1}{c}{$u_{0}>0$} &
\multicolumn{1}{c}{$u_{0}<0$} 

}
\startdata
$\chi^2_{\rm tot}$/dof          &      3494.6/3689         &          3476.0/3689       &       3474.9/3685           &      3475.0/3685        \\
$t_{0}$ (${\rm HJD'}$)          & 8621.421 $\pm$ 0.027     &    8621.337 $\pm$ 0.034    &  8621.224 $\pm$ 0.060       & 8621.226 $\pm$ 0.064    \\
$u_{0}$                         &    0.589 $\pm$ 0.018     &       0.613 $\pm$ 0.020    &     0.587 $\pm$ 0.028       &   -0.607 $\pm$ 0.021    \\
$t_{\rm E}$ (days)              &   28.506 $\pm$ 0.571     &      27.715 $\pm$ 0.617    &    27.979 $\pm$ 0.661       &   27.273 $\pm$ 0.641    \\
$s$                             &    0.491 $\pm$ 0.007     &       1.892 $\pm$ 0.030    &     1.845 $\pm$ 0.054       &    1.857 $\pm$ 0.051    \\
$q$ ($10^{-3}$)                 &    2.143 $\pm$ 0.253     &       2.485 $\pm$ 0.343    &     2.883 $\pm$ 0.680       &    3.004 $\pm$ 0.763    \\
$\langle {\rm log}\,q \rangle$  &   -2.665 $\pm$ 0.051     &      -2.603 $\pm$ 0.059    &    -2.534 $\pm$ 0.101       &   -2.518 $\pm$ 0.104    \\
$\alpha$ (rad)                  &    5.753 $\pm$ 0.010     &       2.766 $\pm$ 0.005    &     2.827 $\pm$ 0.057       &   -2.855 $\pm$ 0.069    \\
$\rho$ ($10^{-2}$)              &    1.179 $\pm$ 0.891     &       2.680 $\pm$ 1.650    &     3.188 $\pm$ 2.172       &    3.409 $\pm$ 2.384    \\
$\pi_{{\rm E}, N}$              &                          &                            &     0.537 $\pm$ 0.440       &   -0.850 $\pm$ 0.590    \\ 
$\pi_{{\rm E}, E}$              &                          &                            &     0.152 $\pm$ 0.128       &   -0.049 $\pm$ 0.057    \\
$ds/dt$ (yr$^{-1}$)             &                          &                            &     0.047 $\pm$ 0.959       &    0.143 $\pm$ 1.026    \\
$d\alpha/dt$ (yr$^{-1}$)        &                          &                            &     0.196 $\pm$ 0.474       &   -0.221 $\pm$ 0.563    \\
$f_{\rm S, OGLE}$               &    1.399 $\pm$ 0.070     &       1.502 $\pm$ 0.081    &     1.407 $\pm$ 0.111       &    1.473 $\pm$ 0.084    \\  
$f_{\rm B, OGLE}$               &    0.022 $\pm$ 0.070     &      -0.083 $\pm$ 0.081    &     0.010 $\pm$ 0.111       &   -0.055 $\pm$ 0.084    
\enddata
\label{tab:kb0298parms}
%\vspace{0.05cm}
\tablecomments{As discussed in Section~\ref{sec:anal-kb190298}, we accept the
  event parameters
from the ``Wide Standard'' solution for this event, but
incorporate the $\bpi_\e$ constraints from the parallax-plus-orbital-motion
solutions in the Bayesian analysis of Section~\ref{sec:phys-kb190298}.}
\end{deluxetable}

\begin{deluxetable}{lrrrrrr}
%%\tabletypesize{\scriptsize}
\tabletypesize{\footnotesize}
\tablecaption{Standard 2L1S Models for KMT-2019-BLG-1216}
\tablewidth{0pt}
\tablehead{
\multicolumn{1}{l}{Parameters} &
\multicolumn{1}{c}{Inner} &
\multicolumn{1}{c}{Outer} &
\multicolumn{1}{c}{Off-axis}
}
\startdata
$\chi^2_{\rm tot}$/dof          &        1236.1/1374       &         1236.3/1374       &     1260.6/1374                \\
$t_{0}$ (${\rm HJD'}$)          &  8658.443 $\pm$ 0.239    &   8658.448 $\pm$ 0.235    &  8660.978 $\pm$ 0.292          \\
$u_{0}$                         &     0.189 $\pm$ 0.036    &      0.175 $\pm$ 0.036    &     0.125 $\pm$ 0.007          \\
$t_{\rm E}$ (days)              &    88.729 $\pm$ 15.031   &     94.466 $\pm$ 16.881   &   138.780 $\pm$ 8.103          \\
$s$                             &     1.118 $\pm$ 0.018    &      1.074 $\pm$ 0.022    &     0.990 $\pm$ 0.003          \\
$q$ ($10^{-4}$)                 &     2.438 $\pm$ 0.538    &      2.323 $\pm$ 0.511    &    84.727 $\pm$ 10.535         \\
$\langle {\rm log}\,q \rangle$  &    -3.617 $\pm$ 0.099    &     -3.637 $\pm$ 0.098    &    -2.066 $\pm$ 0.055          \\
$\alpha$ (rad)                  &     1.561 $\pm$ 0.014    &      1.561 $\pm$ 0.014    &     0.013 $\pm$ 0.032          \\
$\rho$ ($10^{-4}$)              &     4.409 $\pm$ 1.897    &      4.184 $\pm$ 1.750    &     1.757 $\pm$ 0.662          \\
$f_{\rm S, OGLE}$               &     0.044 $\pm$ 0.010    &      0.040 $\pm$ 0.010    &     0.025 $\pm$ 0.002          \\  
$f_{\rm B, OGLE}$               &     0.087 $\pm$ 0.010    &      0.090 $\pm$ 0.009    &     0.105 $\pm$ 0.002                      
\enddata
\label{tab:kb1216parms}
%\vspace{0.05cm}
%%\tablecomments{
%%}
\end{deluxetable}

\begin{deluxetable}{lrrrrrr}
\tabletypesize{\footnotesize}
\tablecaption{Parallax 2L1S Models for KMT-2019-BLG-1216}
\tablewidth{0pt}
\tablehead{
\multicolumn{1}{l}{Parameters} &
\multicolumn{2}{c}{Inner} &
\multicolumn{2}{c}{Outer} \\
\multicolumn{1}{c}{} & 
\multicolumn{1}{c}{$u_{0} > 0$} &
\multicolumn{1}{c}{$u_{0} < 0$} &
\multicolumn{1}{c}{$u_{0} > 0$} &
\multicolumn{1}{c}{$u_{0} < 0$} 
}
\startdata
$\chi^2_{\rm tot}$/dof          &       1226.3/1370         &       1226.4/1370         &       1226.3/1370         &       1226.5/1370         \\
$t_{0}$ (${\rm HJD'}$)          &  8659.089 $\pm$ 0.345     &  8659.053 $\pm$ 0.340     &  8659.115 $\pm$ 0.338     &  8659.048 $\pm$ 0.325     \\
$u_{0}$                         &     0.110 $\pm$ 0.030     &    -0.122 $\pm$ 0.035     &     0.119 $\pm$ 0.035     &    -0.137 $\pm$ 0.037     \\
$t_{\rm E}$ (days)              &   137.465 $\pm$ 33.221    &   134.147 $\pm$ 36.134    &   129.232 $\pm$ 36.368    &   120.992 $\pm$ 30.846    \\
$s$                             &     1.083 $\pm$ 0.013     &     1.086 $\pm$ 0.016     &     1.037 $\pm$ 0.022     &     1.047 $\pm$ 0.023     \\
$q$ ($10^{-4}$)                 &     2.264 $\pm$ 0.593     &     2.197 $\pm$ 0.594     &     2.209 $\pm$ 0.616     &     2.319 $\pm$ 0.585     \\
$\langle {\rm log}\,q \rangle$  &    -3.644 $\pm$ 0.116     &    -3.661 $\pm$ 0.121     &    -3.658 $\pm$ 0.120     &    -3.639 $\pm$ 0.114     \\
$\alpha$ (rad)                  &     1.602 $\pm$ 0.022     &    -1.600 $\pm$ 0.020     &     1.603 $\pm$ 0.022     &    -1.599 $\pm$ 0.020     \\
$\rho$ ($10^{-4}$)              &     2.537 $\pm$ 1.266     &     2.696 $\pm$ 1.384     &     2.772 $\pm$ 1.350     &     3.089 $\pm$ 1.514     \\
$\pi_{{\rm E}, N}$              &     0.122 $\pm$ 0.101     &     0.381 $\pm$ 0.299     &     0.132 $\pm$ 0.103     &     0.453 $\pm$ 0.316     \\ 
$\pi_{{\rm E}, E}$              &     0.543 $\pm$ 0.194     &     0.523 $\pm$ 0.238     &     0.535 $\pm$ 0.199     &     0.490 $\pm$ 0.255     \\
$ds/dt$ (yr$^{-1}$)             &    -0.010 $\pm$ 0.880     &    -0.010 $\pm$ 0.920     &    -0.050 $\pm$ 0.870     &     0.010 $\pm$ 0.890     \\
$d\alpha/dt$ (yr$^{-1}$)        &     0.090 $\pm$ 0.860     &    -0.190 $\pm$ 0.890     &     0.050 $\pm$ 0.870     &    -0.290 $\pm$ 0.910     \\
$f_{\rm S, OGLE}$               &     0.023 $\pm$ 0.007     &     0.025 $\pm$ 0.008     &     0.026 $\pm$ 0.008     &     0.029 $\pm$ 0.009     \\  
$f_{\rm B, OGLE}$               &     0.108 $\pm$ 0.007     &     0.106 $\pm$ 0.008     &     0.106 $\pm$ 0.008     &     0.102 $\pm$ 0.009     
\enddata
\label{tab:kb1216parallax}
%\vspace{0.05cm}
%\tablecomments{
%}
\end{deluxetable}

\begin{deluxetable}{lrrrrrr}
\tabletypesize{\footnotesize}
\tablecaption{Standard 2L1S Model for KMT-2019-BLG-2783}
\tablewidth{0pt}
\tablehead{
\multicolumn{1}{l}{Parameters} &
\multicolumn{1}{c}{Values}
}
\startdata
$\chi^2_{\rm tot}$/dof          &        2113.9/2103       \\
$t_{0}$ (${\rm HJD'}$)          &   8765.395 $\pm$ 0.037   \\
$u_{0}$                         &      0.057 $\pm$ 0.003   \\
$t_{\rm E}$ (days)              &     23.669 $\pm$ 1.065   \\
$s$                             &      0.814 $\pm$ 0.007   \\
$q$ ($10^{-3}$)                 &      3.262 $\pm$ 0.762   \\
$\langle {\rm log}\,q \rangle$  &     -2.483 $\pm$ 0.101   \\
$\alpha$ (rad)                  &      6.277 $\pm$ 0.023   \\
$\rho$ ($10^{-3}$)              &      5.131 $\pm$ 3.515    \\
$f_{\rm S, KMTC}$               &      0.138 $\pm$ 0.008    \\  
$f_{\rm B, KMTC}$               &      0.062 $\pm$ 0.007              
\enddata
\label{tab:kb2783parms}
%\vspace{0.05cm}
%\tablecomments{
%}
\end{deluxetable}

\begin{deluxetable}{lrrrrrr}
%%\tabletypesize{\scriptsize}
\tabletypesize{\footnotesize}
\tablecaption{Survey-Only 2L1S Models for OGLE-2019-BLG-0249}
\tablewidth{0pt}
\tablehead{
%\multicolumn{1}{l}{Parameters} &
%\multicolumn{4}{c}{Survey data sets}   \\
\multicolumn{1}{c}{Parameters} & 
\multicolumn{2}{c}{Planetary} & 
\multicolumn{2}{c}{Binary}   \\
\multicolumn{1}{c}{} & 
\multicolumn{1}{c}{Close} &
\multicolumn{1}{c}{Wide}  &
\multicolumn{1}{c}{Close} &
\multicolumn{1}{c}{Wide} 
}
\startdata
$\chi^2_{\rm tot}$/dof           &      1672.5/1682       &      1670.7/1682         &      1733.2/1682         &       1751.1/1682         \\
$t_{0}$ (${\rm HJD'}$)           &  8607.442 $\pm$ 0.004  &  8607.436 $\pm$ 0.004    &  8607.446 $\pm$ 0.004    &   8607.488 $\pm$ 0.004    \\
$u_{0}$ ($10^{-3}$)              &    30.934 $\pm$ 0.288  &    31.221 $\pm$ 0.290    &    32.388 $\pm$ 0.341    &     28.418 $\pm$ 0.245    \\
$t_{\rm E}$ (days)               &    76.387 $\pm$ 0.624  &    76.596 $\pm$ 0.618    &    76.912 $\pm$ 0.615    &     84.974 $\pm$ 0.689    \\
$s$                              &     0.581 $\pm$ 0.010  &     1.663 $\pm$ 0.030    &     0.231 $\pm$ 0.003    &      5.657 $\pm$ 0.110    \\
$q$ ($10^{-3}$)                  &     5.918 $\pm$ 0.320  &     6.166 $\pm$ 0.335    &   153.281 $\pm$ 8.331    &    247.573 $\pm$ 17.120   \\
$\langle {\rm log}\,q \rangle$   &    -2.226 $\pm$ 0.023  &    -2.209 $\pm$ 0.024    &    -0.813 $\pm$ 0.024    &     -0.603 $\pm$ 0.030    \\               
$\alpha$ (rad)                   &     5.287 $\pm$ 0.003  &     5.292 $\pm$ 0.003    &     2.974 $\pm$ 0.005    &      2.960 $\pm$ 0.004    \\
$\rho$ ($10^{-3}$)               &     9.635 $\pm$ 1.582  &    10.692 $\pm$ 1.466    &    17.631 $\pm$ 0.751    &     16.683 $\pm$ 0.645    \\
$f_{\rm S, OGLE}$                &     0.718 $\pm$ 0.007  &     0.722 $\pm$ 0.007    &     0.720 $\pm$ 0.007    &      0.715 $\pm$ 0.007     \\  
$f_{\rm B, OGLE}$                &     0.219 $\pm$ 0.007  &     0.216 $\pm$ 0.007    &     0.217 $\pm$ 0.006    &      0.220 $\pm$ 0.007               
\enddata
\label{tab:ob0249parms}
%\vspace{0.05cm}
%\tablecomments{
%}
\end{deluxetable}

\begin{deluxetable}{lrrrrrrr}
\tabletypesize{\scriptsize}
%%\tabletypesize{\footnotesize}
\tablecaption{Survey+Followup 2L1S Models for OGLE-2019-BLG-0249}
\tablewidth{0pt}
\tablehead{
%\multicolumn{1}{l}{Parameters} &
%\multicolumn{6}{c}{Survey + Followup data sets}  \\
\multicolumn{1}{c}{Parameters} & 
\multicolumn{3}{c}{Close} & 
\multicolumn{3}{c}{Wide}  \\
\multicolumn{1}{c}{} & 
\multicolumn{1}{c}{Standard} &
\multicolumn{1}{c}{$u_{0} > 0$} & 
\multicolumn{1}{c}{$u_{0} < 0$} & 
\multicolumn{1}{c}{Standard} &
\multicolumn{1}{c}{$u_{0} > 0$} & 
\multicolumn{1}{c}{$u_{0} < 0$}  
}
\startdata
$\chi^2_{\rm tot}$/dof          &     2101.7/2081        &         1828.1/2077      &     1828.2/2077        &       2100.7/2081       &      1831.4/2077        &       1831.4/2077           \\
$t_{0}$ (${\rm HJD'}$)          & 8607.430 $\pm$ 0.003   &   8607.415 $\pm$ 0.011   & 8607.414 $\pm$ 0.011   &   8607.426 $\pm$ 0.003  &  8607.807 $\pm$ 0.017   &   8607.805 $\pm$ 0.019      \\
$u_{0}$ ($10^{-3}$)             &   31.133 $\pm$ 0.269   &     30.859 $\pm$ 0.366   &  -30.911 $\pm$ 0.369   &     31.363 $\pm$ 0.277  &    39.021 $\pm$ 0.490   &    -39.064 $\pm$ 0.512      \\
$t_{\rm E}$ (days)              &   75.628 $\pm$ 0.585   &     76.264 $\pm$ 0.645   &   76.309 $\pm$ 0.687   &     75.923 $\pm$ 0.597  &    76.592 $\pm$ 0.598   &     76.649 $\pm$ 0.647      \\
$s$                             &    0.554 $\pm$ 0.004   &      0.543 $\pm$ 0.010   &    0.545 $\pm$ 0.010   &      1.753 $\pm$ 0.014  &     1.777 $\pm$ 0.014   &      1.775 $\pm$ 0.015      \\
$q$ ($10^{-3}$)                 &    6.643 $\pm$ 0.195   &      7.461 $\pm$ 0.270   &    7.455 $\pm$ 0.268   &      6.983 $\pm$ 0.214  &     7.819 $\pm$ 0.229   &      7.805 $\pm$ 0.237      \\
$\langle {\rm log}\,q \rangle$  &   -2.177 $\pm$ 0.013   &     -2.127 $\pm$ 0.015   &   -2.127 $\pm$ 0.016   &     -2.156 $\pm$ 0.013  &    -2.106 $\pm$ 0.013   &     -2.107 $\pm$ 0.013      \\
$\alpha$ (rad)                  &    5.295 $\pm$ 0.002   &      5.298 $\pm$ 0.012   &   -5.298 $\pm$ 0.012   &      5.302 $\pm$ 0.003  &     5.302 $\pm$ 0.004   &     -5.302 $\pm$ 0.005      \\
$\rho$ ($10^{-3}$)              &    5.138 $\pm$ 1.308   &      5.925 $\pm$ 1.231   &    6.084 $\pm$ 1.197   &      6.375 $\pm$ 1.057  &     7.369 $\pm$ 0.803   &      7.413 $\pm$ 0.804      \\
$\pi_{{\rm E}, N}$              &                        &      0.016 $\pm$ 0.076   &    0.008 $\pm$ 0.082   &                         &    -0.003 $\pm$ 0.068   &     -0.005 $\pm$ 0.075      \\
$\pi_{{\rm E}, E}$              &                        &      0.061 $\pm$ 0.019   &    0.059 $\pm$ 0.020   &                         &     0.056 $\pm$ 0.019   &      0.056 $\pm$ 0.018      \\
$ds/dt$ (yr$^{-1}$)             &                        &     -0.339 $\pm$ 0.688   &   -0.249 $\pm$ 0.744   &                         &     0.048 $\pm$ 0.528   &      0.051 $\pm$ 0.527      \\
$d\alpha/dt$ (yr$^{-1}$)        &                        &      0.419 $\pm$ 1.419   &   -0.569 $\pm$ 1.469   &                         &    -0.002 $\pm$ 0.303   &     -0.026 $\pm$ 0.308      \\
$f_{\rm S, OGLE}$               &    0.727 $\pm$ 0.007   &      0.719 $\pm$ 0.008   &    0.720 $\pm$ 0.008   &      0.730 $\pm$ 0.007  &     0.724 $\pm$ 0.007   &      0.724 $\pm$ 0.007      \\  
$f_{\rm B, OGLE}$               &    0.211 $\pm$ 0.006   &      0.218 $\pm$ 0.007   &    0.218 $\pm$ 0.007   &      0.208 $\pm$ 0.006  &     0.213 $\pm$ 0.007   &      0.213 $\pm$ 0.007   
\enddata
\label{tab:ob0249followup}
%\vspace{0.05cm}
%\tablecomments{
%}
\end{deluxetable}

\begin{deluxetable}{lrrrrrrr}
%%\tabletypesize{\scriptsize}
\tabletypesize{\footnotesize}
\tablecaption{Standard \& Parallax 2L1S Models for OGLE-2019-BLG-0679}
\tablewidth{0pt}
\tablehead{
\multicolumn{1}{l}{Parameters} &
\multicolumn{1}{c}{Standard} &
\multicolumn{1}{c}{$u_{0} > 0$} & 
\multicolumn{1}{c}{$u_{0} < 0$} 
}
\startdata
$\chi^2_{\rm tot}$/dof          &     4194.4/4438        &         4164.9/4434      &     4163.2/4434         \\
$t_{0}$ (${\rm HJD'}$)          & 8660.604 $\pm$ 0.056   &   8660.775 $\pm$ 0.085   & 8660.766 $\pm$ 0.075    \\
$u_{0}$                         &    0.831 $\pm$ 0.016   &      0.804 $\pm$ 0.026   &   -0.817 $\pm$ 0.027    \\
$t_{\rm E}$ (days)              &   32.608 $\pm$ 0.469   &     33.252 $\pm$ 0.734   &   32.939 $\pm$ 0.729    \\
$s$                             &    2.216 $\pm$ 0.023   &      2.161 $\pm$ 0.034   &    2.167 $\pm$ 0.035    \\
$q$ ($10^{-3}$)                 &    4.785 $\pm$ 0.146   &      4.072 $\pm$ 0.508   &    3.837 $\pm$ 0.395    \\
$\langle {\rm log}\,q \rangle$  &   -2.319 $\pm$ 0.013   &     -2.387 $\pm$ 0.049   &   -2.410 $\pm$ 0.044    \\
$\alpha$ (rad)                  &    0.537 $\pm$ 0.003   &      0.551 $\pm$ 0.011   &   -0.570 $\pm$ 0.017    \\
$\rho$ ($10^{-3}$)              &    9.204 $\pm$ 6.906   &     19.579 $\pm$ 7.847   &   17.335 $\pm$ 8.553    \\
$\pi_{{\rm E}, N}$              &                        &      0.108 $\pm$ 0.078   &   -0.398 $\pm$ 0.192    \\
$\pi_{{\rm E}, E}$              &                        &     -0.293 $\pm$ 0.087   &   -0.358 $\pm$ 0.101    \\
$ds/dt$ (yr$^{-1}$)             &                        &      0.368 $\pm$ 0.413   &    0.554 $\pm$ 0.332    \\
$d\alpha/dt$ (yr$^{-1}$)        &                        &      0.177 $\pm$ 0.193   &   -0.369 $\pm$ 0.205    \\
$f_{\rm S, OGLE}$               &    1.824 $\pm$ 0.067   &      1.729 $\pm$ 0.106   &    1.766 $\pm$ 0.109    \\  
$f_{\rm B, OGLE}$               &    0.203 $\pm$ 0.067   &      0.295 $\pm$ 0.106   &    0.255 $\pm$ 0.109               
\enddata
\label{tab:ob0679parms}
%\vspace{0.05cm}
\tablecomments{As discussed in Section~\ref{sec:anal-ob190679}, we adopt
  the ``Standard'' parameters for this event.}
\end{deluxetable}

\begin{deluxetable}{lrrrrrr}
\tabletypesize{\scriptsize}
%%\tabletypesize{\footnotesize}
\tablecaption{Planetary Models for OGLE-2019-BLG-0344}
\tablewidth{0pt}
\tablehead{
%\multicolumn{1}{l}{Parameters} &
%\multicolumn{4}{c}{Planetary} \\
\multicolumn{1}{c}{Parameters} & 
\multicolumn{2}{c}{Inner} &
\multicolumn{2}{c}{Outer} \\
\multicolumn{1}{c}{} & 
\multicolumn{1}{c}{free $\rho$} &
\multicolumn{1}{c}{$\rho = 0$} &
\multicolumn{1}{c}{free $\rho$} &
\multicolumn{1}{c}{$\rho = 0$} 
}
\startdata
$\chi^2_{\rm tot}$/dof          &      1612.8/1621             &      1620.2/1622          &    1612.1/1621               &    1619.9/1622          \\
$t_{0}$ (${\rm HJD'}$)          & 8567.532 $\pm$ 0.014         & 8567.529 $\pm$ 0.014      & 8567.533 $\pm$ 0.014         & 8567.528 $\pm$ 0.013    \\
$u_{0}$                         &    0.103 $\pm$ 0.005         &    0.102 $\pm$ 0.004      &    0.103 $\pm$ 0.005         &    0.103 $\pm$ 0.005    \\
$t_{\rm E}$ (days)              &   14.413 $\pm$ 0.470         &   14.108 $\pm$ 0.366      &   14.578 $\pm$ 0.496         &   14.227 $\pm$ 0.426    \\
$s$                             &    0.621 $\pm$ 0.044         &    0.540 $\pm$ 0.015      &    1.459 $\pm$ 0.114         &    1.696 $\pm$ 0.069    \\
$q$ ($10^{-2}$)                 &    2.374 $\pm$ 0.528         &    2.958 $\pm$ 0.457      &    2.567 $\pm$ 0.601         &    3.058 $\pm$ 0.584    \\
$\langle {\rm log}\,q \rangle$  &   -1.627 $\pm$ 0.096         &   -1.530 $\pm$ 0.067      &   -1.595 $\pm$ 0.102         &   -1.526 $\pm$ 0.083    \\
$\alpha$ (rad)                  &    4.714 $\pm$ 0.023         &    4.706 $\pm$ 0.021      &    4.713 $\pm$ 0.022         &    4.713 $\pm$ 0.025    \\
$\rho$ ($10^{-2}$)              &    5.809 $\pm$ 1.166         &                           &    6.269 $\pm$ 1.097         &                         \\
$f_{\rm S, OGLE}$               &    0.365 $\pm$ 0.021         &    0.380 $\pm$ 0.018      &    0.365 $\pm$ 0.021         &    0.388 $\pm$ 0.019    \\  
$f_{\rm B, OGLE}$               &    0.063 $\pm$ 0.021         &    0.048 $\pm$ 0.018      &    0.062 $\pm$ 0.021         &    0.040 $\pm$ 0.019               
\enddata 
%\vspace{0.05cm}
\label{tab:ob0344parms}
%\tablecomments{
%}
\end{deluxetable}

\begin{deluxetable}{lrrrrrr}
\tabletypesize{\scriptsize}
%%\tabletypesize{\footnotesize}
\tablecaption{Close Binary-Star Models for OGLE-2019-BLG-0344}
\tablewidth{0pt}
\tablehead{
%\multicolumn{1}{l}{Parameters} &
%\multicolumn{6}{c}{Binary Close} \\
\multicolumn{1}{c}{Parameters} & 
\multicolumn{2}{c}{Local A} &
\multicolumn{2}{c}{Local B} &
\multicolumn{2}{c}{Local C} \\
\multicolumn{1}{c}{} & 
\multicolumn{1}{c}{free $\rho$} &
\multicolumn{1}{c}{$\rho = 0$} &
\multicolumn{1}{c}{free $\rho$} &
\multicolumn{1}{c}{$\rho = 0$} &
\multicolumn{1}{c}{free $\rho$} &
\multicolumn{1}{c}{$\rho = 0$} 
}
\startdata
$\chi^2_{\rm tot}$/dof          &       1617.2/1621       &       1620.7/1622       &     1605.9/1621         &     1626.1/1622         &     1637.8/1621         &    1637.9/1622          \\
$t_{0}$ (${\rm HJD'}$)          &  8567.444 $\pm$ 0.019   &  8567.450 $\pm$ 0.016   &  8567.687 $\pm$ 0.027   &  8567.616 $\pm$ 0.021   &  8568.157 $\pm$ 0.025   & 8568.166 $\pm$ 0.026    \\
$u_{0}$                         &     0.113 $\pm$ 0.006   &     0.110 $\pm$ 0.005   &     0.125 $\pm$ 0.005   &     0.111 $\pm$ 0.006   &     0.049 $\pm$ 0.002   &    0.048 $\pm$ 0.002    \\
$t_{\rm E}$ (days)              &    13.916 $\pm$ 0.389   &    13.846 $\pm$ 0.361   &    13.794 $\pm$ 0.367   &    13.697 $\pm$ 0.380   &    13.771 $\pm$ 0.320   &   13.764 $\pm$ 0.328    \\
$s$                             &     0.315 $\pm$ 0.013   &     0.301 $\pm$ 0.007   &     0.362 $\pm$ 0.017   &     0.302 $\pm$ 0.010   &     0.462 $\pm$ 0.011   &    0.467 $\pm$ 0.010    \\
$q$                             &     0.527 $\pm$ 0.162   &     0.515 $\pm$ 0.135   &     0.804 $\pm$ 0.168   &     1.841 $\pm$ 0.824   &     0.333 $\pm$ 0.025   &    0.329 $\pm$ 0.025    \\
$\langle {\rm log}\,q \rangle$  &    -0.273 $\pm$ 0.131   &    -0.288 $\pm$ 0.116   &    -0.079 $\pm$ 0.079   &     0.281 $\pm$ 0.173   &    -0.474 $\pm$ 0.032   &   -0.480 $\pm$ 0.032    \\
$\alpha$ (rad)                  &     2.412 $\pm$ 0.035   &     2.419 $\pm$ 0.028   &     3.999 $\pm$ 0.037   &     3.882 $\pm$ 0.061   &     1.245 $\pm$ 0.016   &    1.248 $\pm$ 0.016    \\
$\rho$ ($10^{-2}$)              &     4.936 $\pm$ 2.014   &                         &     8.504 $\pm$ 0.539   &                         &     0.114 $\pm$ 0.103   &                         \\
$f_{\rm S, OGLE}$               &     0.397 $\pm$ 0.020   &     0.399 $\pm$ 0.019   &     0.412 $\pm$ 0.018   &     0.412 $\pm$ 0.018   &     0.360 $\pm$ 0.011   &    0.360 $\pm$ 0.011    \\  
$f_{\rm B, OGLE}$               &     0.031 $\pm$ 0.020   &     0.029 $\pm$ 0.018   &     0.016 $\pm$ 0.017   &     0.016 $\pm$ 0.017   &     0.069 $\pm$ 0.011   &    0.068 $\pm$ 0.011               
\enddata
\label{tab:ob0344bin}
%\vspace{0.05cm}
%\tablecomments{
%}
\end{deluxetable}

\begin{deluxetable}{lrrrr}
%%\tabletypesize{\scriptsize}
\tabletypesize{\footnotesize}
\tablecaption{1L2S Models for OGLE-2019-BLG-0344}
\tablewidth{0pt}
\tablehead{
%\multicolumn{1}{l}{Parameters} &
%\multicolumn{2}{c}{1L2S} \\ 
\multicolumn{1}{c}{Parameters} & 
\multicolumn{1}{c}{free $\rho$} &
\multicolumn{1}{c}{$\rho = 0$} 
}
\startdata
$\chi^2_{\rm tot}$/dof            &    1609.7/1620       &    1613.7/1622          \\
$t_{0,1}$ (${\rm HJD'}$)          & 8566.544 $\pm$ 0.071 & 8566.528 $\pm$ 0.069    \\
$t_{0,2}$ (${\rm HJD'}$)          & 8568.511 $\pm$ 0.069 & 8568.506 $\pm$ 0.065    \\
$u_{0,1}$                         &    0.082 $\pm$ 0.006 &    0.077 $\pm$ 0.004    \\
$u_{0,2}$                         &    0.092 $\pm$ 0.012 &    0.080 $\pm$ 0.005    \\
$t_{\rm E}$ (days)                &   15.215 $\pm$ 0.440 &   15.394 $\pm$ 0.430    \\
$\rho_{1}$ ($10^{-2}$)            &    4.432 $\pm$ 2.775 &                         \\
$\rho_{2}$ ($10^{-2}$)            &    7.721 $\pm$ 3.885 &                         \\
$q_{\rm F}$                       &    1.032 $\pm$ 0.189 &    1.054 $\pm$ 0.180    \\
$f_{\rm S, OGLE}$                 &    0.328 $\pm$ 0.015 &    0.320 $\pm$ 0.014    \\
$f_{\rm B, OGLE}$                 &    0.101 $\pm$ 0.015 &    0.108 $\pm$ 0.014    
\enddata
\label{tab:ob0344-1l2s}
%\vspace{0.05cm}
%\tablecomments{
%}
\end{deluxetable}

\begin{deluxetable}{lrrrrrr}
%%\tabletypesize{\scriptsize}
\tabletypesize{\footnotesize}
\tablecaption{2L1S Standard Models for KMT-2019-BLG-0304}
\tablewidth{0pt}
\tablehead{
\multicolumn{1}{l}{Parameters} &
\multicolumn{2}{c}{Close} &
\multicolumn{2}{c}{Wide} \\
\multicolumn{1}{c}{} & 
\multicolumn{1}{c}{free $\rho$} &
\multicolumn{1}{c}{$\rho = 0$} &
\multicolumn{1}{c}{free $\rho$} &
\multicolumn{1}{c}{$\rho = 0$} 
}
\startdata
$\chi^2_{\rm tot}$/dof          &      2394.3/2387        &      2394.9/2388        &      2393.8/2387       &      2394.8/2388        \\
$t_{0}$ (${\rm HJD'}$)          & 8574.014 $\pm$ 0.346    & 8574.020 $\pm$ 0.361    & 8574.019 $\pm$ 0.341   & 8574.039 $\pm$ 0.327    \\
$u_{0}$                         &    0.083 $\pm$ 0.024    &    0.098 $\pm$ 0.048    &    0.099 $\pm$ 0.029   &    0.093 $\pm$ 0.030    \\
$t_{\rm E}$ (days)              &  166.981 $\pm$ 45.725   &  144.948 $\pm$ 62.216   &  140.494 $\pm$ 38.089  &  151.043 $\pm$ 45.234   \\
$s$                             &    0.713 $\pm$ 0.033    &    0.722 $\pm$ 0.052    &    1.546 $\pm$ 0.051   &    1.569 $\pm$ 0.051    \\
$q$ ($10^{-2}$)                 &    1.220 $\pm$ 0.340    &    1.415 $\pm$ 0.445    &    1.494 $\pm$ 0.422   &    1.496 $\pm$ 0.435    \\
$\langle {\rm log}\,q \rangle$  &   -1.917 $\pm$ 0.122    &   -1.863 $\pm$ 0.138    &   -1.838 $\pm$ 0.139   &   -1.828 $\pm$ 0.127    \\
$\alpha$ (rad)                  &    1.536 $\pm$ 0.025    &    1.537 $\pm$ 0.024    &    1.538 $\pm$ 0.024   &    1.536 $\pm$ 0.023    \\
$\rho$ ($10^{-3}$)              &    1.667 $\pm$ 1.008    &                         &    1.954 $\pm$ 1.211   &                         \\
$f_{\rm S, KMTC}$               &    0.009 $\pm$ 0.003    &    0.011 $\pm$ 0.006    &    0.011 $\pm$ 0.004   &    0.010 $\pm$ 0.003    \\  
$f_{\rm B, KMTC}$               &    0.180 $\pm$ 0.002    &    0.177 $\pm$ 0.004    &    0.176 $\pm$ 0.002   &    0.177 $\pm$ 0.002    \\               
$t_{\rm eff} (u_{0} t_{\rm E})$ &   14.098 $\pm$ 0.796    &   14.581 $\pm$ 1.025    &   14.393 $\pm$ 0.887   &   14.492 $\pm$ 0.820    \\        
$t_{*} (\rho t_{\rm E})$        &    0.299 $\pm$ 0.164    &                         &    0.299 $\pm$ 0.164   &                         \\           
$t_{q} (q t_{\rm E})$           &    2.044 $\pm$ 0.411    &    2.018 $\pm$ 0.466    &    2.142 $\pm$ 0.447   &    2.305 $\pm$ 0.463    \\          
$f_{\rm S}t_{\rm E}$            &    1.578 $\pm$ 0.084    &    1.631 $\pm$ 0.137    &    1.633 $\pm$ 0.101   &    1.635 $\pm$ 0.094    \\           
\enddata
\label{tab:kb0304parms}
%\vspace{0.05cm}
%\tablecomments{
%}
\end{deluxetable}

\begin{deluxetable}{lrrrr}
%%\tabletypesize{\scriptsize}
\tabletypesize{\footnotesize}
\tablecaption{1L2S Models for KMT-2019-BLG-0304}
\tablewidth{0pt}
\tablehead{
%\multicolumn{1}{l}{Parameters} &
%\multicolumn{2}{c}{1L2S} \\
\multicolumn{1}{c}{Parameters} &
\multicolumn{1}{c}{free $\rho$} &
\multicolumn{1}{c}{$\rho = 0$} 
}
\startdata
$\chi^2_{\rm tot}$/dof              &     2389.5/3687          &     2393.9/3689          \\
$t_{0,1}$ (${\rm HJD'}$)            & 8573.946 $\pm$ 0.478     & 8573.928 $\pm$ 0.491     \\
$t_{0,2}$ (${\rm HJD'}$)            & 8574.478 $\pm$ 0.048     & 8574.520 $\pm$ 0.039     \\
$u_{0,1}$                           &    0.202 $\pm$ 0.066     &    0.207 $\pm$ 0.067     \\
$u_{0,2}$ ($10^{-3}$)               &    3.732 $\pm$ 1.432     &    3.411 $\pm$ 1.024     \\
$t_{\rm E}$ (days)                  &   95.002 $\pm$ 25.114    &   92.397 $\pm$ 24.558    \\
$\rho_{1}$ ($10^{-2}$)              &    4.834 $\pm$ 3.066     &                          \\
$\rho_{2}$ ($10^{-3}$)              &    4.989 $\pm$ 2.087     &                          \\
$q_{\rm F}$ ($10^{-2}$)             &    3.234 $\pm$ 0.451     &    3.399 $\pm$ 0.460     \\
$f_{\rm S, KMTC}$                   &    0.019 $\pm$ 0.007     &    0.020 $\pm$ 0.008     \\  
$f_{\rm B, KMTC}$                   &    0.169 $\pm$ 0.007     &    0.169 $\pm$ 0.007     \\
$t_{\rm eff,1} (u_{0,1} t_{\rm E})$ &   19.637 $\pm$ 1.650     &   19.799 $\pm$ 1.686     \\        
$t_{\rm eff,2} (u_{0,2} t_{\rm E})$ &    0.362 $\pm$ 0.075     &    0.324 $\pm$ 0.039     \\        
$t_{*,1} (\rho_{1} t_{\rm E})$      &    4.429 $\pm$ 3.152     &                          \\
$t_{*,2} (\rho_{2} t_{\rm E})$      &    0.518 $\pm$ 0.149     &                          \\
$f_{\rm S}t_{\rm E}$                &    1.865 $\pm$ 0.222     &    1.895 $\pm$ 0.235    
\enddata
\label{tab:kb0304-1l2s}
%\vspace{0.05cm}
%\tablecomments{
%}
\end{deluxetable}

 \begin{deluxetable}{lrrrrrrrr}
 \tablecolumns{9} \rotate \tablewidth{0pc}
 \tablecaption{\textsc {CMD Parameters}}
 \tablehead{\colhead{Name} & 
\colhead{$(V-I)_{\rm S}$} &
\colhead{$(V-I)_{\rm cl}$} &
\colhead{$(V-I)_{\rm S,0}$} &
\colhead{$I_{\rm S}$} &
\colhead{$I_{\rm cl}$} &
\colhead{$I_{\rm cl,0}$} &
\colhead{$I_{\rm S,0}$} &
\colhead{$\theta_*\ (\muas)$} }
%\hline
 \startdata
 KMT-2019-BLG-0298 & 3.67$\pm$0.06 & 3.56$\pm$0.03 & 1.11$\pm$0.07 & 17.69$\pm$0.06 & 17.53$\pm$0.03 & 14.42 & 14.58$\pm$0.07 & 6.117$\pm$0.530 \\
 KMT-2019-BLG-1216 & N.A.          & 1.91$\pm$0.03 & 1.04$\pm$0.14 & 21.52$\pm$0.25 & 15.78$\pm$0.05 & 14.61 & 20.35$\pm$0.25 & 0.396$\pm$0.047 \\
 KMT-2019-BLG-2783 & 1.31$\pm$0.11 & 1.70$\pm$0.03 & 0.67$\pm$0.11 & 20.18$\pm$0.06 & 15.32$\pm$0.05 & 14.59 & 19.45$\pm$0.08 & 0.387$\pm$0.051 \\
 OGLE-2019-BLG-0249& 2.71$\pm$0.01 & 2.65$\pm$0.02 & 1.10$\pm$0.02 & 18.47$\pm$0.01 & 16.75$\pm$0.04 & 14.62 & 16.34$\pm$0.04 & 2.695$\pm$0.155 \\
 OGLE-2019-BLG-0679& N.A.          & 4.20$\pm$0.03 & 0.94$\pm$0.10 & 17.48$\pm$0.06 & 17.99$\pm$0.05 & 14.41 & 13.90$\pm$0.08 & 7.004$\pm$0.958 \\
 OGLE-2019-BLG-0344& 1.93$\pm$0.04 & 2.29$\pm$0.04 & 0.70$\pm$0.06 & 18.80$\pm$0.03 & 15.99$\pm$0.05 & 14.59 & 17.40$\pm$0.06 & 1.026$\pm$0.090 \\
 KMT-2019-BLG-0304 & N.A.          & 4.58$\pm$0.05 & 0.74$\pm$0.07 & 22.89$\pm$0.36 & 19.07$\pm$0.08 & 14.60 & 18.42$\pm$0.37 & 0.680$\pm$0.128 \\
\hline
 \enddata
 \tablecomments{$(V-I)_{\rm cl,0}=1.06$}
 \label{tab:cmd}
 \end{deluxetable}

%OGLE-2018-BLG-0866& 2.50$\pm$0.07 & 2.38$\pm$0.02 & 1.18$\pm$0.07 & $\pm$ & 16.15$\pm$0.04 & 14.46 & $\pm$ & $\pm$ \\

\begin{deluxetable}{lrrrrrrrr}
\tabletypesize{\footnotesize}
\tablecaption{Bayesian Estimates}
\tablewidth{0pt}
\tablehead{
\multicolumn{1}{l}{Events} &
\multicolumn{1}{c}{$M_{\rm host} (M_{\odot})$} & 
\multicolumn{1}{c}{$M_{\rm planet} (M_{\rm J})$} & 
\multicolumn{1}{c}{$a_{\perp} (\rm au)$} & 
\multicolumn{1}{c}{$D_{\rm L} (\rm kpc)$} & 
\multicolumn{1}{c}{Gal.Mod.}  &
\multicolumn{1}{c}{$\chi^2$} 
}
\startdata
KB190298 $(u_{0}>0)$        &    $0.686_{-0.356}^{+0.395}$  & $1.787_{-0.926}^{+1.029}$  &  $5.763_{-2.603}^{+3.165}$ &  $6.555_{-1.854}^{+1.218}$  & 1.00 & 1.00 \\
KB190298 $(u_{0}<0)$        &    $0.710_{-0.343}^{+0.370}$  & $1.850_{-0.895}^{+0.966}$  &  $5.525_{-2.129}^{+2.741}$ &  $6.921_{-1.344}^{+1.032}$  & 0.72 & 0.95 \\
{\bf Adopted}              &    $0.70\pm 0.37$         & $1.81\pm 0.96$         &  $5.67\pm 2.70$       &  $6.71\pm 1.39$  \\
\hline
KB191216 (Inner, standard) &    $0.525_{-0.247}^{+0.271}$  & $0.134_{-0.063}^{+0.069}$  &  $2.987_{-1.295}^{+1.838}$ &  $3.573_{-1.338}^{+1.836}$  & 1.00 & 1.00\\
KB191216 (Outer, standard) &    $0.521_{-0.239}^{+0.265}$  & $0.127_{-0.059}^{+0.065}$  &  $2.859_{-1.233}^{+1.751}$ &  $3.441_{-1.272}^{+1.782}$  & 0.93 & 0.95\\ 
KB191216 (Inner, $u_{0}>0$) &    $0.413_{-0.172}^{+0.229}$  & $0.101_{-0.042}^{+0.056}$  &  $2.560_{-0.944}^{+1.253}$ &  $2.895_{-0.852}^{+1.134}$  & 0.54 & 1.00 \\ % 0.54
KB191216 (Inner, $u_{0}<0$) &    $0.361_{-0.162}^{+0.252}$  & $0.088_{-0.039}^{+0.061}$  &  $2.323_{-0.896}^{+1.318}$ &  $2.607_{-0.810}^{+1.230}$  & 0.93 & 0.95 \\ % 0.88
KB191216 (Outer, $u_{0}>0$) &    $0.422_{-0.176}^{+0.226}$  & $0.103_{-0.043}^{+0.055}$  &  $2.591_{-0.957}^{+1.262}$ &  $2.931_{-0.864}^{+1.140}$  & 0.64 & 1.00 \\ % 0.64
KB191216 (Outer, $u_{0}<0$) &    $0.370_{-0.169}^{+0.253}$  & $0.090_{-0.041}^{+0.062}$  &  $2.359_{-0.927}^{+1.355}$ &  $2.649_{-0.846}^{+1.272}$  & 1.00 & 0.90 \\ % 0.90
{\bf Adopted}              &    $0.39\pm 0.21$         & $0.094\pm 0.050$      &  $2.44\pm 1.12$        &  $2.74\pm 1.02$  \\ % 296
\hline
KB192783                   &    $0.339_{-0.197}^{+0.254}$  & $1.160_{-0.672}^{+0.871}$  &  $1.850_{-0.930}^{+0.937}$ &  $5.913_{-2.118}^{+1.224}$  & 1.00 & 1.00 \\
{\bf Adopted}              &    $0.34\pm 2.25$         & $1.16\pm 0.77$        &  $1.85\pm 0.93$        &  $5.91\pm 1.67$  \\
\hline
OB190249 (Close, $u_{0}>0$) &    $0.912_{-0.204}^{+0.170}$  & $7.129_{-1.598}^{+1.326}$  &  $1.837_{-0.438}^{+0.457}$ &  $6.333_{-1.026}^{+0.672}$  & 0.89 & 1.00 \\ % 0.89
OB190249 (Close, $u_{0}<0$) &    $0.912_{-0.210}^{+0.165}$  & $7.124_{-1.645}^{+1.288}$  &  $1.840_{-0.429}^{+0.419}$ &  $6.357_{-1.002}^{+0.648}$  & 1.00 & 0.95 \\ % 0.95
OB190249 (Wide,  $u_{0}>0$) &    $0.865_{-0.206}^{+0.179}$  & $7.090_{-1.684}^{+1.467}$  &  $5.201_{-0.782}^{+0.938}$ &  $6.633_{-0.636}^{+0.684}$  & 0.46 & 0.19 \\ % 0.09
OB190249 (Wide,  $u_{0}<0$) &    $0.850_{-0.216}^{+0.163}$  & $6.954_{-1.767}^{+1.330}$  &  $5.079_{-0.848}^{+0.989}$ &  $6.687_{-0.654}^{+0.672}$  & 0.46 & 0.19 \\ % 0.09
{\bf Adopted}              &    $0.91\pm 0.19$         & $7.12\pm 1.47$         & $ 1.84\pm 0.44$       &  $6.37\pm 0.82$  \\    % 202
\hline
OB190679                   &    $0.665_{-0.350}^{+0.407}$  & $3.337_{-1.758}^{+2.039}$  &  $6.988_{-3.034}^{+3.385}$ &  $5.631_{-1.980}^{+1.452}$  & 1.00 & 1.00 \\
{\bf Adopted}              &    $0.66\pm 0.38$         & $3.34\pm 1.90$        &  $6.99\pm 3.21$        &  $5.63\pm 1.71$ \\
\hline
\enddata 
\label{tab:physall}
%\tablecomments{
%}
\end{deluxetable}

\begin{deluxetable}{llccl}
\tablecolumns{5} \tablewidth{0pc}
\tablecaption{\textsc{AnomalyFinder Planets in KMT Sub-prime Fields for 2019}}
\tablehead{\colhead{Event Name} &
\colhead{KMT Name} &
\colhead{$\log q$} &
\colhead{$s$} &
\colhead{Reference} }
\startdata
OB190960$^{a}$ & KB191591 & $-4.87$ & 1.00 & \citet{ob190960} \\
KB191806$^{a}$ & KB191806 & $-4.72$ & 0.94 & \citet{logqlt-4} \\ %AF
KB191367$^{a}$ & KB191367 & $-4.30$ & 0.94 & \citet{logqlt-4} \\ %AF
KB191216$^{a}$ & KB191216 & $-3.62$ & 1.11 & This Work \\ %by-eye Han
KB190298       & KB190298 & $-2.53$ & 1.85 & This Work \\ %AF
KB192783       & KB192783 & $-2.48$ & 0.81 & This Work \\ %AF
OB190468c$^{b}$& KB192696 & $-2.46$ & 0.85 & \citet{ob190468} \\
OB190679       & KB192688 & $-2.36$ & 2.18 & This Work \\ %by-eye Spitzer
OB190362$^{c}$ & KB190075 & $-2.13$ & 0.90 & \citet{ob190362} \\
OB190249$^{c}$ & KB190109 & $-2.12$ & 0.54 & This Work \\ %by-eye Spitzer
OB190299       & KB192735 & $-2.00$ & 0.99 & \citet{ob190299} \\
OB190468b$^{b,c}$ & KB192696 & $-1.98$ & 0.72 & \citet{ob190468} \\
\hline
\hline
OB191470$^{c,d,e}$& KB192814 & $-2.32$ & 1.17 & \citet{ob191470} \\
KB190414$^{c,f}$  & KB190414 & $-2.25$ & 0.35 & \citet{kb190414} \\
KB190304$^{c,f}$  & KB190304 & $-1.84$ & 1.57 & This Work \\
OB190344$^{c,f,g}$& KB190149 & $-1.52$ & 1.70 & This Work \\
\enddata
\tablecomments{Event names are abbreviations for, e.g.,
OGLE-2019-BLG-0960 and KMT-2019-BLG-1216.
a: Inconsequential $s$ degeneracy.
b: 2-planet system.
c: $s$ degeneracy. 
d: large $q$ degeneracy.
e: planet in binary system.
f: 1L2S/2L1S degeneracy.
g: planet/binary degeneracy.
}
\label{tab:all2019events}
\end{deluxetable}

 \begin{deluxetable}{llrrr}
 \tablecolumns{5} \tablewidth{0pc}
 \tablecaption{\textsc{Breakdown of Detections by 4 Questions}}
 \tablehead{\colhead{Q\&A} & 
   \colhead{Year} &
  \colhead{ }   &
  \colhead{ }   &
  \colhead{ }  }
%\hline
 \startdata
Field Type?           & 2018 &  19 &  14\\
Prime, Subprime       & 2019 &  13 &  12\\
& Total &  32 &  26\\
 \hline
Identified By?        & 2018 &  11 &  22\\
AnomalyFinder, Eye    & 2019 &  12 &  13\\
& Total &  23 &  35\\
 \hline
Caustic Crossing?     & 2018 &  16 &  17\\
Yes, No               & 2019 &  13 &  12\\
& Total &  29 &  29\\
 \hline
Image Perturbed?      & 2018 &  17 &  14 &   2\\
Major, Minor, Central & 2019 &   9 &  12 &   4\\
& Total &  26 &  26 &   6\\
 \hline
 \enddata
% \tablecomments{a: M}
 \label{tab:sum}
 \end{deluxetable}

\clearpage

\end{document}